\documentclass[conference,compsoc]{IEEEtran} \IEEEoverridecommandlockouts

\usepackage{cite}
\usepackage{amsmath,amssymb,amsfonts}
\usepackage{graphicx}
\usepackage{textcomp}
\usepackage{url}
\usepackage{tikz}
\usepackage{amsmath}
\usepackage[T1]{fontenc}
\usepackage{graphicx}
\usepackage{xspace}
\usepackage{tcolorbox}
\usepackage{url}
\usepackage{float}
 \usepackage{acronym}
\usepackage{adjustbox}
\usepackage{latexsym}
\usepackage{courier}

\usepackage{algpseudocode}
\usepackage{comment}
\usepackage{longtable}
\usepackage{tabularx}
\usepackage{cancel}
\usepackage{amssymb}
\usepackage{colortbl}
\usepackage{array}
\usepackage{multirow}
\usepackage[colorlinks=true,linkcolor=blue,urlcolor=violet,citecolor=blue]{hyperref}
\usepackage{breakurl}
\usepackage{hyperref}
\usepackage{cleveref}

\usepackage{minted}
\usepackage[linesnumbered,ruled,vlined]{algorithm2e}
 \usepackage{filecontents}
\usepackage{subcaption}
\usepackage{xcolor}

\algnewcommand\algorithmicmatch{\textbf{match}}
\algnewcommand\algorithmiccase{\textbf{case}}
\algdef{SE}[MATCH]{Match}{EndMatch}[1]{\algorithmicmatch\ #1\ \algorithmicreturn}{\algorithmicend\ \algorithmicmatch}%
\algdef{SE}[CASE]{Case}{EndCase}[1]{\algorithmiccase\ #1}{\algorithmicend\ \algorithmiccase}%
\algtext*{EndMatch}%
\algtext*{EndCase}%

\definecolor{Mycolor2}{HTML}{00F9DE}
\renewcommand\fbox{\fcolorbox{black}{lightgray}}
\newcommand{\sysname}[0]{\textsf{DeepProv}\xspace}

\def\BibTeX{{\rm B\kern-.05em{\sc i\kern-.025em b}\kern-.08em
    T\kern-.1667em\lower.7ex\hbox{E}\kern-.125emX}}

\begin{document}

\title{\sysname{}: Behavioral Characterization and Repair of Neural Networks via Inference Provenance Graph Analysis\\}

\author{
\IEEEauthorblockN{
Firas Ben Hmida\IEEEauthorrefmark{1}, 
Abderrahmen Amich\IEEEauthorrefmark{2}, 
Ata Kaboudi\IEEEauthorrefmark{3}, 
Birhanu Eshete\IEEEauthorrefmark{4}
}
\IEEEauthorblockA{
\IEEEauthorrefmark{1}fbhmida@umich.edu, \;
\IEEEauthorrefmark{2}aamich@umich.edu, \;
\IEEEauthorrefmark{3}kaboudi@umich.edu, \;
\IEEEauthorrefmark{4}birhanu@umich.edu
}
\IEEEauthorblockA{
Department of Computer and Information Science, \\
University of Michigan-Dearborn, Michigan, USA
}
}

\maketitle

\begin{abstract}
Deep neural networks (DNNs) are increasingly being deployed in high-stakes applications, from self-driving cars to biometric authentication. However, their unpredictable and unreliable behaviors in real-world settings require new approaches to characterize and ensure their reliability.\\
This paper introduces \sysname{}, a novel and customizable system designed to capture and characterize the runtime behavior of DNNs during inference by using their underlying graph structure. Inspired by system audit provenance graphs, \sysname{} models the computational information flow of a DNN’s inference process through Inference Provenance Graphs (IPGs). These graphs provide a detailed structural representation of the behavior of DNN, allowing both empirical and structural analysis. \sysname{} uses these insights to systematically repair DNNs for specific objectives, such as improving robustness, privacy, or fairness.\\
We instantiate \sysname{} with adversarial robustness as the goal of model repair and conduct extensive case studies to evaluate its effectiveness. Our results demonstrate its effectiveness and scalability across diverse classification tasks, attack scenarios, and model complexities. \sysname{} automatically identifies repair actions at the node and edge-level within IPGs, significantly enhancing the robustness of the model. In particular, applying \sysname{} repair strategies to just a single layer of a DNN yields an average 55\% improvement in adversarial accuracy. Moreover, \sysname{} complements existing defenses, achieving substantial gains in adversarial robustness.
Beyond robustness, we demonstrate the broader potential of \sysname{} as an adaptable system to characterize DNN behavior in other critical areas, such as privacy auditing and fairness analysis. 
\end{abstract}


\section{Introduction}\label{sec: intro}
Deep neural networks (DNNs) are increasingly embedded in high-stakes domains such as self-driving cars~\cite{Tesla-AI,DL-autnonmous17}, biometric authentication~\cite{Facial-rec-attack16}, medical diagnostics~\cite{CheXNet17,DeepCC2019}, and cyber-attack detection ~\cite{Intrusion-ML21,malconv18}.
Their ability to model complex patterns from large amounts of data has fueled their widespread adoption. Yet, this success is shadowed by a persistent challenge: DNNs often exhibit unreliable behavior in real-world environments, jeopardizing the safety and robustness of systems that depend on them. This vulnerability calls for more than just explanations of model decisions: it demands a systematic understanding of how DNNs compute their outputs so as to enable fine-grained behavioral characterization and systematic model repair.

Existing research on model repair, runtime monitoring, and explanation-based attribution has made important strides toward improving the reliability of DNNs. However, these approaches either (i) focus narrowly on weight-level interventions with limited generalizability \cite{Sotoudeh2021ProvableRO,arachane,Liang2023RepairingDN,Schumi_2023,Metarepair}, (ii) monitor runtime behavior only for detection rather than systematic repair \cite{Runtime-Mon-neuron-activation19,Runtime-Mon-In2zUnknown21,Runtime-Mon-Safety20}, or (iii) provide per-instance explanations without capturing inference dynamics at scale \cite{Zhang_2019,wiegreffe2019attention,adebayo2020debugging,10917556,LIME,SHAP,LEMNA}. What remains missing is a model-centric mechanism to capture, analyze, and characterize how information flows through a DNN during inference in a way that facilitates systematic model repair.

We posit that capturing and analyzing inference provenance offers a new foundation for understanding the computational dynamics of a DNN during inference. To achieve this, it is essential to characterize the inference-time behavior of a DNN on a distribution of inputs, using its underlying graph structure, typically a directed acyclic graph (DAG).
The central hypothesis of this paper is that systematically capturing and characterizing inference provenance can provide deep insights into the behavior of a DNN under different conditions, such as benign versus adversarial inputs, members versus non-members of the training data, or across different demographic groups of the training data. Testing this hypothesis requires addressing three key challenges:
$(1)$ defining inference provenance in the context of a DNN: what constitutes inference provenance, and how it can be represented?, $(2)$ capturing and characterizing inference provenance in a DNN: how can we systematically capture and analyze provenance artifacts?, and $(3)$ leveraging inference provenance for model repair: how can provenance-based insights guide systematic model repair for goals such as adversarial robustness, privacy, or fairness?

\textbf{\sysname{} Intuition \& Overview}: To address these challenges, we develop \sysname{}, a customizable system designed to capture and characterize the computational dynamics of a DNN during inference using {\em Inference Provenance Graphs (IPGs)}, and systematically repair the DNN. Inspired by system audit provenance graphs ~\cite{Poirot19,Holmes19,Sleuth17,unicorn-ndss20,whole-sys-prov-usenix15}, \sysname{} models a DNN as a runtime-auditable system, where nodes represent computational units (neurons) and edges encode the flow of information (weights) between nodes. Using the underlying graph representation of a DNN, \sysname{} enables reasoning about the computational information flow during inference, providing a foundation for systematic model repair. Given a pre-trained DNN, for a specified model repair goal (e.g., adversarial robustness), \sysname{} addresses the three challenges as follows:

\textbf{Challenge-1}: {\em inference provenance representation  in the context of a DNN's inference}: An IPG in \sysname{} captures a DAG representation of the computational information flow relationships between input features, internal neuron activations, and the final output of a DNN (\S \ref{subsec:iag-extract}). IPG nodes represent computational units (neurons) and edges carry the information flow (weights) between nodes.

\textbf{Challenge-2}: {\em characterization of inference provenance}: \sysname{} characterizes IPGs using empirical and structural artifacts. Empirical characterization involves activation-based measurements under settings relevant to the repair goal (\S \ref{subsec:empirical-cxn}), while structural characterization learns graph representations of IPGs across settings (e.g., adversarial vs. benign inputs for robustness enhancing repair) (\S \ref{subsec:structural-cxn}). Node- and edge-level IPG attribution is then performed on the graph representation of the IPG to identify IPG nodes and edges that contribute to the DNN's inference provenance.

\textbf{Challenge-3}: {\em analysis to achieve the model repair goal(s)}: \sysname{} enables systematic model repair by identifying and evaluating node- and edge-level repair actions based on IPG-driven insights for a repair goal (\S \ref{subsec:action-gen}, \S\ref{subsec:act-eval}). 

\textbf{Case Studies on Adversarial Robustness}: To validate \sysname{}, we conducted extensive case studies using adversarial robustness as a model repair goal. Across diverse types of attacks~\cite{PGD,APGD-DLR,andriushchenko2020square,uesato2018adversarial,FGSM} and model architectures (e.g., DenseNN, CNN, ResNet) on image classification and malware detection, \sysname{} consistently  distinguishes between benign and adversarial inference behavior. Notably, applying \sysname{} repair actions to just one layer of the studied models improved adversarial accuracy by $\approx 55\%$ on average. In some cases, deactivating a single node improves robustness by 20\%, highlighting the power of IPG-driven model repair. Moreover, \sysname{} enhanced state-of-the-art adversarial robustness methods \cite{bartoldson2024adversarialrobustnesslimitsscalinglaw}, demonstrating its ability to complement existing defenses.
Beyond robustness enhancing model repair, in \S \ref{subsec:adaptability}, we also demonstrate \sysname{}'s potential for other repair goals, such as privacy-enhancing and fairness-enhancing model repair, suggesting its broader applicability. 

\textbf{Note on Scope}: \sysname{} is a general-purpose system for inference-time behavioral analysis and repair. In our case studies in \S \ref{sec: eval}, we use adversarial robustness as the repair goal to demonstrate its effectiveness.

\textbf{Contributions}: \sysname{} is the first system to introduce IPGs to characterize DNNs and enable systematic model repair. We make the following contributions:

$\bullet$ A task- and model-agnostic system for IPG-driven characterization of a DNN’s computational dynamics, grounded in empirical inference provenance (what neurons fire) and structural inference provenance (which paths within an IPG are critical).

 $\bullet$ Empirical validation of IPG-based characterizations for systematic model repair, demonstrated through extensive case studies on adversarial robustness as a model repair goal, and demonstration of \sysname{}'s adaptability to privacy auditing and fairness analysis.

$\bullet$ Reproducible implementation of \sysname{} is available at: {\url{https://github.com/um-dsp/DeepProv}}.

\section{Background}\label{sec: bground}
\subsection{ML Setting and Notation}
\textbf{Training}: Our ML setting is a  supervised ML for a DNN classifier trained over a set of labeled training samples $\mathcal{D} = (X_i,y_i): i: 1...n$ where $X_i$ is a $d$-dimensional feature vector and its corresponding label $y_i \in Y$  is a $k$-dimensional output space. Training a model $f_{\theta}$ parameterized by a parameter vector $\theta$ aims to minimize the loss $J(\theta) = \frac{1}{n}\sum_{1=1}^{n}\mathcal{L}(f_{\theta},X_i,y_i)$. The loss minimization problem is typically solved using mini-batch-based stochastic gradient descent (SGD)~\cite{SGD98} via an iterative update of $\theta$ as: 
  \begin{equation}\label{eq:gradient-descent}
     \theta_{i+1} = \theta_{i} - \eta \cdot\sum\limits_{(x,y)\in B} \Delta_{\theta}\mathcal{L}(f_{\theta_{i}}(x),y)
  \end{equation}

 \noindent where $\Delta_{\theta}$ is the gradient of $J$, $\eta$ is {\em learning rate}, and $B$ is the mini-batch size. For classification tasks, the typical loss function is the cross-entropy loss computed as: $\mathcal{L}(f_{\theta}(x),y) = -log(f_{\theta}(x)_{y})$.
 
 \textbf{Inference}: An inference on an input $x$ is  $ y = f_{\theta}(x) = argmax(y \in Y$), where $y$ is a $k$-dimensional vector and each dimension represents the probability of $x$ belonging to the corresponding class. For simplicity, from now on we will use $f$ to refer to $f_{\theta}$. 

\subsection{Case Study Attacks: Adversarial Examples}
Because we instantiate \sysname{} under two settings (benign and adversarial) with adversarial robustness as repair goal, here we briefly introduce adversarial examples. 
Algorithmic details of the various attacks we used for our case studies appear in the Appendix (\ref{subsec:attacks}).

Given an input $x$ correctly classified by a model $f$ to its true class $y_{true}$, an adversary performs a non-random perturbation $\delta$ to obtain an adversarial example $x' = x+\delta$ such that $||\delta|| < \epsilon$, where $||.||$ is a distance metric (e.g., $L_{1}$, $L_{2}$, $L_{\infty}$) and $\epsilon$ is the maximum allowable perturbation. The attack is successful when $f$ misclassifies $x'$ as $f(x') =y' \ne y_{true}$ for the case of non-targetd attacks. Misclassification can be targeted ($f(x')=y_{target} \ne y_{true}$) or untargeted ($f(x')=y' \ne y_{true}$). Adversarial examples can be crafted in a {\em white-box} setting (e.g., ~\cite{FGSM,BIM,PGD,CW}) where the adversary has access to model details (e.g., gradient of loss function) or {\em black-box} setting (e.g., ~\cite{MIM,HSJA20,uesato2018adversarial}) where the adversary typically has only oracle access to the model.
In our case studies, we use both white-box and black-box attacks on DNNs trained for image classification and malware detection. 

\subsection{Graph Neural Networks} \label{subsec:GNNs}
\textbf{Overview:}
Graph Neural Networks (GNNs) are a class of DNNs designed to perform inference on data structured as graphs. Unlike traditional DNNs that process Euclidean data (e.g., images, text, or sequential data), GNNs are well suited for learning over graph representations, where entities (nodes) and their relationships (edges) provide crucial contextual information. GNNs facilitate inference at the node-level, edge-level, and graph-level. The core mechanism of GNN representation learning is based on \emph{message passing}, an iterative process where nodes aggregate information from their neighbors to refine their feature representations. Given a graph $G = (V, E)$ with nodes $V$ and edges $E$, message passing follows these steps:(1) each node $i$ starts with an initial feature vector $\mathbf{x}_i$; (2) for each edge $(j, i) \in E$, the neighboring node $j$ sends a message $\mathbf{m}_{j \to i}$ to node $i$, (3) node $i$ aggregates incoming messages using an aggregation function $\sigma$ (e.g., \textit{mean}, \textit{sum}, \textit{max}), and (4) the aggregated message is used to update the node’s representation.

Formally, the message passing function is defined as:
\begin{equation}
    \mathbf{m}_{j \to i} = \text{message}(\mathbf{x}_j, \text{edge\_index}, \text{size}, \sigma)
\end{equation}
where: $\mathbf{x}_j$ represents the feature vector of node $j$, \texttt{edge\_index} encodes the graph’s connectivity, \texttt{size} is the total number of nodes in the graph, and $\sigma$ is an aggregation function that combines messages from multiple neighbors. By iteratively propagating information through message passing, GNNs generate node embeddings that incorporate both local and global structures, enabling powerful learning capabilities for a wide range of applications.

\textbf{Homogeneous vs. Heterogeneous GNNs:} Homogeneous GNNs operate on graphs where all nodes and edges share the same type and feature dimensions. They are well-suited for dense-layer DNNs, where each node represents an activation value and maintains a uniform feature representation across the graph. Homogeneous GNNs facilitate effective information propagation in graphs where all entities share a common semantic meaning, such as citation networks or social graphs.
Heterogeneous GNNs are designed for graphs with multiple node and edge types, where nodes may have varying feature dimensions across layers. A key application of heterogeneous GNNs is in convolutional neural networks (CNNs) with multi-dimensional feature maps, where nodes represent different feature channels extracted from convolutional layers.

In the structural characterization of IPGs described in \S \ref{subsec:structural-cxn}, we use GNNs to enrich the attribution of inference results by incorporating relevant subgraph structures captured via IPGs.
\section{Related Work}\label{sec: related}
We position \sysname{} with respect to previous work on: model repair, runtime monitoring, and explanation-based inference attribution.

\textbf{DNN Repair Methods.} Recent approaches to model repair aim to correct faulty behavior in DNNs without requiring full retraining. For instance, \cite{Sotoudeh2021ProvableRO} propose layer-wise fine-tuning to improve model robustness under input perturbations. However, their repairs are limited to single-layer interventions and often degrade performance on unseen inputs. In contrast, \sysname{} demonstrates that multi-layer repair—guided by IPGs can yield superior generalization and robustness. Archane \cite{arachane} adjusts DNN weights post-training to fix targeted misclassifications (e.g., correcting class A → B errors). While effective in narrow settings, it suffers from scalability issues in high-dimensional search spaces and lacks generalizability across broader error types. BIRDNN \cite{Liang2023RepairingDN} selectively retrains components of a model based on runtime behavior states, \cite{Schumi_2023}  handles errors without assessing overall performance, and MetaRepair \cite{Metarepair} relies on prior repair experiences for training-based fixes. \emph{In contrast, \sysname{} does not modify model weights or require retraining; instead, it operates directly on activation values to repair defective nodes while improving overall performance of the model}.

\textbf{Run-time Monitoring of Models.} Another related thread explores the use of runtime signals to monitor model behavior.
Chen et.al \cite{Runtime-Mon-neuron-activation19} monitor neuron activation patterns to detect inputs that may be misclassified, comparing incoming activations with those from the training distribution. Lukina et. al \cite{Runtime-Mon-In2zUnknown21} introduce human-in-the-loop monitoring for novel class detection, relying on model confidence scores and thresholding. Ferreira \cite{Runtime-Mon-Safety20} develop runtime monitors for safety in perception systems, especially in autonomous driving. While these methods focus on detecting anomalies or distribution shifts, \emph{\sysname{} goes further by using runtime activation data to enable systematic repair, not just detection. \sysname{} departs from these approaches in that it is (a) domain- and attack-agnostic and (b) combines empirical and structural inference provenance for systematic model repair}.
\sysname{} focuses on the runtime behavior of models by carefully recording the activation values for each layer at inference time. Additionally, \cite{Runtime-Mon-neuron-activation19} not only encompasses the entire model's activation patterns, but also ensures that pattern characterization is conducted at a high level. In contrast, \cite{Runtime-Mon-In2zUnknown21} provides a quantitative, rather than qualitative, analysis of inputs to monitor the ML system, presenting a distinct evaluation method. \sysname{} extends its analysis by examining each node in the DNN and characterizing them based on their activation values, offering a comprehensive runtime characterization.
In terms of detection, these methods exhibit variations in their objectives; their primary objective is to improve the accuracy and precision of the model. \emph{On the other hand, \sysname{} aims to identify DNN components (neurons and weights) that should be repaired to enhance the model's robustness for a given goal}.

\textbf{Explanation-based Inference Attribution.}
A growing body of work explores interpretability and explanation in machine learning.
The Manifold framework \cite{Zhang_2019} offers a visual analysis technique to aid interpretation, debugging, and comparison of various ML models. Another line of research \cite{wiegreffe2019attention}, proposes alternative methods to evaluate the explanatory power of attention mechanisms in recurrent neural network models: it emphasizes the importance of defining explanation and suggests novel tests, including adversarial attention training, to assess the utility of attention for explainability. Adebayo et.al \cite{adebayo2020debugging} categorize bugs into data, model, and test-time contamination, evaluating several explanation methods' ability to address these issues. Collectively, these studies underscore the complexity of developing universally applicable explanation methods, suggesting a concrete approach to leverage such tools for enhancing model transparency and robustness.
\emph{Unlike this body of work that uses per-example attribution to explain single predictions, \sysname{} uses attribution across a population of settings (e.g., benign and adversarial IPGs)  to capture and analyze provenance of model inference}.

Zhang et al. \cite{10917556} propose Critical Inference Graphs (CIGs) that identify subgraphs of a DNN critical to specific predictions. Their approach uses Layer-wise Relevance Propagation (LRP) to identify important nodes for benign inputs and builds one-class classifiers to detect anomalies. However, this method ignores nodes deemed irrelevant by LRP and focuses solely on classification, missing broader patterns in runtime behavior. \emph{In contrast, \sysname{} includes all active nodes in its IPGs, regardless of their individual attribution score, allowing a comprehensive empirical and structural characterization of the behavior of a DNN in diverse settings. This whole-graph perspective is critical to understanding the full range of inference dynamics and enables interventions that generalize across model repair goals such as robustness, privacy, and fairness.} 
 \section{\sysname{} }\label{sec: approach}

\begin{figure*}[t!]
    \centering   
    \includegraphics[scale=0.47]{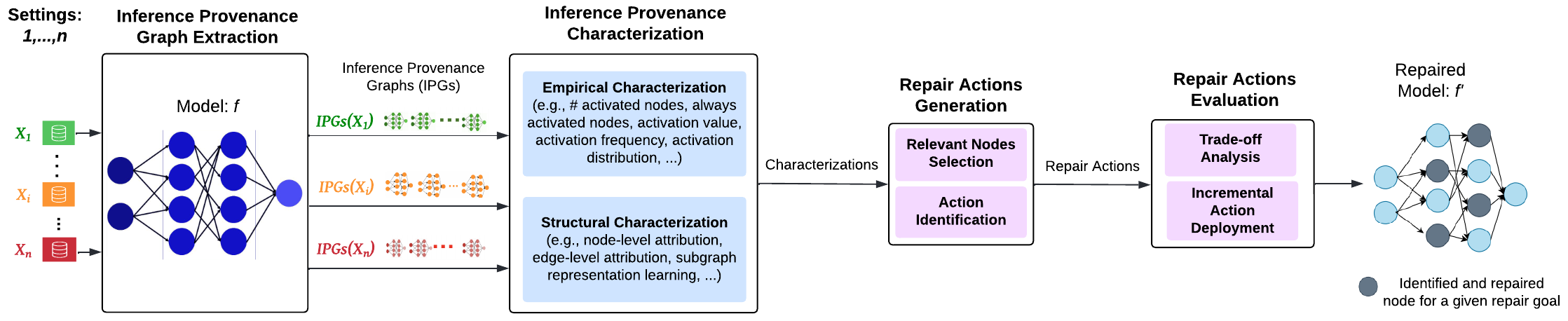}
    \caption{\sysname{} system overview.}
    \label{fig:sys}
\end{figure*}

\subsection{Overview}
Figure \ref{fig:sys} presents an overview of the \sysname{} system. We begin with a DNN $f$, trained on a dataset $X_{train}$ for a classification task. The behavior of $f$ is evaluated across $n$ different settings, $X_1, ..., X_n$, each corresponding to a particular setting of the input distribution such as benign, adversarial, or demographically grouped inputs (e.g., male vs. female). The DNN is represented as a directed acyclic graph (DAG), where nodes represent computation units (e.g., neurons or layers) and edges represent data flow. 

\textbf{Step 1: Inference Provenance Extraction (\S \ref{subsec:iag-extract})}. \sysname{} begins by extracting IPGs for each individual sample in each setting. This results in $n$ sets of IPGs, one for each setting: $IPGs(X_{1}), ..., IPGs(X_{n})$. Each IPG is a subgraph of the DAG representation of $f$, which captures the activation graph instance of $f$ on an input. These IPGs represent the best approximation of the runtime behavior of $f$ on inputs that span multiple settings.\\
\textbf{Step 2: Inference Provenance Characterization (\S \ref{subsec:empirical-cxn}, \S \ref{subsec:structural-cxn})}. Next, \sysname{} characterizes the runtime behavior of $f$ using two complementary schemes: empirical inference provenance (\S \ref{subsec:empirical-cxn}) and structural inference provenance (\S \ref{subsec:structural-cxn}). The empirical inference provenance captures quantitative signals such as neuron activation magnitudes, statistical trends between input groups, and other value-based indicators of model behavior. Structural inference provenance focuses on the topological aspects of the IPGs (e.g., which paths are activated, how node/edge connectivity varies across settings, and what structural deviations are associated with problematic inferences). This dual analysis allows \sysname{} to holistically characterize how and why the DNN behaves differently across the input settings of interest.\\
\textbf{Step 3: Repair Actions Generation  (\ref{subsec:action-gen})}.
Based on inference provenance characterization, \sysname{} identifies \emph{repair actions} that modify or deactivate nodes or edges to mitigate undesirable model behaviors. Importantly, \sysname{} does not modify the model weights or retrain the model. Instead, it generates candidate repairs at the activation level, making the repair actions lightweight and non-invasive.\\
\textbf{Step 4: Repair Actions Evaluation (\S \ref{subsec:act-eval})}. The generated candidate repair actions are evaluated in terms of their \emph{feasibility} (do they preserve the core functionality of the DNN?) and \emph{effectiveness} (do they improve performance for the repair goal?). Evaluation is guided by a metric we refer to as \emph{Model Repair Effectiveness (MRE)}, which is tailored to the specific repair goal. 

\textbf{A Customizable System}. \sysname{} is model- and task-agnostic, provided that $f$ can be represented as a DAG. The specific repair goal determines the number and type of settings (e.g., different input distributions), the provenance patterns of interest, and the appropriate MRE metric. When the goal is \emph{robustness-enhancing model repair}, the focus of our case studies in \S\ref{sec: eval}, we have two settings ($n = 2$): benign and adversarial. \sysname{} extracts, characterizes, and analyzes IPGs for benign and adversarial inputs, using $f$'s accuracy under attack as the MRE metric. For \emph{privacy enhancing model repair}, \sysname{} operates on IPGs of members and non-members of $f$'s training data in the context of privacy-motivated attacks such as membership inference~\cite{MemInfeAttack}. Similar to the robustness-enhancing repair goal, we have two settings. However, the MRE metric is membership inference attack accuracy. When \emph{fairness-enhancing repair} is the goal, \sysname{} extracts IPGs for data spanning multiple protected classes (e.g., gender or race). The MRE metric could be the demographic parity score or other relevant fairness metrics such as error parity. Depending on the protected class, the number of settings could vary (e.g., for race: black, white, Asian, Hispanic).
Across these model repair goals, types and number of settings, specifics of the repair goals, and associated metrics for each repair goal can be customized to meet the model deployer's needs, making \sysname{} customizable to various contexts and requirements.

\subsection{IPG Extraction}\label{subsec:iag-extract}

\begin{figure}[t!]
  
    \centering   
    \includegraphics[scale=0.4]{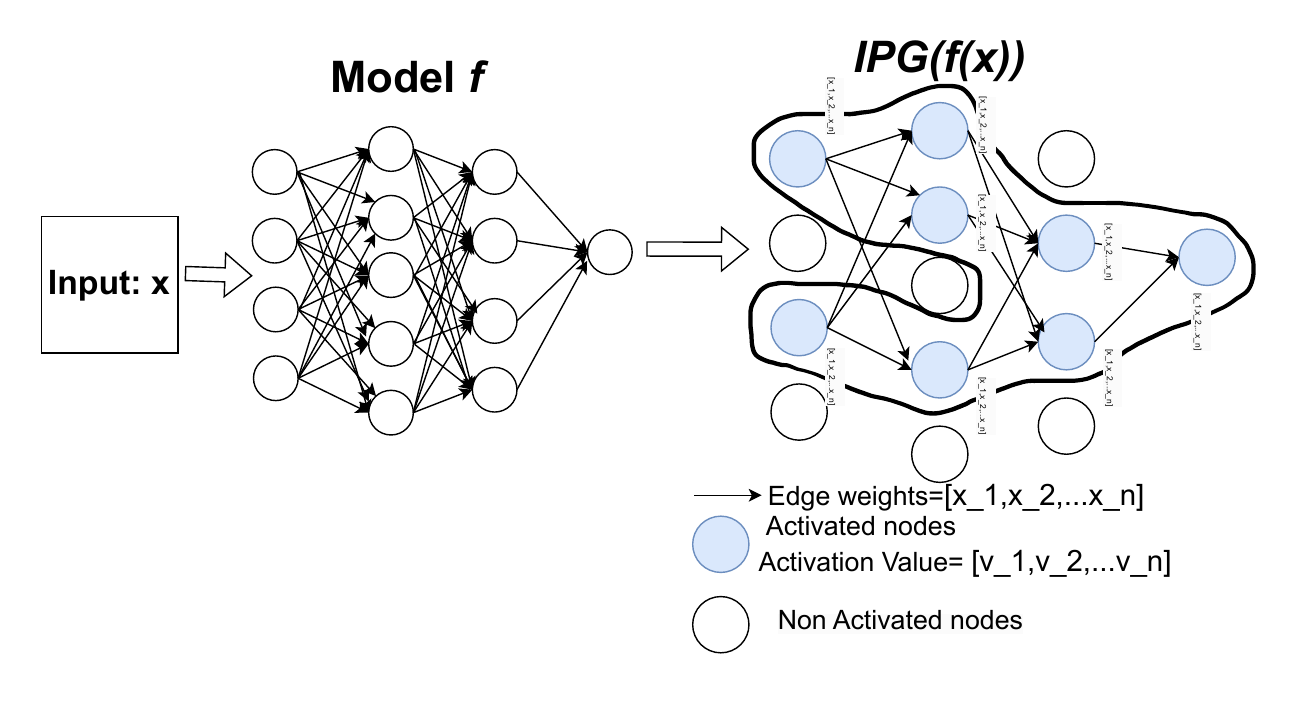}
    \caption{An illustration of IPG extraction.}
    \label{fig:IPG}
\end{figure}

 IPG extraction in \sysname{}  is inspired by audit log-based provenance graphs in Systems/network security. In operating systems (OSs), audit logs record various events such as file accesses, process creations, and network connections, providing a detailed history of system activities. Provenance graphs (PGs) represent the lineage of data and processes, capturing causal links and information flow dependencies in a graph structure to narrate the history of system execution~~\cite{Sleuth17,Holmes19,Poirot19,Prov-SOK23,unicorn-ndss20,whole-sys-prov-usenix15}. Modern OSs use auditing hooks and system call tracing to monitor interactions between active system entities (e.g., processes) and passive system  entities (e.g., files, sockets). By systematically connecting system entity nodes via system/application-level event edges, PGs provide a comprehensive view of system events, facilitating precise threat detection and forensic analysis through causal analysis of complex intrusion behaviors.

\textbf{IPGs}:
Similarly, in DNNs, we model inference provenance via IPGs to capture the lineage of computations during inference. IPGs depict how input data propagates through the input$\rightarrow$hidden$\rightarrow$output layers of the DNN to produce inference output. Each node in an IPG represents an intermediate computation or transformation, while an edge represents computational information flow between nodes/layers. 
For example, in convolutional neural networks (CNNs) trained for image classification, the IPG captures how pixels of the input image are processed through convolutional layers, pooling layers, and fully connected layers to ultimately result in the predicted class in the final layer.

Once a model is trained and learned parameters are fixed, given an input, the underlying model structure and the learned parameters dictate the internal computations to produce an inference result. Because the model itself is a DAG, to extract the IPG of an input, we leverage the DAG representation as an abstraction of the model's {\em computational dynamics}, akin to a system's runtime behavior. \\
\indent Figure \ref{fig:IPG} illustrates \sysname{}'s IPG extraction scheme. Given an input $x$ and a model $f$, when $f$ computes an inference result on $x$, \sysname{} extracts node {\em activation values} and associated edge weights for each activated node at each layer. A node is considered activated based on its activation value, typically determined by the activation function used. 

\indent The highlighted (activated) nodes and their edges capture the IPG ($f(x)$) of our illustrative example in Figure \ref{fig:IPG}. By definition, IPG ($f(x)$) is a subgraph of $f$'s underlying DAG representation. For each activated node, we store an {\em activation value} that captures the computational output produced by the node. The dimension of activation value depends on $f$'s model architecture and layer parameters. For dense and linear layers, the activation value is a single scalar. For more complex architectures such as CNNs, the activation value's dimension is dictated by the layer parameters. For instance, a Conv1D layer parameters include $C_{out}$, which represents the number of nodes in the layer and $1\times L_{out}$, which denotes the shape of the node activation values. 
\scalebox{0.8}{$
  L_{\text{out}}
  = \left\lfloor
      \frac{L_{\text{in}} + 2\,\text{padding}
            - \text{dilation}\,(\text{kernel\_size}-1) - 1}
           {\text{stride}} + 1
    \right\rfloor
$}

For a Conv2D layer, the dimension is defined by $C_{\text{out}}$ as the number
of channels and $H_{\text{out}}\times W_{\text{out}}$ as the spatial activation
shape. The Conv2D layers' main parameters are
$\big(C_{\text{in}}, H_{\text{in}}, W_{\text{in}}, C_{\text{out}}\big)$,
where $\big(C_{\text{in}}, H_{\text{in}}, W_{\text{in}}\big)$ match the
previous layer's output.%

\scalebox{0.75}{$
\begin{aligned}
  H_{\text{out}} &=
    \left\lfloor
      \frac{H_{\text{in}} + 2\,\text{padding}[0]
            - \text{dilation}[0]\big(\text{kernel\_size}[0]-1\big) - 1}
           {\text{stride}[0]} + 1
    \right\rfloor \\
  W_{\text{out}} &=
    \left\lfloor
      \frac{W_{\text{in}} + 2\,\text{padding}[1]
            - \text{dilation}[1]\big(\text{kernel\_size}[1]-1\big) - 1}
           {\text{stride}[1]} + 1
    \right\rfloor
\end{aligned}
$}

To determine whether or not a node is activated, Zhang et al. \cite{10917556} utilize Layer-wise Relevance Propagation (LRP) for subgraph extraction. \sysname{} relies on a binary decision criterion, by determining whether a node is activated or not when signals traverse particular layers. Our approach deliberately considers all signals within the DNN to capture and analyze both empirical and structural values of each node to inference. This comprehensive strategy allows \sysname{} to account for the nuanced dynamics of node activity, ensuring a thorough examination of the DNN's behavior across different settings.

By systematically capturing and analyzing these IPGs, 
\sysname{} provides details of the DNN's computational dynamics, which is essential for effective model repair and enhancement.Based on our IPG-based modeling of inference provenance, \sysname{} can be generalized to support any type of DNN that is amenable to a representation of multidimensional matrices of parameters (i.e., weights). Each DNN consists of a sequence of layers, where each layer possesses its own set of parameters and dimensional properties. Consequently, each layer can be modeled as several superposed nodes as illustrated in Figure \ref{fig:IPG} within a graph, the edges representing the connections to its preceding and subsequent layers. Once this computational graph is constructed, \sysname{} can seamlessly integrate with any graph-based monitoring system designed to track and analyze the execution of graph computations.

\subsection{Empirical Characterization of IPGs}\label{subsec:empirical-cxn}
DNNs are highly complex, both in their structural design and computational behavior. Due to this complexity, it becomes essential to compute statistical metrics and perform empirical analyses to gain insights into the overall inference behavior of these models. Understanding the "big picture" of how DNNs function at scale is what motivates the need for systematic empirical investigation.
Empirical inference provenance is computed based on quantities aggregated by traversing {\em activation values} at the overall IPG, at each layer, or each node of the IPG. To this end, we capture a suite of IPG-based metrics to quantify the inference provenance on $IPGs(X_{1}), ..., IPGs(X_{n})$. Across these $n$ settings of $f$, we capture empirical inference artifacts such as the number of activated nodes, always activated nodes, average activation value per node, average activation frequency per node, and activation value differences among the $n$ settings of $f$. Below, we detail the characterization metrics we use as empirical proxies to capture $f$'s inference provenance.

\textbf{Average Node Activation Value.}
Beyond activation status, another characteristic of IPGs is the variation in activation values of nodes across different settings of $f$. 
To quantify this empirical provenance, we compute the average activation value of each node across all settings. This metric allows us to compare the average activation value of a given node of an IPG across $n$ settings. Establishing the per-node average activation values, along with other metrics, can help identify outlier IPGs within a setting (e.g., benign IPGs) or across settings (e.g., between benign and adversarial IPGs).

\textbf{Node Activation Frequency.}
To understand how often a node is activated across multiple IPGs, we track the activation frequency of each node under each setting. This tracking enables us to infer whether transformations on inputs based on a setting (e.g., adversarial perturbations)  manifest at the node activation frequency level. If so, such empirical correlation can capture the cause-effect link between setting-relevant manipulations and activation frequency. Considering our previous example model repair goal of adversarial robustness, node-level activation frequencies can serve as reliable empirical provenance to determine set of nodes that are frequently activated on adversarial data while they are less activated on benign data or vice-versa.

\begin{figure*}[!t]
    \centering   
    \includegraphics[scale=0.33]{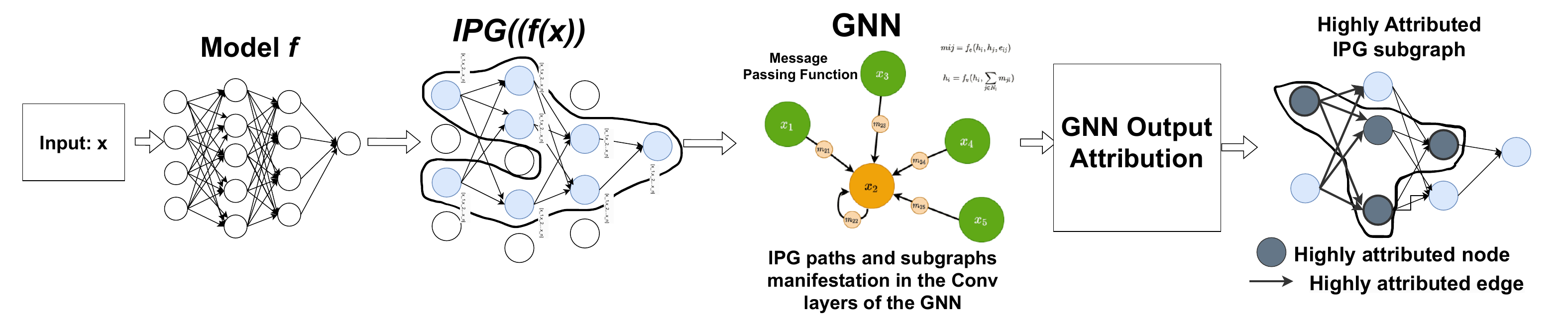}
    \caption{An illustration of \sysname{}'s structural characterization of IPGs.}
    \label{fig:strutural explan}
\end{figure*}

\subsection{\textbf{Structural Characterization of IPGs}}\label{subsec:structural-cxn}
While the empirical IPG artifacts captured in \S \ref{subsec:empirical-cxn} provide quantitative insights into inference provenance, they are inherently oblivious to the underlying graph structure of IPGs. This oversight can lead to gaps in inference characterization. We note that relying on graph properties such as connectedness, degree, and centrality is insufficient to fill this characterization gap.
Therefore, to achieve a comprehensive characterization, we develop a complementary characterization scheme based on the graph structure of IPGs. For this purpose, we find GNNs suitable in two ways: ($1$) GNNs allow us to analyze IPGs locally through message passing functions (see \S \ref{subsec:GNNs}) and globally at the graph level, and ($2$) GNNs can be trained on IPG data from multiple settings to learn graph representation of inference provenance dynamics derived from IPG structures. 

Figure \ref{fig:strutural explan} illustrates \sysname{}'s structural characterization using a single input $x$. Given an IPG on input $x$ computed as $IPG(f(x))$, a GNN trained on settings of interest (e.g., benign and adversarial IPGs) predicts the label of an IPG as $GNN(IPG(f(x)))$. The GNN, built upon convolutional layers facilitating message passing between nodes, naturally models structural features of IPGs by considering paths between nodes, with each path representing a message conveyed to the destination node. The GNN structure is dictated by the DNN architecture that we monitor. For instance, for the DenseNet DNN architecture, we employ a homogeneous Graph Neural Network (GNN) with two convolutional layers and one linear layer. In contrast, for the CNN-ResNet DNN architecture, we utilize a heterogeneous GNN with two heterogeneous convolutional layers, where the number of linear layers corresponds to the number of distinct dimensions in the DNN layers.
Our selected GNN architectures achieved 99\% accuracy in classifying different IPG settings. Furthermore, our approach leverages these GNNs to interpret IPGs using explanation methods for GNNs, identifying specific graph structures that contribute most significantly to the GNN’s prediction of a particular class label. Information propagates through multiple convolutional layers of the GNN from nodes not directly connected to the target nodes. Importantly, the GNN learns IPG subgraphs to identify correlations between IPG substructures and their inference results (e.g., benign or adversarial). 

Once we obtain $GNN(IPG(f(x)))$, next we leverage inference attribution~\cite{kokhlikyan2020captum} to explain a GNN's prediction of an IPG. By doing so, we now have proxies for structural characterization of IPGs as generalized through IPG graph representation learning. Across settings, by extracting the structural attributions of a GNN's predictions on IPGs, we identify nodes and edges that are exclusively relevant to a particular setting (e.g., adversarial or benign). Additionally, we examine the typical activation behavior (e.g., activation value) of highly attributed nodes for IPGs in one setting (e.g., benign) compared to other settings (e.g., adversarial), and how these differences among settings guide model repair goals (e.g., robustness enhancement).

Through GNN prediction attribution, we fine-tune our characterization approach to focus on highly attributed IPG features, i.e., IPG subgraphs that contain nodes and edges with strong attribution signals. Such subgraphs exhibit different patterns when the input to $f$ is in a nominal setting (e.g., benign) versus other settings (e.g., adversarial). The right-most graph in Figure \ref{fig:strutural explan} illustrates how a GNN prediction attribution results in a subgraph of the IPG, highlighted via {\em highly attributed nodes and associated edges} of the IPG. Combining empirical and structural inference provenance characterization, \sysname{} offers comprehensive IPG-based inference provenance for targeted model repair. In \S \ref{sec: eval}, we evaluate the effectiveness of our holistic IPG-driven characterization with adversarial robustness enhancement as the model repair goal.

\subsection{Repair Actions Generation}\label{subsec:action-gen}
Leveraging \sysname{}'s IPG-driven characterization, we now focus on how to guide a model deployer generate model repair actions given a repair goal (e.g., enhancing robustness to adversarial attacks, ensuring privacy and fairness). 
This involves first \textit{identifying relevant nodes to act on} and then \textit{selecting suitable actions for each node}. The choice of nodes to act on and repair actions for each node depends on the model repair goal. For instance, in the context of adversarial robustness goal, if we find that a node is only activated in a particular undesirable setting (e.g., when an input is adversarial or when privacy is compromised), a repair action might be to nullify its activation. We use Algorithm \ref{alg:Action-gen} to explain how to identify model repair actions. The algorithm is designed to be instantiated for various repair goals such as adversarial robustness, privacy, and fairness by abstracting the concepts of desired and  undesired settings. It ensures that the technical details remain applicable across different model repair goals.

\textbf{Relevant Nodes Identification:} As shown in Algorithm \ref{alg:Action-gen}, the node selection mechanism depends on activation frequency ($Freq$) and activation value ($Act$) of IPGs across settings. To account for the structural characterization, we also examine whether a node has a high-attribution value ($Att_N$) or is linked to a highly attributed edge ($Att_E$) in different settings. We first look for nodes to nullify ($N_n$), which are only activated in undesired settings (line 10). These nodes are triggered only when $f$ behaves undesirably (e.g., under adversarial input or privacy breach). After that, we consider nodes influential in the undesired settings (i.e., with high attribution scores in the GNN decision). These are values of attribution scores obtained from the GNN explainer for every node in an IPG.  We set a threshold based on empirical observations of the distribution of attribution values of nodes, and based on this threshold we consider a node as an influential node. The threshold is fixed by the model deployer based on a box plot of attributions across all nodes and edges. These values represent attributions values for each node or edge.  If such nodes are not influential in desired settings (e.g., benign, privacy-preserving), we call them \textit{undesirable nodes} and select them as nodes to act on with priority ($N_p$). Otherwise, they are deemed regular candidate nodes ($N_r$) for actions, given their significance in both undesired (e.g., adversarial) and desired (e.g., benign) settings. In general, we focus on nodes with different activation values across settings (line 13).

\begin{algorithm}[!t]
\caption{Repair Actions Generation}
\label{alg:Action-gen}
\scriptsize
\KwResult{$N_{n}$: nodes to nullify, $N_{p}$:  nodes to prioritize for repair actions, $N_{r}$: regular candidates}
\SetAlgoLined

\textbf{Input:}\\
$Freq$: Node Activation Frequency;\\
$Act$: Node Activation Value;\\
$Att_N$: Nodes with high attribution (can be None);\\
$Att_E$: Edges with high attribution (can be None);\\
\textbf{Hyperparameters:}\\
$p$: norm order, can be ($1$, $2$, or $\infty$);\\
$\alpha$: action sensitivity;

\tcp{Select repair relevant nodes for actions}
\ForEach{$\mathrm{Node} \in \mathrm{Nodes}$}{
  \tcp{Nodes that are only activated in undesirable settings}
  \If{$Freq[\mathrm{Desired}][\mathrm{Node}] = 0 \& \& Freq[\mathrm{Undesired}][\mathrm{Node}] > 0$}{
    Insert($\mathrm{Node}$, $N_n$)\;
  }

  \tcp{ Calculate activation value difference}
  $\Delta_{\mathrm{Undesired}-\mathrm{Desired}}
    =  \& \& Act[\mathrm{Undesired}][\mathrm{Node}]
      - Act[\mathrm{Desired}][\mathrm{Node}] \rVert_p$\;

  \If{$\Delta_{\mathrm{Undesired}-\mathrm{Desired}} > 0  \& \& \mathrm{Node} \in Att_N[\mathrm{Desired}]$}{
    \eIf{$\mathrm{Node} \in Att_N[\mathrm{Undesired}]$}{
      Insert($\mathrm{Node}$, $N_p$)\;
    }{
      Insert($\mathrm{Node}$, $N_r$)\;
    }
  }
}

\ForEach{$\mathrm{Node} \in N_p \cup N_r$}{
  \tcp{Next lines are explained in the text}
  $\beta=\operatorname{Avg}\!\left(\left|
    Act[\mathrm{Desired}][\mathrm{Node}]
    - Act[\mathrm{Undesired}][\mathrm{Node}]
  \right|\right)$\;

  \tcp{Perform repair actions given an IPG of
sample $x$}
  $\delta = Act_x[\mathrm{Node}]
    - \operatorname{Reference\_Dist\_Agg}\!\left(
        Act[\mathrm{Desired}][\mathrm{Node}]
      \right)$\;

  \If{$\beta \ge \delta$}{
    \tcp{Update activation}
    $Act_x[\mathrm{Node}]
      \gets Act_x[\mathrm{Node}]
      - \alpha\ . \delta\ . \operatorname{sign}(\delta)$\;
  }
}

\end{algorithm}

\textbf{Node-Level Repair Actions Identification:} The next challenge is identifying suitable actions for each selected node, given an input $x$. Generally, we want $f$ to behave similarly across settings (e.g., on adversarial data and benign data). Thus, actions are designed to guide the activation values of each selected node to mimic the desired setting's behavior. For example, in our case studies in \S \ref{sec: eval}, actions are aimed at guiding activation values of each selected node to mimic $f$'s activation behavior in a benign setting. 

More precisely, for each node we analyze its activation values distribution in the desired setting (e.g., benign IPGs) to calculate a reference activation value to be used in actions with a sensitivity $\alpha$ (line 24). This \textit{reference value} is the output of $Reference\_Dist\_Agg(...)$ in line 24 (details next). The parameter $\alpha$ ranges from 0 to 1 and it measures how close the new activation value is to the reference value, compared to its original value. In other words, $\alpha$ determines to what extent we want to make changes on the current activation value $Act_x$. When applying actions to bring activation values to a nominal (e.g., benign) setting, the closer alpha is to 1, the closer is the changed activation value to the nominal (e.g., benign) setting. For nodes in $N_n$, the action is nullification, regardless of their activation distribution in the desired (e.g., benign) setting.


\textbf{Reference Distribution Aggregation:} For a node \texttt{N} selected for action, we analyze its activation values in the desired setting's IPGs (e.g., on benign IPGs). Specifically, we compute a statistical aggregation ($Reference\_Dist\_Agg(...)$: line 24) across all samples in the desired setting to fix a baseline activation pattern (i.e., reference value), given a node \texttt{N}. A common approach for aggregation is to use the {\em mean} value, assuming a dense distribution of activation values. We assess the distribution density by evaluating the standard deviation of \texttt{N}’s activation values in their normalized form. If the distribution is sparse, we adopt an alternative aggregation strategy, Kernel Density Estimation (KDE), to identify dense clusters within the data, which signify the most prevalent activation values for \texttt{N}. We then probabilistically select among the identified values to determine the reference value, favoring higher density areas as they represent typical activation patterns. To address node activation value shifts effectively, the reference distribution is controlled by the parameter $\alpha$, with a value in the range [0,1]. The primary objective of this distribution is to correct deviations in node activations. Extending the range for $\alpha$ beyond 1 would result in a shift away from the reference value, potentially altering the node activation dynamics and compromising the intended adjustment.


\subsection{Repair Actions Evaluation}\label{subsec:act-eval}
Towards a systematic action evaluation and enforcement, we first evaluate each action $A_i \in \{A_1,...,A_N\}$ generated by Algorithm \ref{alg:Action-gen}. An action  qualifies as a candidate if it serves the model repair goal. For our case studies in \S \ref{sec: eval} aimed at  robustness-enhancing repair, an action needs to improve the accuracy of the model on adversarial data without sacrificing accuracy on benign data (i.e., MRE = Accuracy). To account for this trade-off, we formulate a metric that we call \textit{Tradeoff\_Score} ($TS$), which can be adapted to the model repair goal at hand and relevant model utility metrics. For our case studies, since adversarial example attacks are aimed to degrade the model's accuracy, we use \textit{Accuracy} as a measure of the model utility. Accordingly, $TS$ computes the difference between the \textit{gain} in accuracy on a target setting (e.g., adversarial data) and the \textit{loss} of accuracy on the nominal setting (e.g., benign data). Formally, $TS$ on a candidate action $A_i$ is defined as:\\
\vspace{-1em}
\begin{equation}
\resizebox{.9\hsize}{!}{$TS(A_i) = A_i(Nominal) - A_0(Nominal) + \frac{1}{n} \sum_{j=1}^n \left(A_i(target_j) - A_0(target_j)\right) $}
\label{eq:TS}
\end{equation}
where $A_i(Nominal)$ is $f$'s accuracy on nominal data (e.g., benign data) after performing $A_i$ and $A_i(target_j)$ is the accuracy on target data (e.g., adversarial samples crafted with attack $j$). We use $i=0$ to refer to the initial accuracy, with no action performed. A positive $TS(A_i)$ suggests that $A_i$ improves the accuracy on target data more than it sacrifices accuracy on nominal data. In Equation \ref{eq:TS}, this means that the \textit{gain} ($2^{nd}$ term) on target (e.g., adversarial) data is higher than the \textit{loss} ($1^{st}$ term) on nominal (e.g., benign) data.

\begin{figure*}[h!]
    \centering
    \subfloat[Step 1  ]{\includegraphics[width=0.32\textwidth]{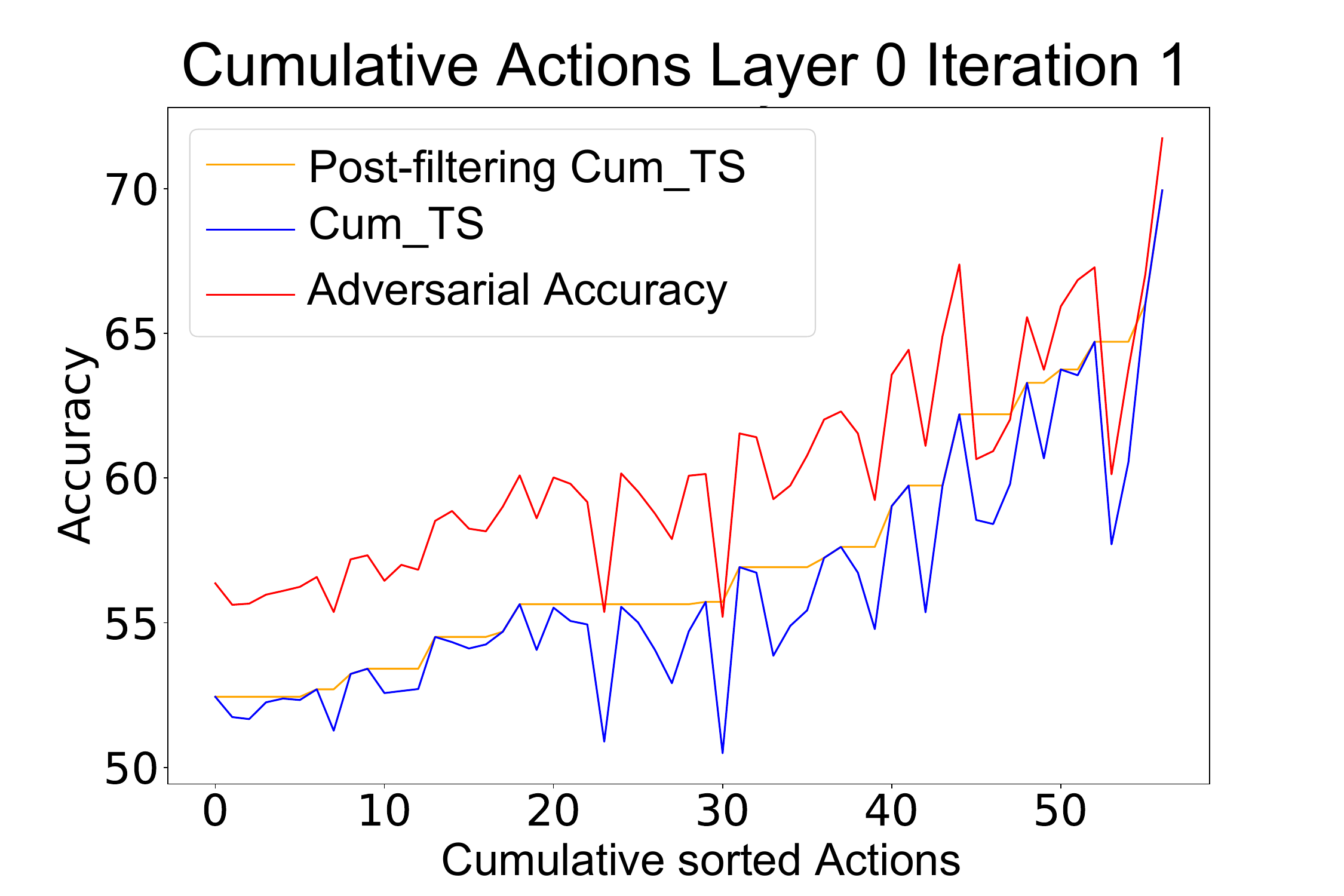}} 
     \subfloat[Step 2]{\includegraphics[width=0.32\textwidth]{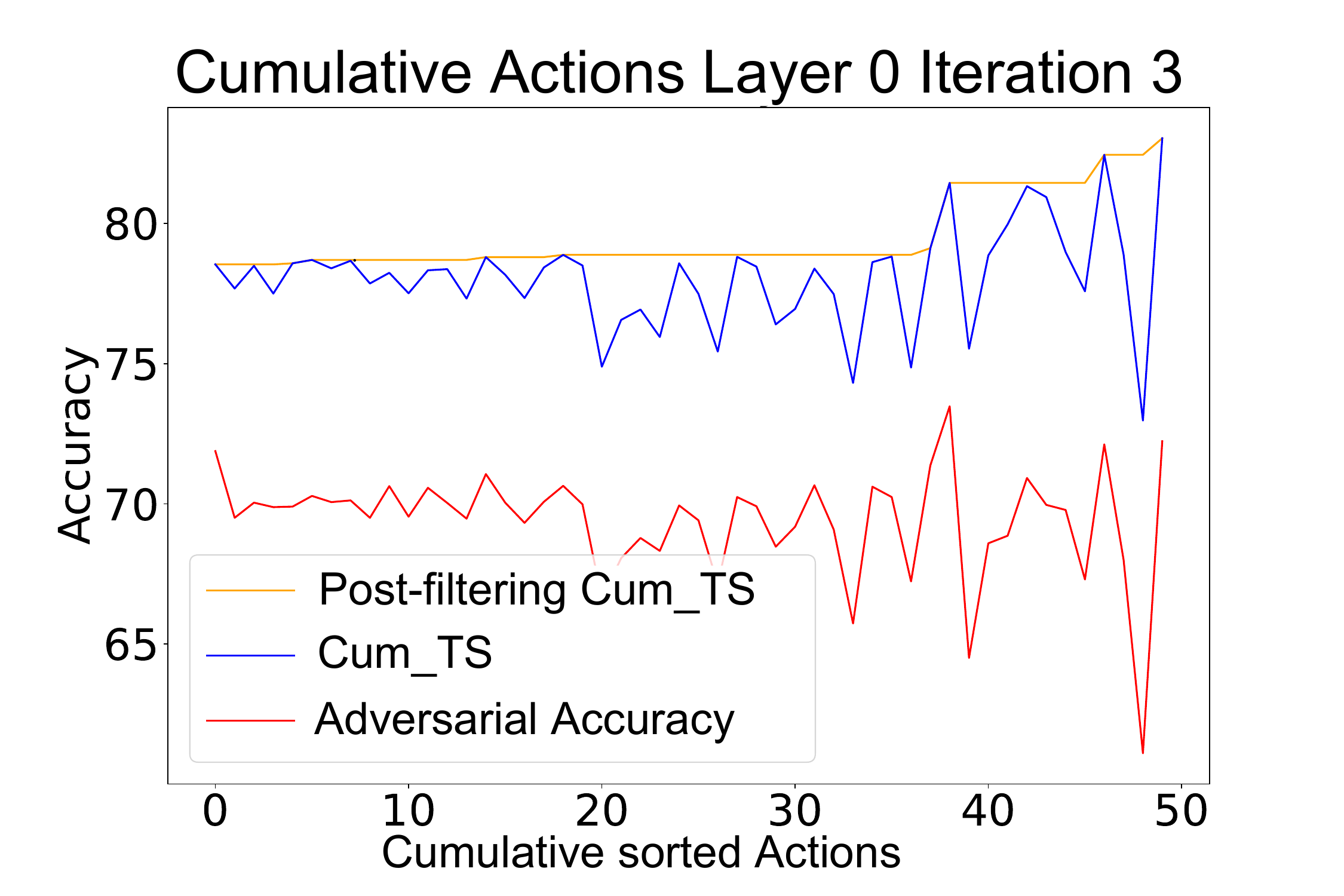}} 
     \subfloat[Step 3]{\includegraphics[width=0.32\textwidth]{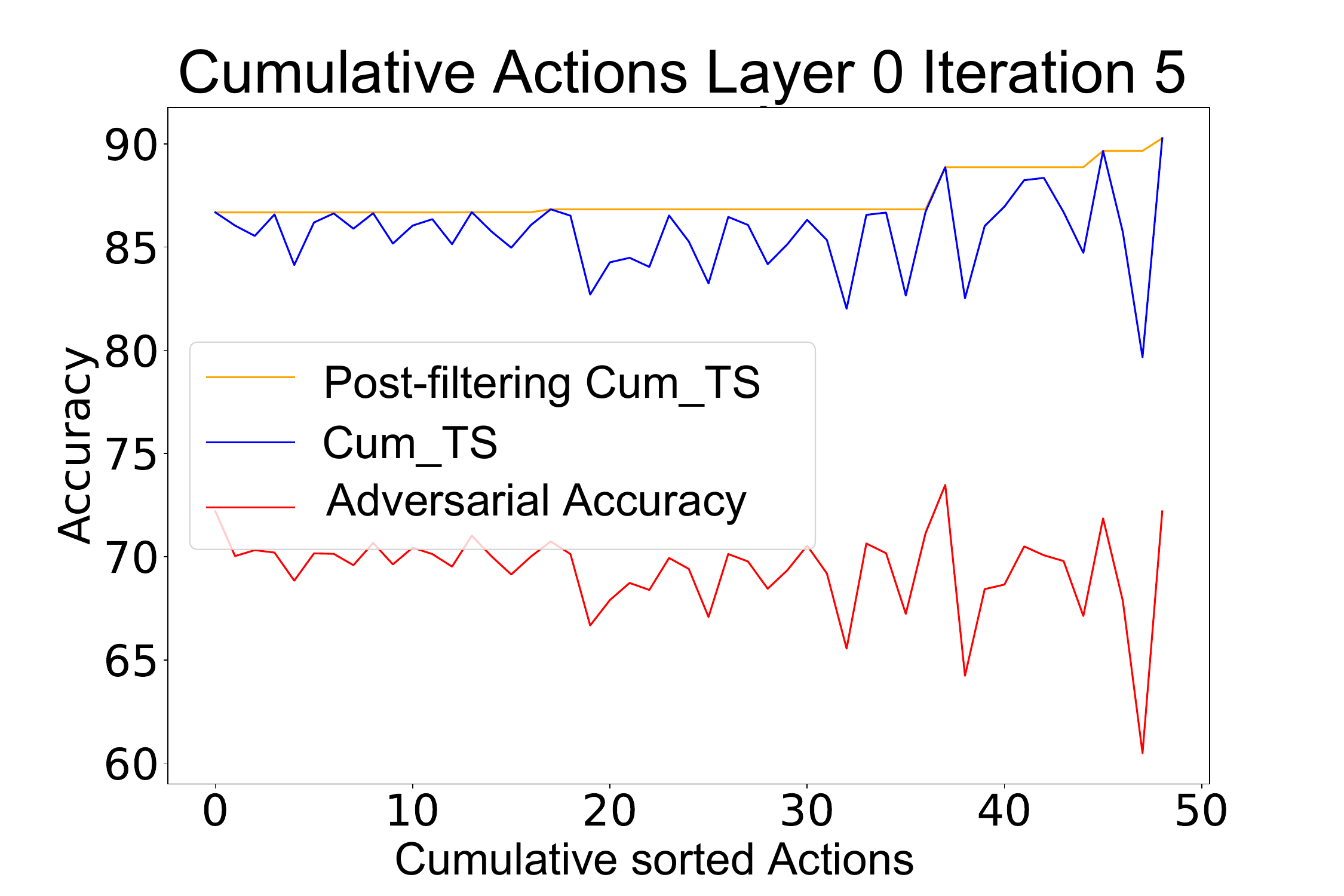}}
    \caption{Cumulative actions analysis and filtering for the EMBER dataset.}
    \label{fig:Cum-ana}
\end{figure*}

Guided by $TS$ scores, we sort all actions per layer beginning with the best action (i.e., $A_1$) and incrementally apply subsequent actions on $f$ while tracking changes in $Cum\_TS(\{A_i\}_i)$. Actions with $TS\leq0$ are ignored, since they cause more accuracy degradation on nominal data than they improve accuracy on target data. \\
Figure \ref{fig:Cum-ana} shows the incremental cumulative actions analysis and filtering in one of our case study datasets: Ember. In Figure \ref{fig:Cum-ana} (a), we observe an increasing $Cum\_TS$ curve along with an increasing accuracy on target (adversarial) data. This increase is a positive indicator that the initial actions sorting is effective, though the fluctuating curve suggests that some actions are causing drops on the cumulative trade-off when performed in this order. Therefore, we filter these actions to realise the post-filtering $Cum\_TS$ (smooth orange line in Figure \ref{fig:Cum-ana} (a). The filtered actions are queued to be tested in future steps. Figure \ref{fig:Cum-ana} (b) shows the next iteration of the same process on the queued actions. The process stops when the post-filtering $Cum\_TS$ converges to a maximum value or the queue is empty at the beginning of the next iteration. The first stopping criterion suggests that the remaining actions do not improve performance, while the second implies that all remaining actions have been tested and qualify as potential enhancement measures. In \S \ref{eval:loop}, we expand on Ember and other datasets in our case studies.

\section{Case Studies: Robustness Enhancing Repair }\label{sec: eval}

We validate \sysname{} with robustness enhancing model repair goal. We conduct comprehensive case studies on four datasets, two for image classification and two for malware detection. These case studies encompass a variety of attacks and DNN architectures, guided by the following research questions:\\
\noindent$\bullet$\textbf{RQ1:} How effective are the empirical and structural characterizations of IPGs in identifying runtime behaviors of a DNN on benign and adversarial settings across attacks?

\noindent$\bullet$\textbf{RQ2:} How effective is \sysname{} in identifying graph-specific robustness enhancing repair actions with minimal impact on a nominal setting (accuracy on benign inputs)?

\noindent$\bullet$\textbf{RQ3:} How generalizable or transferable are the IPG characterization-guided model repair actions across attacks?




\subsection{Case Studies Setup}\label{subsec: setup}
 \textbf{Dataset and Models:} We instantiated \sysname{} on four models covering image classification (MNIST \cite{MNIST} and CIFAR10\cite{CIFAR10}) and Windows PE malware detection (Cuckoo-Traces \cite{cuckoo-data}  and Ember\cite{EMBER2018}).
 The models used are: MNIST-DNN (accuracy = 97.7$\%$), CIFAR10-CNN (accuracy = 81.47 $\%$), CIFAR10-ResNet18 (accuracy = 87.94 $\%$), Cuckoo-DNN (accuracy = 95.48 $\%$), and Ember-DNN (accuracy = 94.56$\%$).
In Table \ref{tab:summary2}, we provide details about the models' architectures (``Layer Type'' column) for Cuckoo and Ember. For MNIST model and CIFAR-10 models, we provide it in the Appendix \S \ref{subsec:modelarchi}

\textbf{Attacks:} For MNIST and CIFAR10, we use 3 white-box attacks: FGSM\cite{FGSM}, PGD \cite{PGD}, and APGD-DLR derived from  comprehensive attack suites auto-attack \cite{APGD-DLR} and three black-box attacks: SPSA \cite{uesato2018adversarial}, Square \cite{andriushchenko2020square}, and SIT \cite{wang2023structure}. For all attacks, we maintain a perturbation
bound $\epsilon \leq  0.3$. For Cuckoo-Traces and Ember, we assume that the adversary lacks knowledge about the target model but is aware of the features used to train it (e.g., API calls, DLL files).  
We call the attack on Cuckoo-DNN {\em Bit-Flip} and the one on Ember is denoted as {\em Emb-Att}. For both attacks, as in previous work \cite{MalGAN17,ExploreAdvEx18, cuckoo-data}, we incrementally perform additive perturbations until the model flips its label to benign.

\textbf{Hyperparameters Configuration:} Algorithm \ref{alg:Action-gen} is guided by two parameters: actions' sensitivity $\alpha$ and norm type $p$. To select $\alpha$, we run the actions evaluation process per-layer using candidate values in $[0.1,1]$. For all models, results showed that $\alpha = 1$ achieves the highest $Cum\_TS$, indicating that the algorithm should update the activation value of a candidate node with the reference value returned by $Reference\_Dist\_Agg()$ to achieve the optimal performance (see Figure \ref{fig:alpha} in the Appendix).
$\beta$ is computed for each node with respect to the description provided in \S \ref{subsec:action-gen}.

\subsection{Effectiveness of IPG-Guided Characterization}
\label{eval:charac}
\begin{figure*}[h!]
    \centering
    \captionsetup[subfloat]{labelformat=empty}
    
    \subfloat[]{\includegraphics[width=0.3
    \textwidth, ]{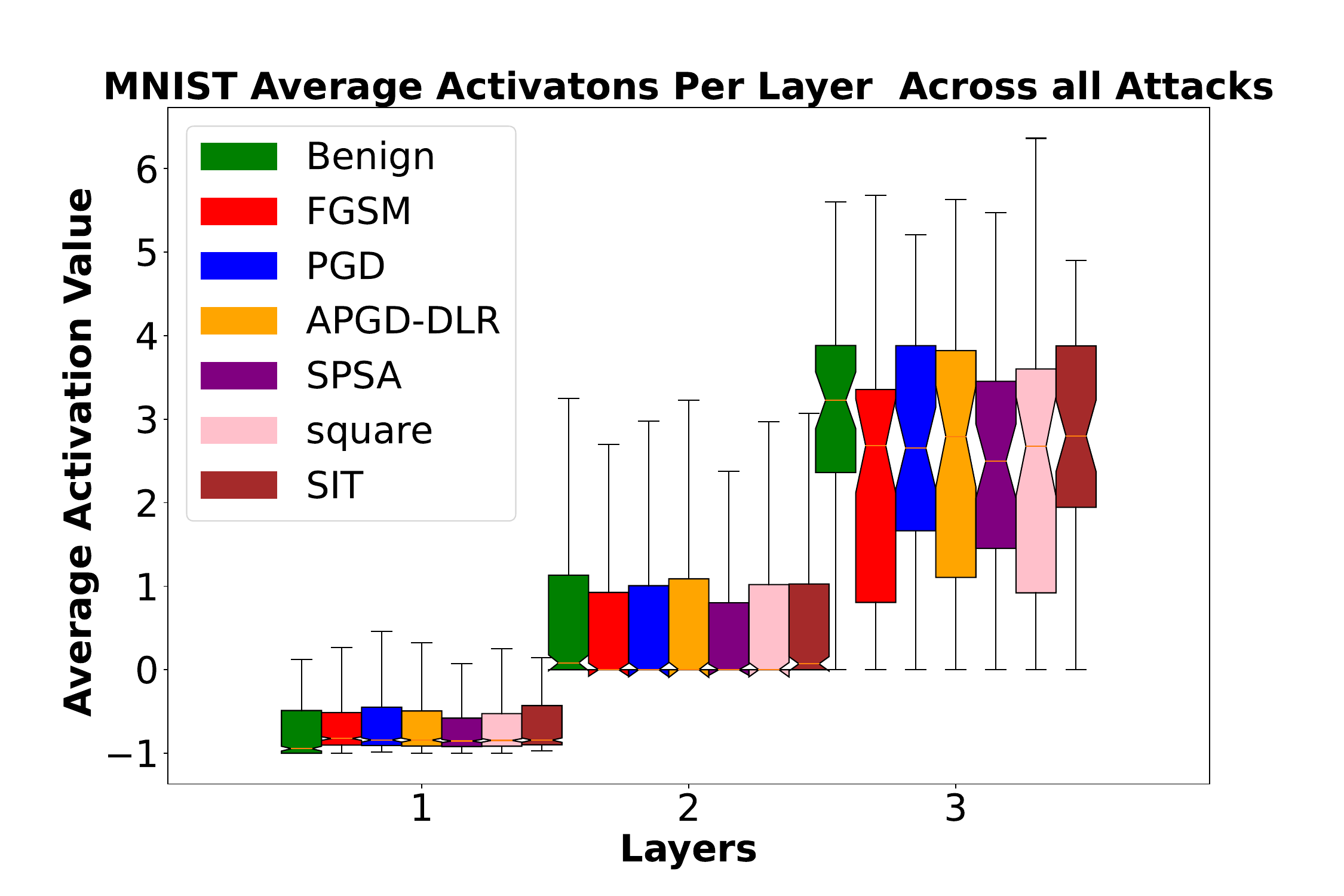}}
     \subfloat[]{\includegraphics[width=0.3\textwidth]{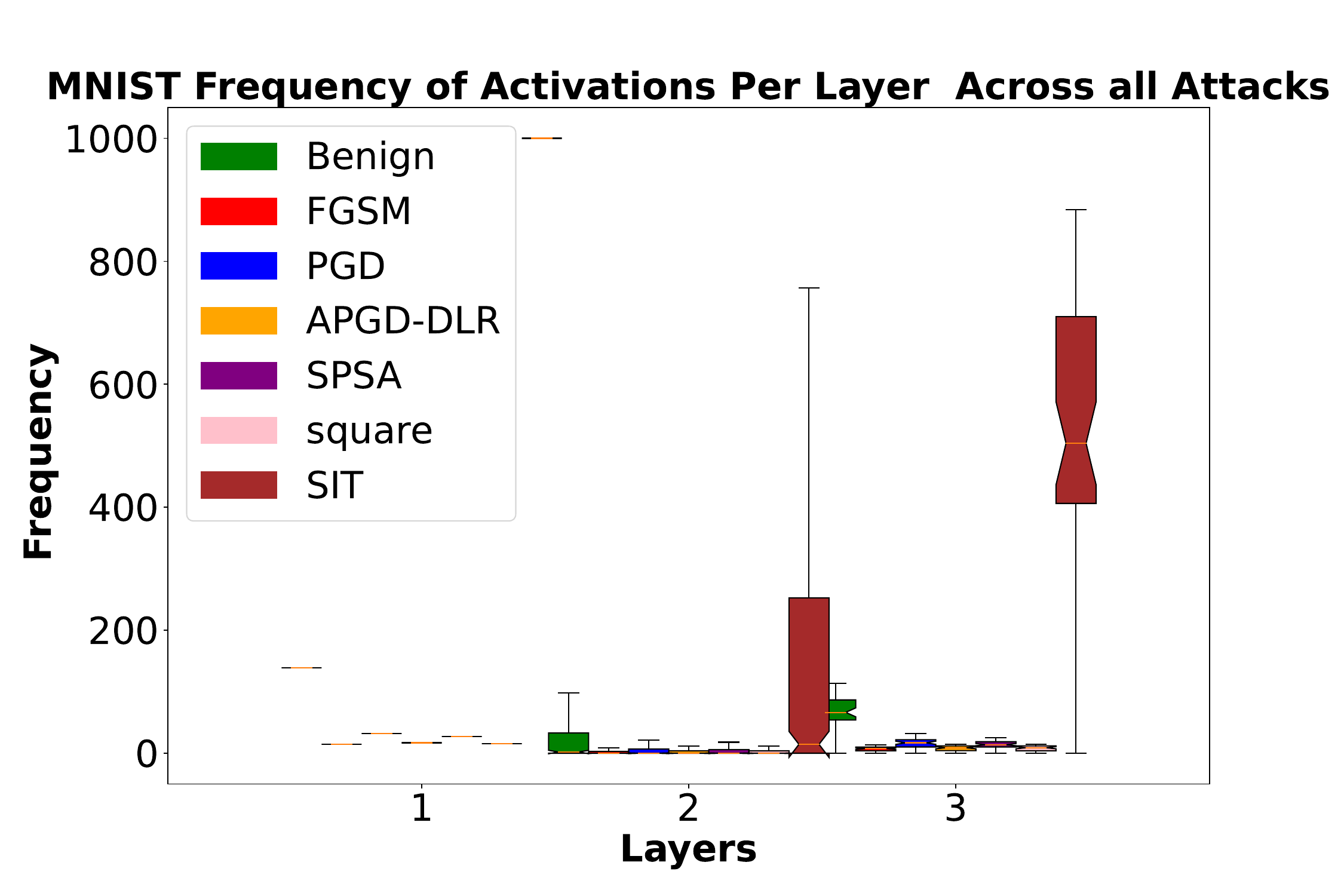}}
     \subfloat[]{\includegraphics[width=0.3\textwidth]{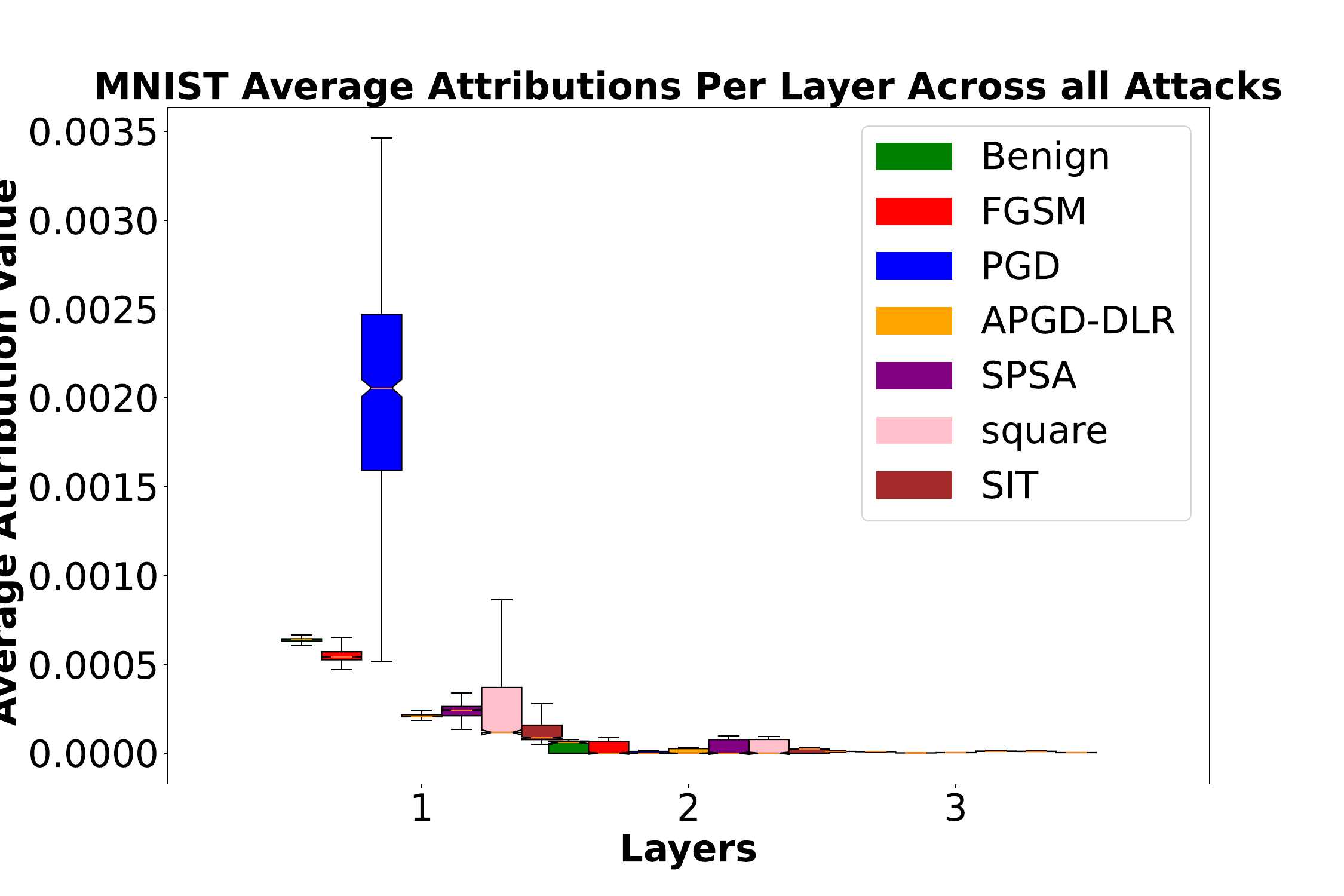}}
    \vspace{-1cm}
        \subfloat[ ]{\includegraphics[width=0.3\textwidth]{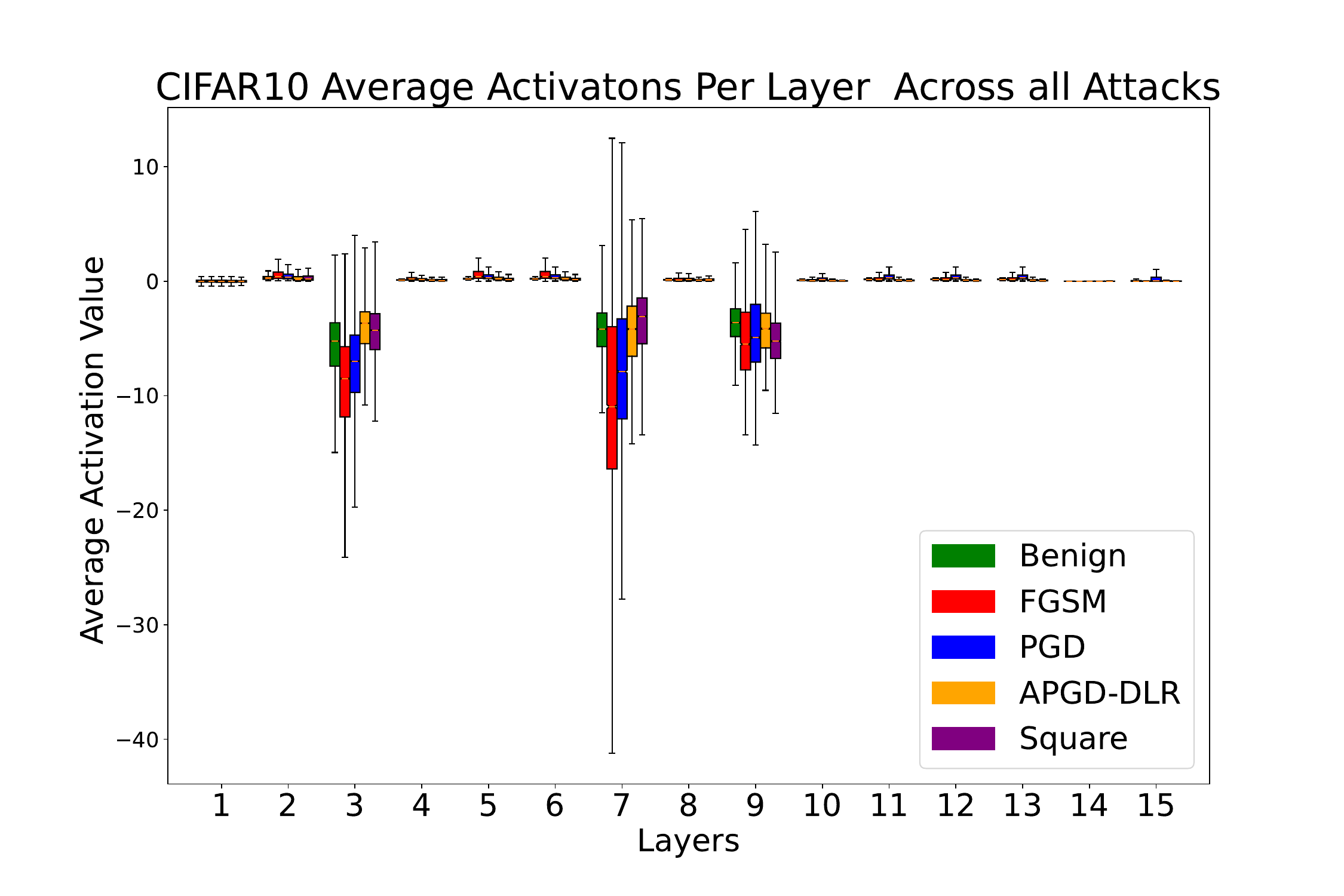}}
     \subfloat[]{\includegraphics[width=0.3\textwidth]{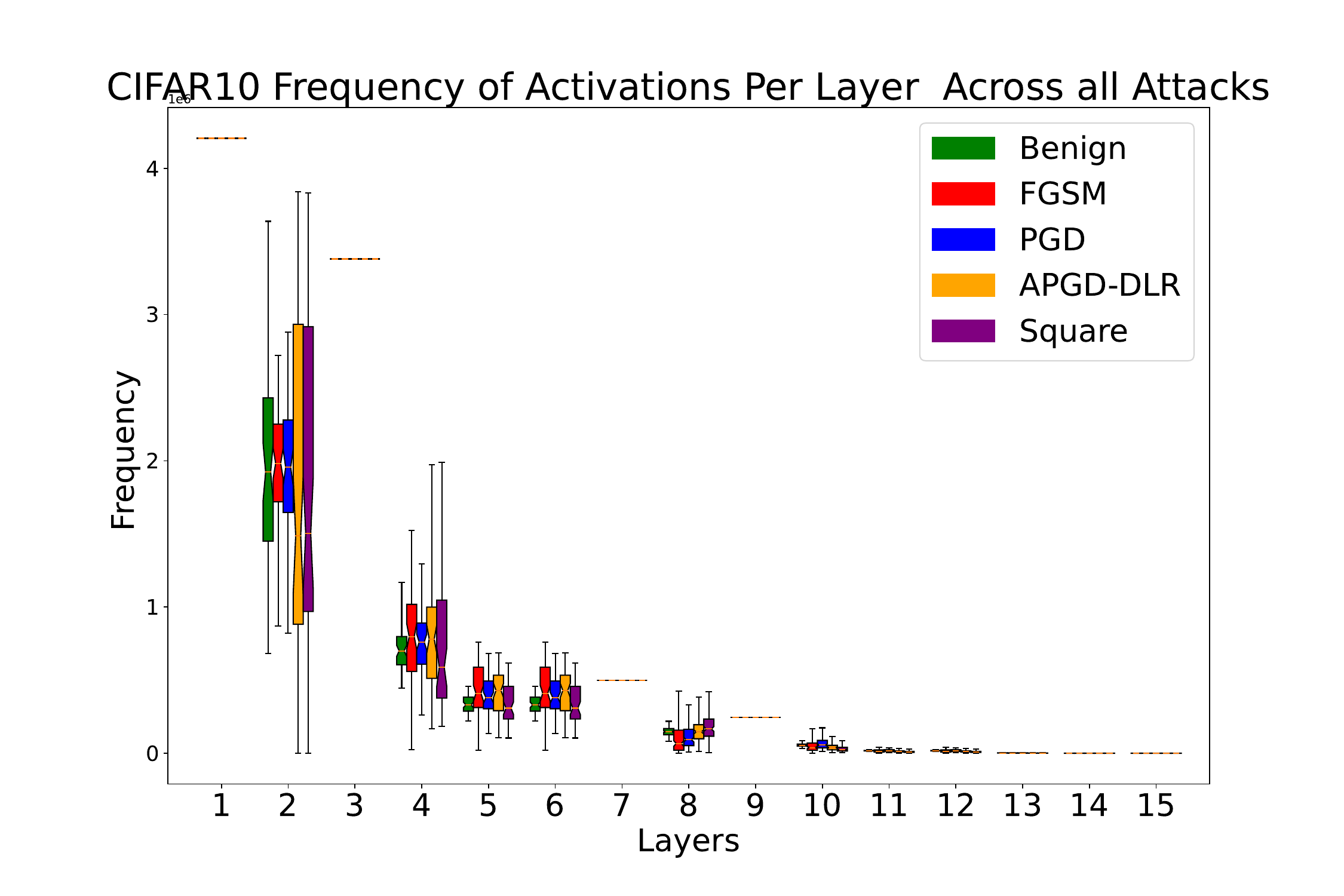}}
     \subfloat[]{\includegraphics[width=0.3\textwidth]{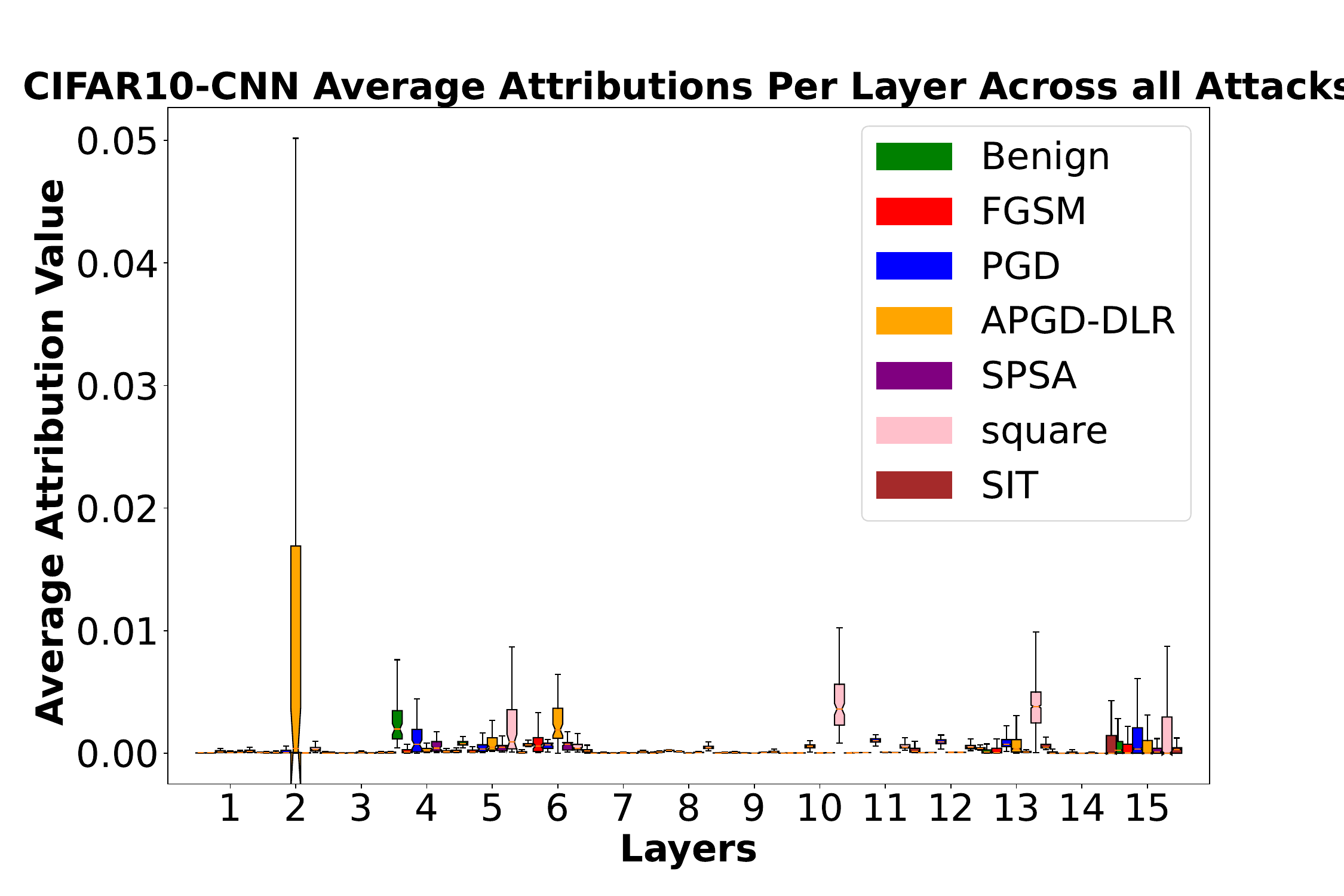}}
      \vspace{-0.7cm}
      \\
    \subfloat[]{\includegraphics[width=0.3\textwidth, ]{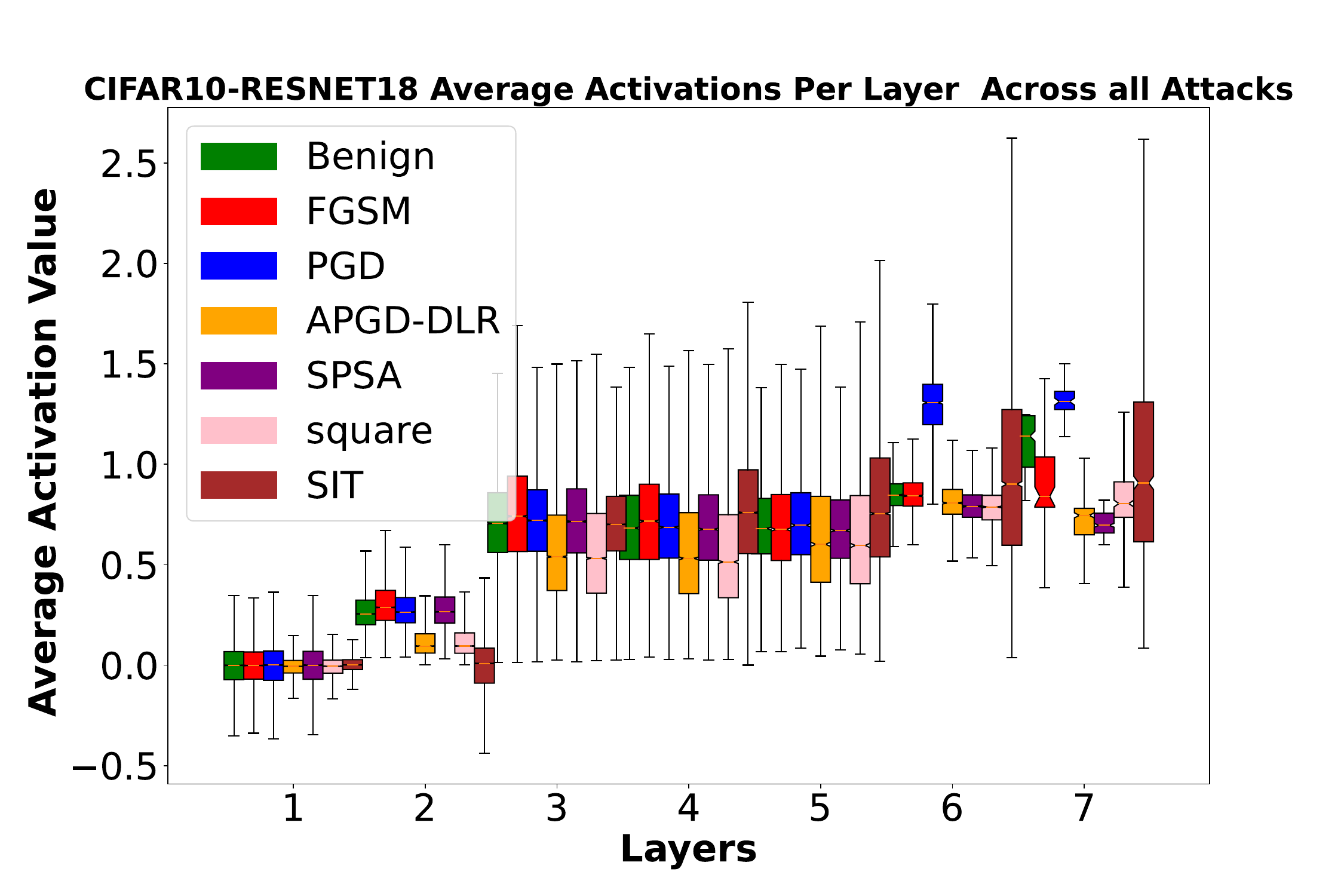}}
     \subfloat[]{\includegraphics[width=0.3\textwidth]{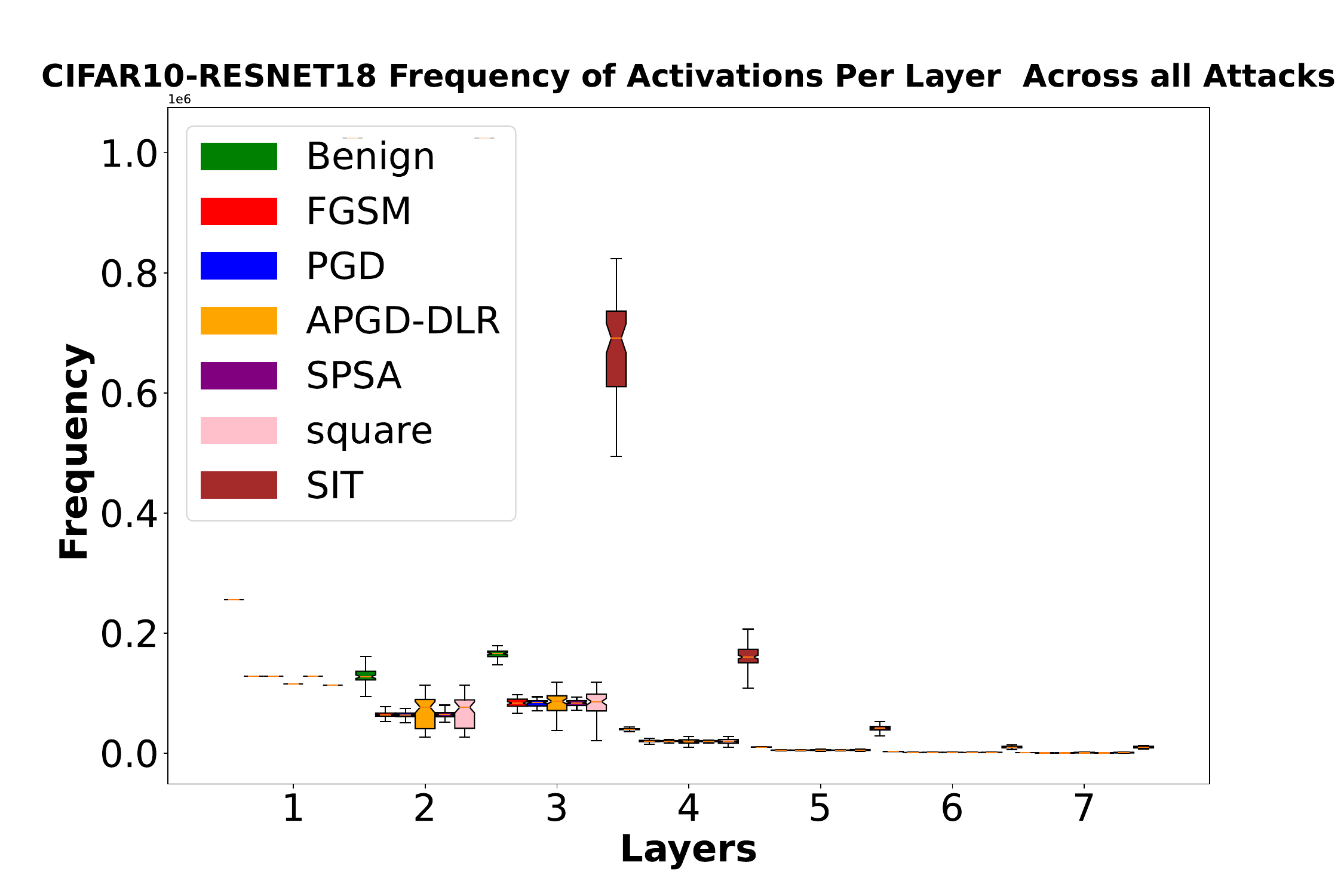}}
     \subfloat[]{\includegraphics[width=0.3\textwidth]{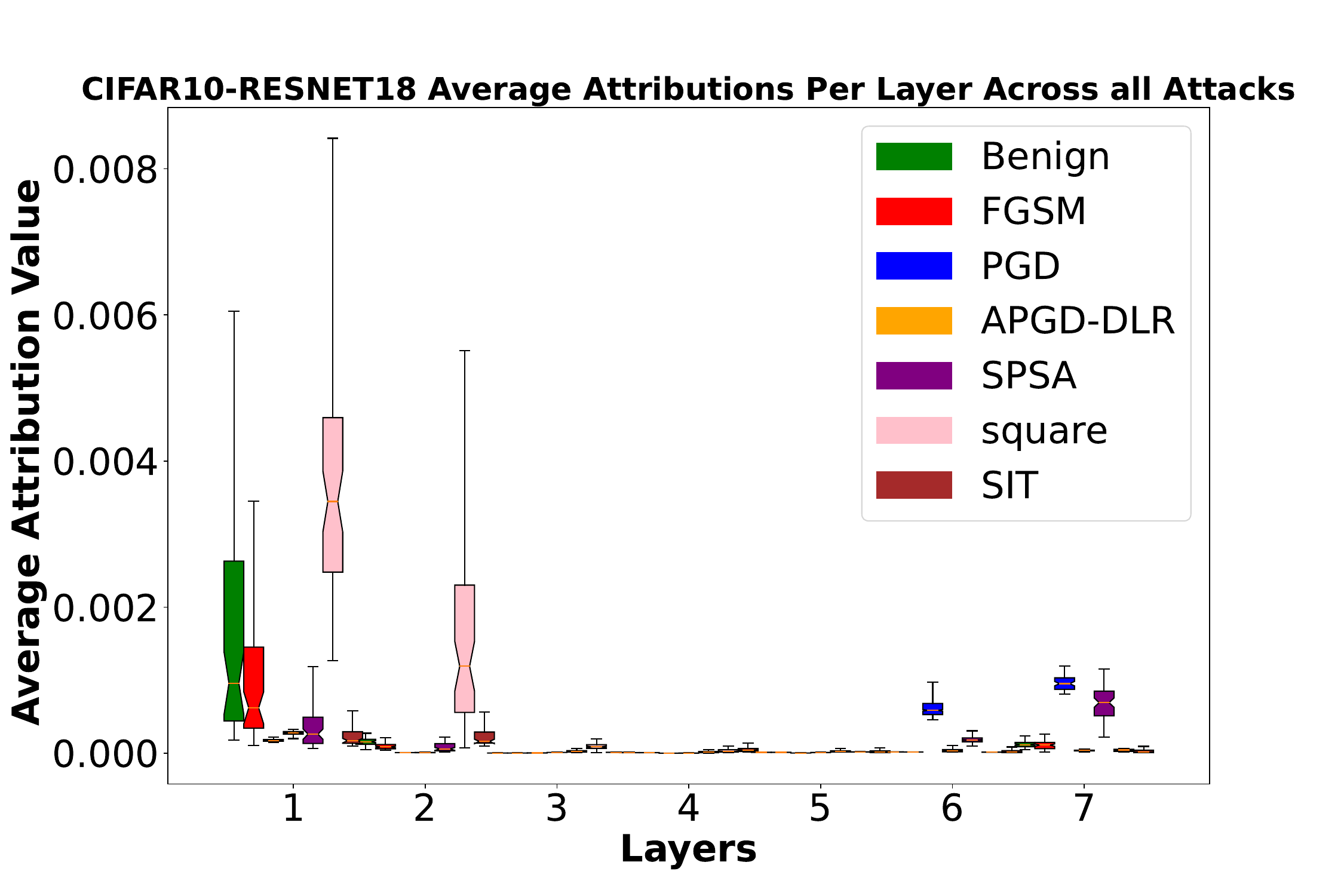}}
    \vspace{-1cm}
    \\
        \subfloat[]{\includegraphics[width=0.3\textwidth]{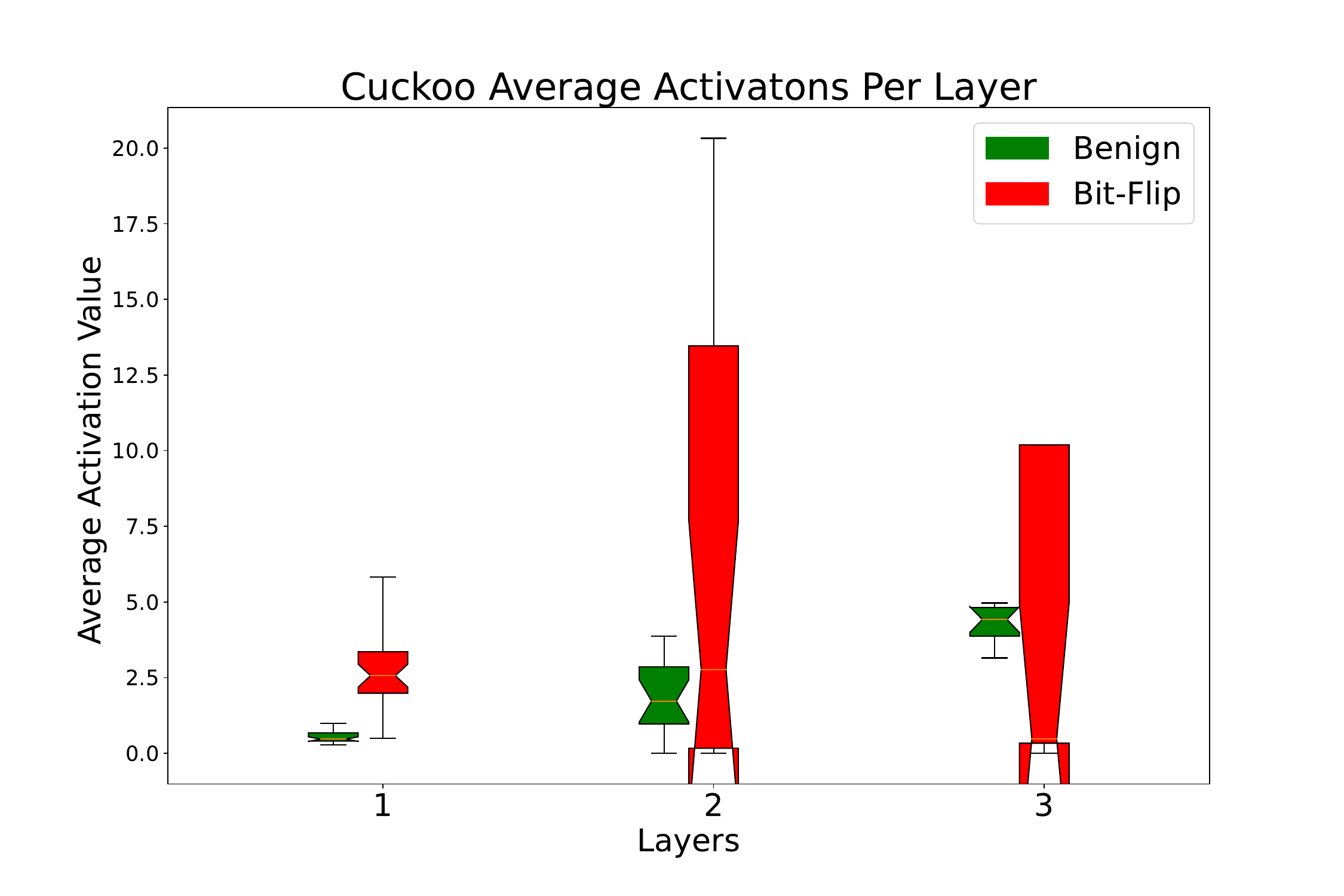}}
     \subfloat[]{\includegraphics[width=0.3\textwidth]{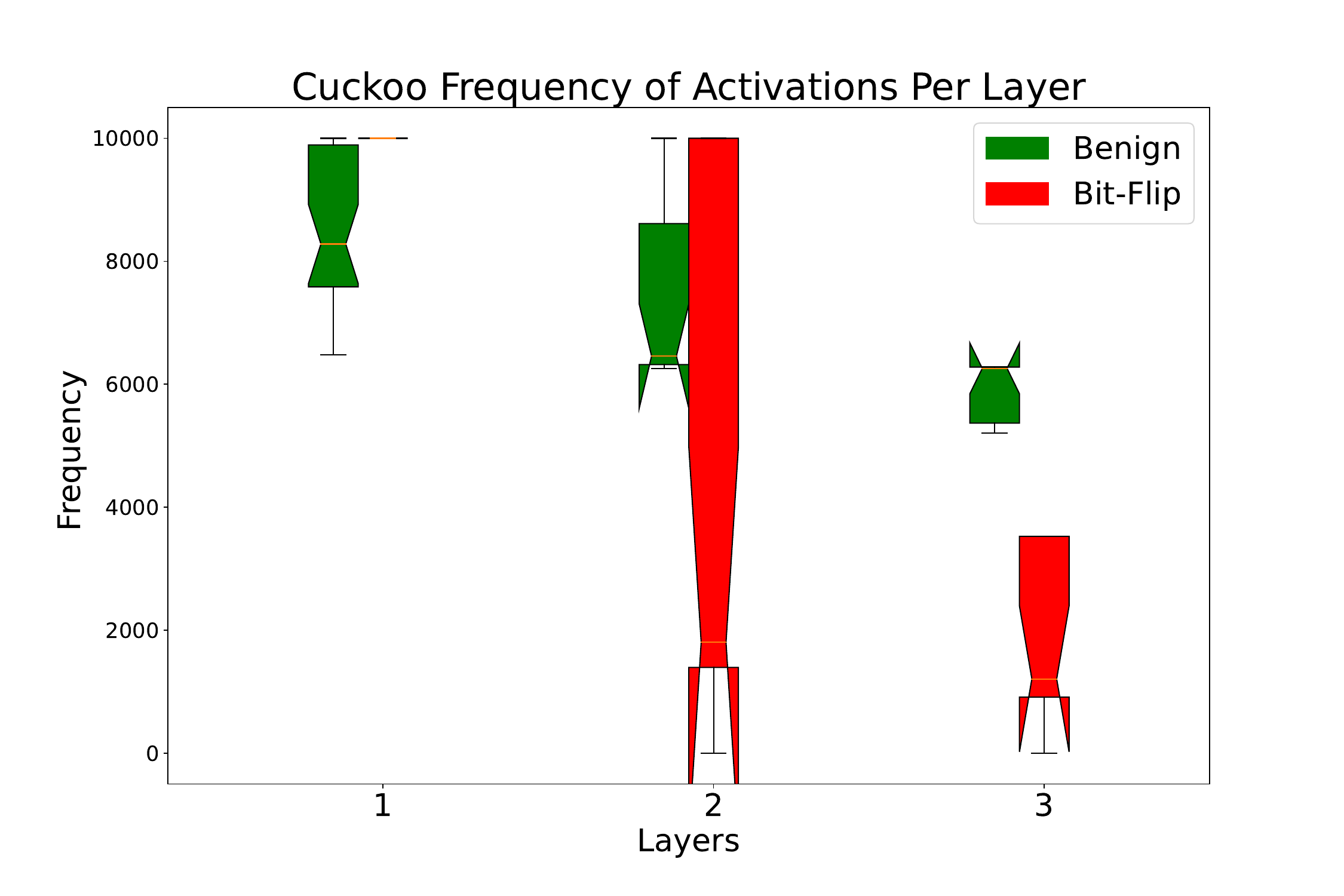}}
     \subfloat[]{\includegraphics[width=0.3\textwidth]{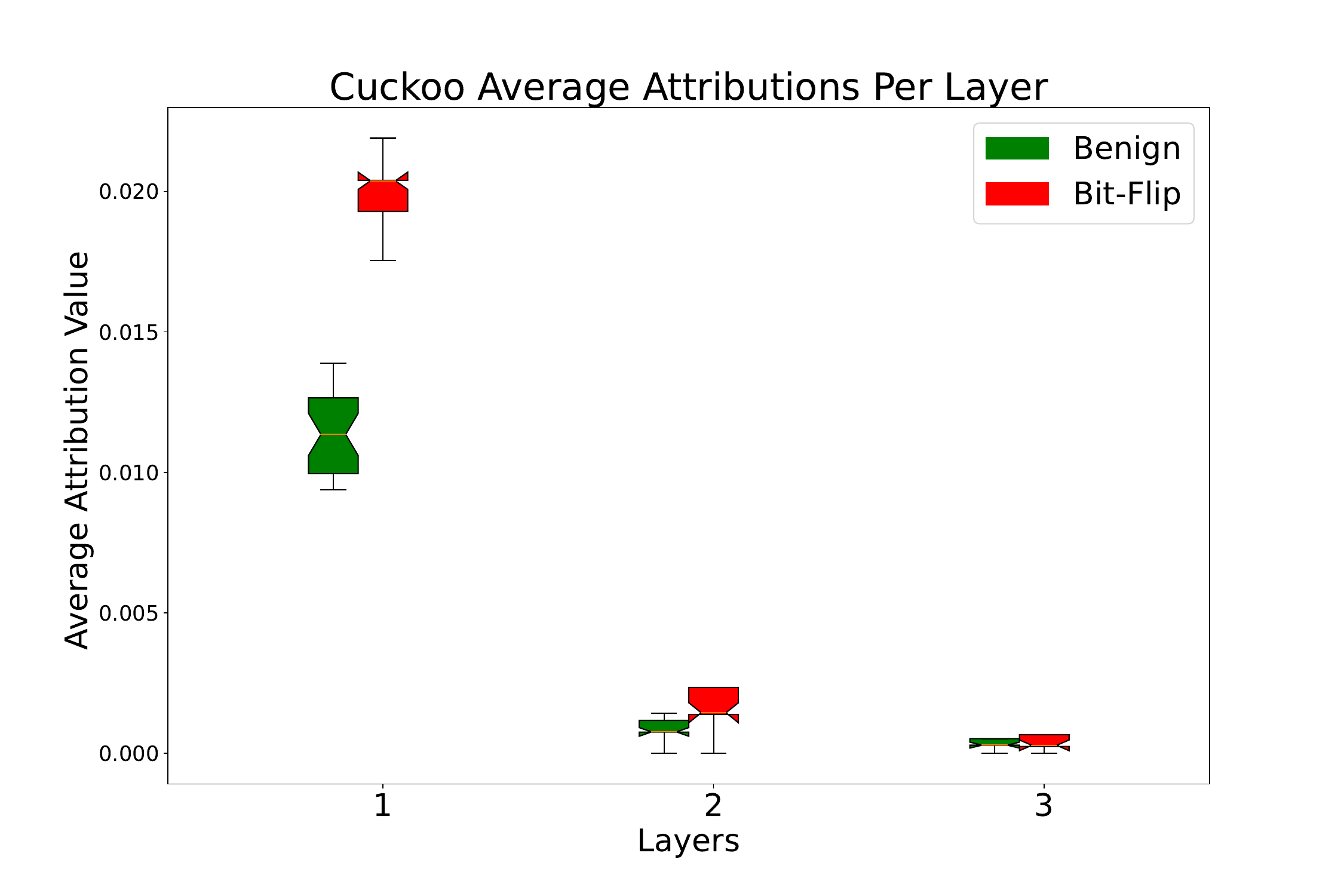}}
      \vspace{-1cm}
     \\
      \subfloat[Empirical: Average Activation Values]{\includegraphics[width=0.3\textwidth]{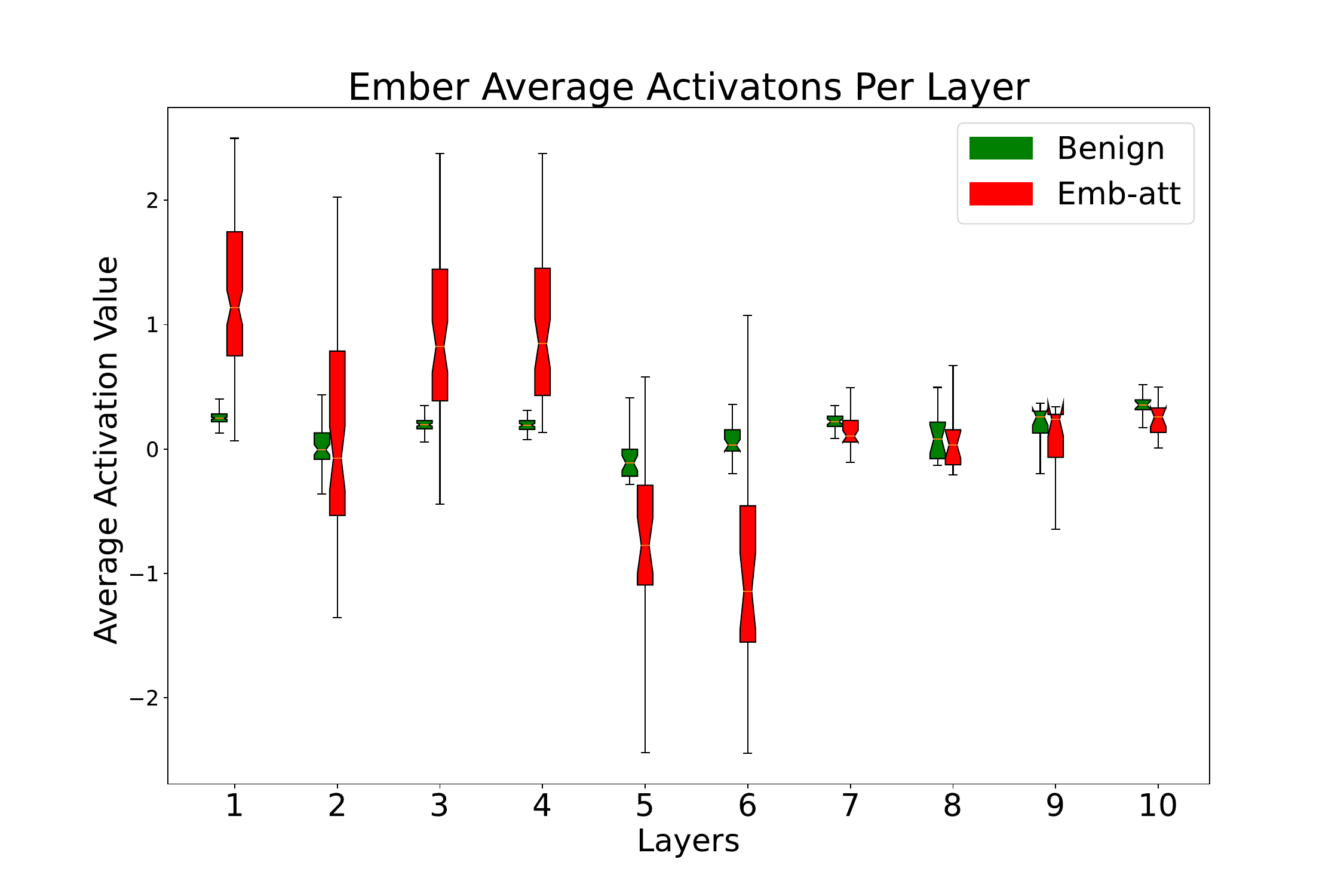}}
     \subfloat[Empirical: Frequency of Activations]{\includegraphics[width=0.3\textwidth]{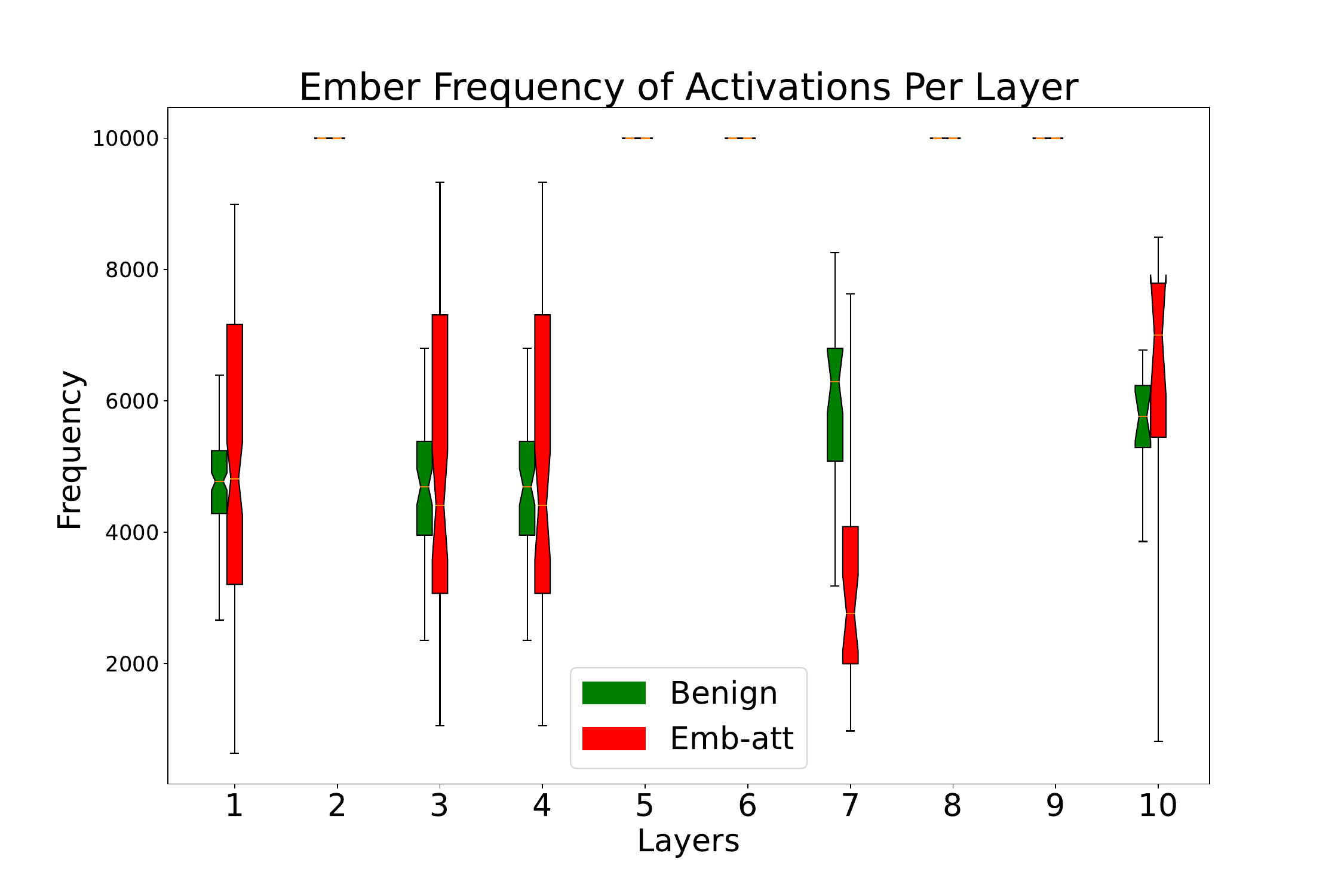}}
     \subfloat[Structural: Average Attribution Values]{\includegraphics[width=0.3\textwidth]{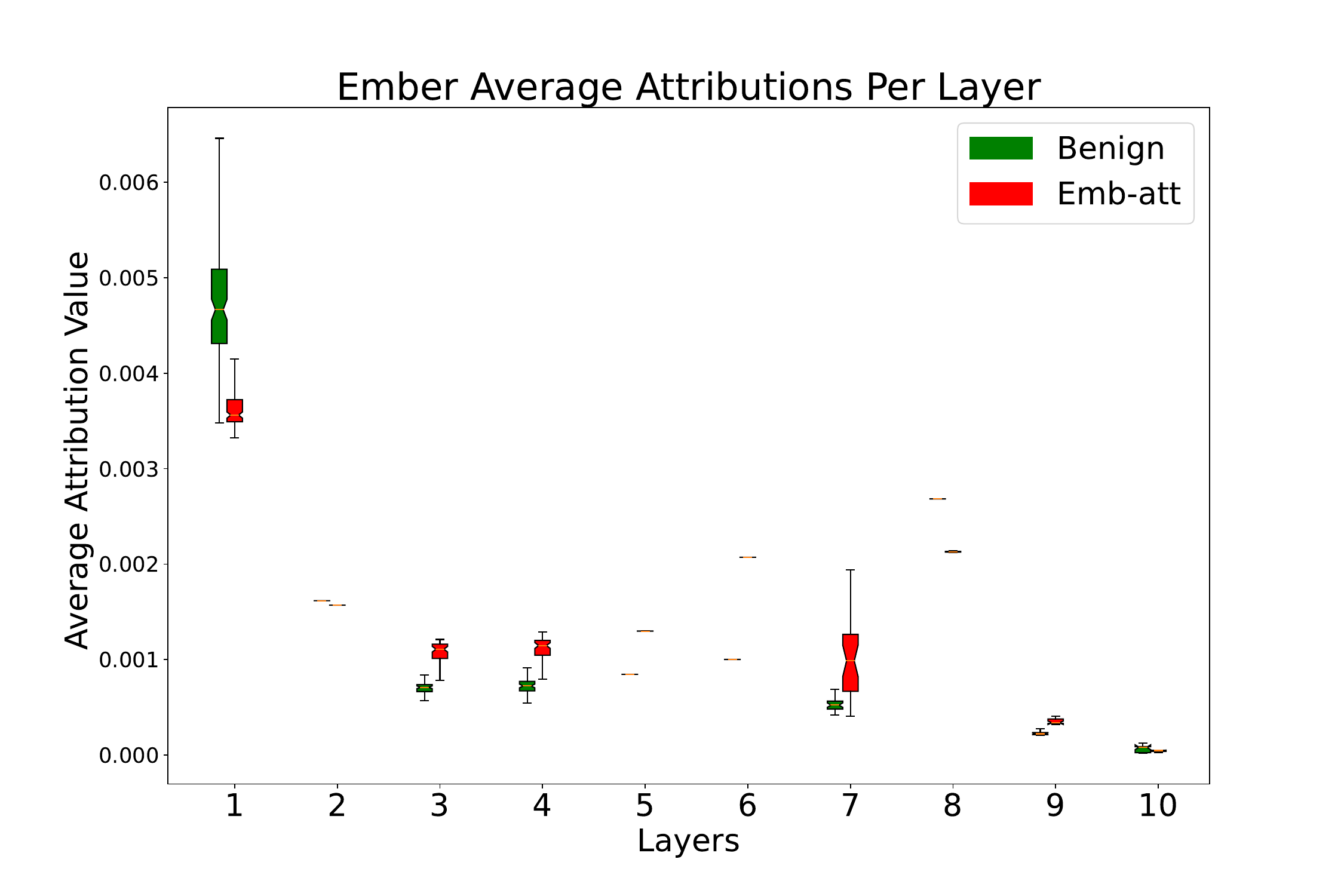}}

      \caption{Benign vs. adversarial characterization insights.  From top-to-bottom, each row refers to a studied model (e.g., MNIST-DNN, etc) with plots for Average Activation per layer, Frequency of Activations per layer, and Average Attribution Values per layer, respectively.}
    \label{fig:charac-all}
\end{figure*}

\textbf{Empirical Characterization:} Figure \ref{fig:charac-all} presents box plots of node \textit{Activation Values} per layer, averaged across all benign IPGs (green) and adversarial IPGs (other colors). It also shows node \textit{Activation Frequency} per layer. We observe a clear distinction between benign and adversarial settings. More precisely, activation value and activation frequency ranges on adversarial data differ from their benign counterparts. These distinctions are more significant in malware detection models, as both models serve the task of binary classification. For MNIST and CIFAR10, it is noteworthy that such distinction is sometimes observable across attacks, reflecting different activation patterns from one attack to another. For instance, layer 3 of MNIST-DNN has different ranges of activation values on test samples generated with FGSM (red) and APGD-DLR (orange). \textbf{\em These observations suggest \sysname{} reveals clear differences in neuron activation values and frequencies between benign and adversarial inputs. These empirical signals provide reliable clues for identifying nodes and layers most affected by adversarial perturbations.}

  \textbf{Structural Characterization:} To assess the GNN-based structural characterization at node and edge levels, we analyze the average attributions per layer across attacks (Figure \ref{fig:charac-all}). 
  The node-level analysis, presented in the form of box plots, reveals distinct node attribution patterns between benign and adversarial IPGs, highlighting the impact of evasion attacks on the contribution of each node to activation patterns. This suggests that the most influential edges in benign IPGs maintain their influence in adversarial settings, albeit with varying color intensity, indicating different edge attribution values. This combined analysis at the node and edge levels underscores the importance of identifying nodes linked by influential edges, leading to the recognition of a novel set of candidate nodes for repair actions.
\begin{table*}[htbp]
\centering
\tabcolsep=0.13cm
\scalebox{0.60}{
\begin{tabular}{|l|lllllll|llllllllllllll|}
\hline
Model & \multicolumn{7}{c|}{MNIST-DNN}      & \multicolumn{14}{c|}{CIFAR10-CNN/ResNet18}                 \\ 
\hline
Test Attack & \multicolumn{1}{r|}{FGSM} & \multicolumn{1}{r|}{PGD} & \multicolumn{1}{r|}{APGD-DLR} & \multicolumn{1}{c|}{SPSA}& \multicolumn{1}{c|}{Square}& \multicolumn{1}{c|}{SIT} & Benign      & \multicolumn{2}{c|}{FGSM} & \multicolumn{2}{c|}{PGD} & \multicolumn{2}{c|}{APGD-DLR} & \multicolumn{2}{c|}{SPSA} & \multicolumn{2}{c|}{Square}  & \multicolumn{2}{c|}{SIT} & \multicolumn{2}{c|}{Benign}      \\ \hline
	
Without Repair           & \multicolumn{1}{r|}{5.89\%}     & \multicolumn{1}{r|}{3.5\%}    & \multicolumn{1}{r|}{40.56\%}          & \multicolumn{1}{r|}{4.14\%}    & \multicolumn{1}{r|}{38.43\%}  & \multicolumn{1}{r|}{53.33\%}      & 97.70\% & \multicolumn{1}{r|}{11.84\%} & \multicolumn{1}{r|}{11.08\%}    & \multicolumn{1}{r|}{3.30\%}   & \multicolumn{1}{r|}{3.14\%}    & \multicolumn{1}{r|}{2.63\%}     & \multicolumn{1}{r|}{4.14\%}        & \multicolumn{1}{r|}{13.5\%}     & \multicolumn{1}{r|}{24.05\%}    & \multicolumn{1}{r|}{3.45\%}     & \multicolumn{1}{r|}{4.14\%}  &  \multicolumn{1}{r|}{0.00\%}     & \multicolumn{1}{r|}{0.78\%}    & 81.47\%  & 87.94\%  \\ \hline

\sysname{} (FGSM)           & \multicolumn{1}{r|}{\cellcolor{Mycolor2}\textbf{73.1\%}}     & \multicolumn{1}{r|}{\cellcolor{Mycolor2}73.84\%}    & \multicolumn{1}{r|}{\cellcolor{Mycolor2} 50.27\%}          & \multicolumn{1}{r|}{\cellcolor{Mycolor2}80.23\%}  & \multicolumn{1}{r|}{40.51\%}   & \multicolumn{1}{r|}{73.12\%}     & 97.11\% & \multicolumn{1}{r|}{\cellcolor{Mycolor2}\textbf{40.25\%}} & \multicolumn{1}{r|}{\cellcolor{Mycolor2}\textbf{64.72\%}}   & \multicolumn{1}{r|}{\cellcolor{Mycolor2}31.00\%} & \multicolumn{1}{r|}{\cellcolor{Mycolor2}65.92	\%}   & \multicolumn{1}{r|}{\cellcolor{Mycolor2} 15.91\%}    & \multicolumn{1}{r|}{23.8\%}   & \multicolumn{1}{r|}{20.78\%}    & \multicolumn{1}{r|}{\cellcolor{Mycolor2}42.29\%}     & \multicolumn{1}{r|}{\cellcolor{Mycolor2} 17.33\%}  & \multicolumn{1}{r|}{\cellcolor{Mycolor2}27.29\%} & \multicolumn{1}{r|}{\cellcolor{Mycolor2} 10.56\%}  & \multicolumn{1}{r|}{\cellcolor{Mycolor2}12.01\%} & 80.84\% &84.46\%  \\

\sysname{} (PGD)            & \multicolumn{1}{r|}{73.76\%}     & \multicolumn{1}{r|}{\textbf{74.56\%}}   & \multicolumn{1}{r|}{49.55\%}          & \multicolumn{1}{r|}{79.65\%}    & \multicolumn{1}{r|}{39.33\%} & \multicolumn{1}{r|}{70.01\%}      & 97.03\% & \multicolumn{1}{r|}{15.79\%}   & \multicolumn{1}{r|}{39.01\%}   & \multicolumn{1}{r|}{\textbf{36.33\%}} & \multicolumn{1}{r|}{\textbf{63.29\%}}   & \multicolumn{1}{r|}{13.52\%}     & \multicolumn{1}{r|}{\cellcolor{Mycolor2} 25.9\%}   & \multicolumn{1}{r|}{19.67\%}     & \multicolumn{1}{r|}{20.00\%}     & \multicolumn{1}{r|}{12.53\%}    & \multicolumn{1}{r|}{25.25\%}   & \multicolumn{1}{r|}{9.53\%}    & \multicolumn{1}{r|}{9.12\%}   & 80.23\%  & 84.67\%   \\

\sysname{} (APGD-DLR)       & \multicolumn{1}{r|}{9.16\%}     & \multicolumn{1}{r|}{6.19\%}    & \multicolumn{1}{r|}{\textbf{62.84\%}}         & \multicolumn{1}{r|}{9.74\%}  & \multicolumn{1}{r|}{\cellcolor{Mycolor2} \textbf56.74\%} & \multicolumn{1}{r|}{68.64\%}      & 97.50\% & \multicolumn{1}{r|}{15.23\%}  & \multicolumn{1}{r|}{42.87\%}    & \multicolumn{1}{r|}{12.32\%}   & \multicolumn{1}{r|}{46.78\%}   & \multicolumn{1}{r|}{\textbf{20.45\%}}  & \multicolumn{1}{r|}{\textbf{26.29\%}}        & \multicolumn{1}{r|}{15.27\%}    & \multicolumn{1}{r|}{27.73\%}       & \multicolumn{1}{r|}{5.78\%}    & \multicolumn{1}{r|}{26.45\%}      & \multicolumn{1}{r|}{4.12\%}    & \multicolumn{1}{r|}{7.12\%}    & 80.12\%   & 85.52\%   \\
\sysname{} (SPSA)         & \multicolumn{1}{r|}{71.21\%}     & \multicolumn{1}{r|}{72.45\%}    & \multicolumn{1}{r|}{25.21\%}          & \multicolumn{1}{r|}{\textbf{87.54\%}}   & \multicolumn{1}{r|}{{27.56\%}}  & \multicolumn{1}{r|}{{65.12\%}}      & 95.89\% & \multicolumn{1}{r|}{22.46\%}  & \multicolumn{1}{r|}{15.79\%}    & \multicolumn{1}{r|}{19.55\%}    & \multicolumn{1}{r|}{14.57\%}    & \multicolumn{1}{r|}{9.41\%}   & \multicolumn{1}{r|}{5.73\%}         & \multicolumn{1}{r|}{\textbf{27.18\%}}  & \multicolumn{1}{r|}{\textbf{44.18\%}}   & \multicolumn{1}{r|}{{10.47\%}}  & \multicolumn{1}{r|}{{26.22\%}}   & \multicolumn{1}{r|}{{4.21\%}}  & \multicolumn{1}{r|}{{6.32\%}}    & 80.21\%  & \cellcolor{Mycolor2} 86.04\% \\ 
\sysname{} (Square)         & \multicolumn{1}{r|}{9.03\%}     & \multicolumn{1}{r|}{5.90\%}    & \multicolumn{1}{r|}{61.68\%}          & \multicolumn{1}{r|}{{10.45\%}}   & \multicolumn{1}{r|}{\textbf{58.30\%}}& \multicolumn{1}{r|}{\textbf{69.41\%}}        & \cellcolor{Mycolor2} 97.57\% & \multicolumn{1}{r|}{12.98\%}  & \multicolumn{1}{r|}{11.32\%}    & \multicolumn{1}{r|}{3.72\%}    & \multicolumn{1}{r|}{10.89\%}    & \multicolumn{1}{r|}{5.00\%}   & \multicolumn{1}{r|}{6.81\%}         & \multicolumn{1}{r|}{{19.26\%}}  & \multicolumn{1}{r|}{{16.19\%}}   & \multicolumn{1}{r|}{\textbf{26.22\%}}  & \multicolumn{1}{r|}{\textbf{37.57\%}} & \multicolumn{1}{r|}{\textbf{6.12\%}}  & \multicolumn{1}{r|}{\textbf{8.34\%}}      & \cellcolor{Mycolor2} 81.03\%  & 85.21 \%  \\ 
\sysname{} (SIT)         & \multicolumn{1}{r|}{73.15\%}     & \multicolumn{1}{r|}{56.82\%}    & \multicolumn{1}{r|}{53.01\%}          & \multicolumn{1}{r|}{{42.16\%}}   & \multicolumn{1}{r|}{\textbf{48.23\%}}   & \multicolumn{1}{r|}{\textbf{78.68\%}}        &  97.01\% & \multicolumn{1}{r|}{11.99\%}  & \multicolumn{1}{r|}{5.12\%}    & \multicolumn{1}{r|}{4.11\%}    & \multicolumn{1}{r|}{5.33\%}    & \multicolumn{1}{r|}{6.18\%}   & \multicolumn{1}{r|}{5.68\%}         & \multicolumn{1}{r|}{{6.57\%}}  & \multicolumn{1}{r|}{{7.24\%}}   & \multicolumn{1}{r|}{\textbf{9.32\%}}  & \multicolumn{1}{r|}{\textbf{4.72\%}} & \multicolumn{1}{r|}{\textbf{17.94\%}}  & \multicolumn{1}{r|}{\textbf{17.65\%}}      & \cellcolor{Mycolor2} 81.03\%  & 85.21 \%  \\ 
\hline
\end{tabular}
}
\caption{\small  {\bf \sysname{} Robustness Analysis Results (Accuracy)}. {\small Each row represents an instance of \sysname{} with an attack for characterization and actions generation. Each instance is tested on multiple attacks (columns).}}
\label{tab:tranf}
\end{table*}
\begin{figure}[t!]
    \vspace{-0.1cm}
    \centering   
    \includegraphics[width=.4\textwidth]{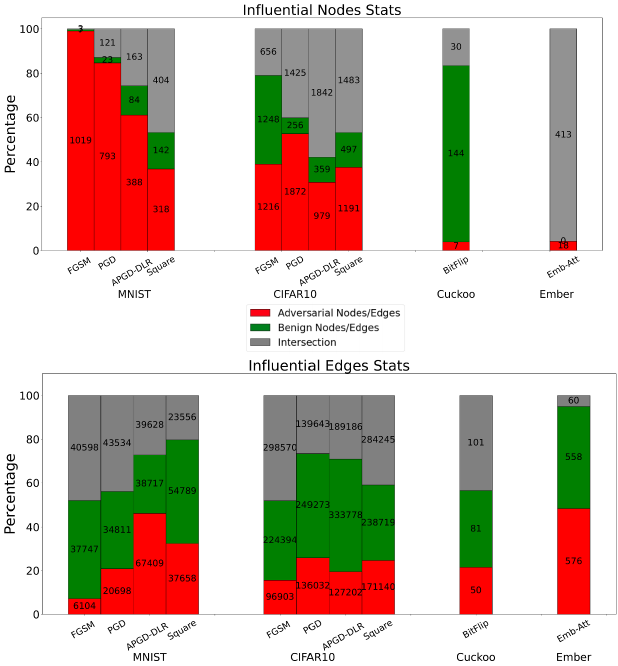}
    \caption{Percentage of influential nodes and edges.}
    \label{fig:charac-stats}
    \vspace{-0.3cm}
\end{figure}

\textbf{Relevant Characterization Statistics:} Figure \ref{fig:charac-stats} shows the number of nodes and edges with high attribution values, referred to as \textbf{Influential Nodes}, for each attack. Additionally, we identify nodes that are highly attributed only in adversarial settings (referred to as \textbf{Adversarial Nodes} and depicted in red) and those highly attributed only in benign settings (referred to as \textbf{Benign Nodes} and depicted in green). Nodes that are influential in both settings are presented in gray. We observe that \sysname{} can effectively identify influential nodes and edges in a DNN, distinguishing between benign and adversarial nodes and edges. This differentiation is crucial for potential model repair actions. Our analysis highlights a significant proportion of adversarial nodes, particularly in MNIST-DNN and CIFAR10-CNN models, where adversarial nodes account for $\approx 98\%$ (FGSM) and $\approx50\%$ (PGD: stronger than FGSM) of the total of influential nodes, respectively. Similarly, for CIFAR10-ResNet18 under the Square attack, adversarial nodes comprise $\approx$80\% of the influential nodes.
\textbf{\em These findings suggest that structural IPG-guided analysis highlights substructures (nodes and edges) critical to adversarial behavior. This structural view enables pinpointing of adversarial-only and benign-only pathways, offering richer repair targets.}

\subsection{Repair Action Efficacy}
\label{eval:loop}

\begin{table*}[t!]
\centering
\scalebox{0.72}{
\begin{tabular}{|l|l|l|l|l|l|l|l|l|l|l|l|l|l|}
\hline
\textbf{\large Model} & \textbf{ \large Layers}  & \multicolumn{6}{c|}{\textbf{\large Accuracy under Attack}}  & \multicolumn{6}{c|}{\begin{tabular}[c]{@{}c@{}}\textbf{\large Accuracy} \textbf{\large on Benign}\end{tabular}} \\ 

\hline

\multirow{7}{*}{ \textbf{ \large MNIST}} &\multicolumn{1}{c|}{ \cellcolor{lightgray} }                                                                                                                                   

& \multicolumn{1}{r|}{\textbf{FGSM}} 
& \multicolumn{1}{r|}{\textbf{PGD}} 
&\multicolumn{1}{r|}{\begin{tabular}[c]{@{}c@{}}\textbf{APGD}\\ \textbf{DLR} \end{tabular}}
& \textbf{SPSA}  
& \textbf{Square}
& \textbf{SIT}

& \multicolumn{1}{r|}{\begin{tabular}[c]{@{}c@{}}\textbf{\sysname{}}\\ \textbf{(FGSM)} \end{tabular}}
&\multicolumn{1}{r|}{\begin{tabular}[c]{@{}c@{}}\textbf{\sysname{}}\\ \textbf{(PGD)} \end{tabular}}
&\multicolumn{1}{r|}{\begin{tabular}[c]{@{}c@{}}\textbf{\sysname{}}\\ \textbf{APGD}\\ \textbf{-DLR} \end{tabular}}
& \multicolumn{1}{r|}{\begin{tabular}[c]{@{}c@{}}\textbf{\sysname{}}\\ \textbf{(SPSA)} \end{tabular}}
& \multicolumn{1}{r|}{\begin{tabular}[c]{@{}c@{}}\textbf{\sysname{}}\\ \textbf{(Square)} \end{tabular}}
& \multicolumn{1}{r|}{\begin{tabular}[c]{@{}c@{}}\textbf{\sysname{}}\\ \textbf{(SIT)} \end{tabular}}
         \\ 

\cline{2-14}

& \multicolumn{1}{c|}{No Actions}   & \multicolumn{1}{r|}{ {5.30\%}}  & \multicolumn{1}{r|}{ {3.50\%}} & \multicolumn{1}{r|}{ {40.56\%}}    & \multicolumn{1}{r|}{ {13.11\%}}   & \multicolumn{1}{r|}{ {38.43\%}}   & \multicolumn{1}{r|}{ {53.33\%}}  & \multicolumn{1}{r|}{ {97.7\%}} & \multicolumn{1}{r|}{ {97.70\%}} & \multicolumn{1}{r|}{ {97.70\%}} & \multicolumn{1}{r|}{ {97.70\%}}     & \multicolumn{1}{r|}{ {97.70 \%}} & \multicolumn{1}{r|}{ {97.70 \%}} \\ 

\cline{2-14}

&Layer 0           & \multicolumn{1}{r|}{\cellcolor{Mycolor2}\textbf {63.70\%}}     & \multicolumn{1}{r|}{ \cellcolor{Mycolor2}\textbf{65.80\%}}    & \multicolumn{1}{r|}{ \cellcolor{Mycolor2}\textbf{63.20\%}}     &    \cellcolor{Mycolor2} \textbf {66.53\%}    &  \cellcolor{Mycolor2}   \textbf {56.10\%}          &  \cellcolor{Mycolor2}   \textbf {70.93\%}                        & \multicolumn{1}{r|}{\cellcolor{Mycolor2}\textbf {97.50\%}}     & \multicolumn{1}{r|}{\cellcolor{Mycolor2}\textbf {97.60\%}}     & \multicolumn{1}{r|}{\cellcolor{Mycolor2}97.20\%}     &      \multicolumn{1}{r|}{\cellcolor{Mycolor2}  96.75\% }        &      \multicolumn{1}{r|}{ \cellcolor{Mycolor2} 97.30\% }      &      \multicolumn{1}{r|}{ \cellcolor{Mycolor2} 97.45\% }               \\ 


& Layer 1                                                                    & \multicolumn{1}{r|}{10.80\%}     & \multicolumn{1}{r|}{4.40\%}    & \multicolumn{1}{r|}{40.30\%}     &            \multicolumn{1}{r|}{36.30\% }  &            \multicolumn{1}{r|}{36.30\% } &            \multicolumn{1}{r|}{65.72\% }                 & \multicolumn{1}{r|}{95.20\%}     & \multicolumn{1}{r|}{94.10\%}     & \multicolumn{1}{r|}{97.40\%}     &          \multicolumn{1}{r|}{87.51\% }      &          \multicolumn{1}{r|}{97.30\% }         &          \multicolumn{1}{r|}{94.06\% }         \\ 

& Layer 2                                                               & \multicolumn{1}{r|}{9.00\%}     & \multicolumn{1}{r|}{11.70\%}    & \multicolumn{1}{r|}{42.00\%}       &              \multicolumn{1}{r|}{23.38\%}     &              \multicolumn{1}{r|}{35.10\%}  &              \multicolumn{1}{r|}{71.99\%}            & \multicolumn{1}{r|}{95.80\%}     & \multicolumn{1}{r|}{94.30\%}     & \multicolumn{1}{r|}{96.90\%}     &          \multicolumn{1}{r|}{ {92.49}\%}  &          \multicolumn{1}{r|}{ {97.60}\%}    &          \multicolumn{1}{r|}{ {97.46}\%}            \\ 

\cline{2-14}

& \begin{tabular}[c]{@{}l@{}}\{0,1,2\}\end{tabular}         & \multicolumn{1}{r|}{\textbf{73.10\%}}     & \multicolumn{1}{r|}{\textbf{73.76\%}}    & \multicolumn{1}{r|}{ \textbf{62.84\%}}    &         \multicolumn{1}{r|}{  \textbf{87.02\%}}   &         \multicolumn{1}{r|}{  \textbf{58.30\%}}    &         \multicolumn{1}{r|}{  \textbf{78.68\%}}                  & \multicolumn{1}{r|}{\textbf{97.11\%} }    & \multicolumn{1}{r|}{\textbf{97.03\%}}     & \multicolumn{1}{r|}{\textbf{97.50\%}}     &   \multicolumn{1}{r|}{ \textbf{ 95.87\%} }      &   \multicolumn{1}{r|}{ \textbf{ 97.10\%} }             &   \multicolumn{1}{r|}{ \textbf{ 97.02\%} }                 \\

& ARLS \cite{bartoldson2024adversarialrobustnesslimitsscalinglaw}                                                                                                                                                                                                                               & \multicolumn{1}{r|}{76.86\%} & \multicolumn{1}{r|}{73.43\%}    & \multicolumn{1}{r|}{79.16\%}         & \multicolumn{1}{r|}{\cellcolor{lightgray}}    & \multicolumn{1}{r|}{\cellcolor{lightgray}}    & \multicolumn{1}{r|}{\cellcolor{lightgray}}      & \multicolumn{1}{r|}{96.42\%} & \multicolumn{1}{r|}{96.5\%}      & \multicolumn{1}{r|}{96.23\%}         & \multicolumn{1}{r|}{\cellcolor{lightgray}}     & \multicolumn{1}{r|}{\cellcolor{lightgray}}               & \multicolumn{1}{r|}{\cellcolor{lightgray}}                                \\ 
\cline{2-14}
 & \begin{tabular}[c]{@{}l@{}}ARLS \cite{bartoldson2024adversarialrobustnesslimitsscalinglaw}   \\ + \sysname{} \end{tabular}                                                                                                                                                                                                                          & \multicolumn{1}{r|}{92.02\%} & \multicolumn{1}{r|}{92.05\%}    & \multicolumn{1}{r|}{85.70\%}         & \multicolumn{1}{r|}{\cellcolor{lightgray}}    & \multicolumn{1}{r|}{\cellcolor{lightgray}}    & \multicolumn{1}{r|}{\cellcolor{lightgray}}      & \multicolumn{1}{r|}{97.59\%} & \multicolumn{1}{r|}{97.32\%}      & \multicolumn{1}{r|}{96.69\%}         & \multicolumn{1}{r|}{\cellcolor{lightgray}}     & \multicolumn{1}{r|}{\cellcolor{lightgray}}               & \multicolumn{1}{r|}{\cellcolor{lightgray}}                                \\ 
\hline
\hline

\multirow{10}{*}{ \begin{tabular}[c]{@{}l@{}}\textbf{ \large CIFAR10} \\ \large \textbf{-ResNet18} \end{tabular}}
& \multicolumn{1}{c|}{No Actions}                                                  & \multicolumn{1}{r|}{11.08\%} & \multicolumn{1}{r|}{7.15\%}    & \multicolumn{1}{r|}{2.95\%}       &    \multicolumn{1}{r|}{24.03
\% }   &    \multicolumn{1}{r|}{4.14\% } &    \multicolumn{1}{r|}{0.79\% }     & \multicolumn{1}{r|}{87.94\%} & \multicolumn{1}{r|}{87.94\%} & \multicolumn{1}{r|}{87.94\%}    & \multicolumn{1}{r|}{87.94\%}    & \multicolumn{1}{r|}{87.94\%}  & \multicolumn{1}{r|}{87.94\%} 
\\ 
\cline{2-14}

& Layer 0                                                                                                 & \multicolumn{1}{r|}{14.30\%} & \multicolumn{1}{r|}{11.10\%}    & \multicolumn{1}{r|}{3.38\%}         & \multicolumn{1}{r|}{33.90\%}  & \multicolumn{1}{r|}{21.45\%}  & \multicolumn{1}{r|}{11.80\%}       & \multicolumn{1}{r|}{86.23\%} & \multicolumn{1}{r|}{87.02\%}      & \multicolumn{1}{r|}{86.79\%}         &   \multicolumn{1}{r|}{87.12 \%}    &   \multicolumn{1}{r|}{85.90 \%}   &   \multicolumn{1}{r|}{86.12 \%}                                   \\ 
&Layer 1                                                                                                 & \multicolumn{1}{r|}{21.04\%} & \multicolumn{1}{r|}{21.00\%}    & \multicolumn{1}{r|}{3.33\%}         & \multicolumn{1}{r|}{\cellcolor{Mycolor2} \textbf{40.07\%}}    & \multicolumn{1}{r|}{26.34\%}  & \multicolumn{1}{r|}{10.71\%}       & \multicolumn{1}{r|}{85.12\%} & \multicolumn{1}{r|}{86.23\%}      & \multicolumn{1}{r|}{87.47\%}         &       \multicolumn{1}{r|}{\cellcolor{Mycolor2} \textbf{86.84\%} }   &       \multicolumn{1}{r|}{85.62\% }   &       \multicolumn{1}{r|}{85.76\% }                                    \\ 


&Layer 2                                                                                                                                                                                                                                & \multicolumn{1}{r|}{35.22\%} & \multicolumn{1}{r|}{13.04\%}    & \multicolumn{1}{r|}{16.88\%}         &  \multicolumn{1}{r|}{33.11 \% }  &  \multicolumn{1}{r|}{25.85 \% }  &  \multicolumn{1}{r|}{\cellcolor{Mycolor2} \textbf{13.31} \% }   & \multicolumn{1}{r|}{84.97\%} & \multicolumn{1}{r|}{86.86\%}      & \multicolumn{1}{r|}{86.12\%}         &      \multicolumn{1}{r|}{86.97\%}    &      \multicolumn{1}{r|}{86.10\%}   &      \multicolumn{1}{r|}{\cellcolor{Mycolor2} \textbf{85.63}\%}                                      \\ 
&Layer 3                                                                                                                                                                                                                                & \multicolumn{1}{r|}{45.34\%} & \multicolumn{1}{r|}{\cellcolor{Mycolor2} \textbf{60.73\%}}    & \multicolumn{1}{r|}{\cellcolor{Mycolor2} \textbf{25.07\%}}         & \multicolumn{1}{r|}{35.78\% } & \multicolumn{1}{r|}{\cellcolor{Mycolor2} \textbf{36.06\%} }  & \multicolumn{1}{r|}{9.99\% }    & \multicolumn{1}{r|}{85.41\%} & \multicolumn{1}{r|}{\cellcolor{Mycolor2} \textbf{85.58\%}}      & \multicolumn{1}{r|}{\cellcolor{Mycolor2} \textbf{85.73\%}}         &   \multicolumn{1}{r|}{  86.51\% }  &   \multicolumn{1}{r|}{ \cellcolor{Mycolor2} \textbf{84.96\%} }   &   \multicolumn{1}{r|}{  85.23\% }                                           \\ 
&Layer 4                                                                                                                                                                                                                         & \multicolumn{1}{r|}{\cellcolor{Mycolor2} \textbf{58.19\%}} & \multicolumn{1}{r|}{46.21\%}    & \multicolumn{1}{r|}{3.47\%}         & \multicolumn{1}{r|}{31.34\%}    & \multicolumn{1}{r|}{20.85\%}  & \multicolumn{1}{r|}{9.99\% }       & \multicolumn{1}{r|}{\cellcolor{Mycolor2} \textbf{85.13\%}} & \multicolumn{1}{r|}{86.12\%}      & \multicolumn{1}{r|}{87.50\%}         &      \multicolumn{1}{r|}{87.32\% }           &      \multicolumn{1}{r|}{84.67\% }       &      \multicolumn{1}{r|}{86.12\% }               \\ 
&Layer 5                                                                                                                                                                                                                                                                                                                                                           & \multicolumn{1}{r|}{ {11.72\%}} & \multicolumn{1}{r|}{7.59\%}    & \multicolumn{1}{r|}{3.33\%}         &   \multicolumn{1}{r|}{25.69\%}    &   \multicolumn{1}{r|}{28.13\%}  & \multicolumn{1}{r|}{9.99\% }      & \multicolumn{1}{r|}{87.72	\%} & \multicolumn{1}{r|}{87.69\%}      & \multicolumn{1}{r|}{87.95\%}         &   \multicolumn{1}{r|}{87.24\%}        &   \multicolumn{1}{r|}{85.82\%}     &   \multicolumn{1}{r|}{86.94\%}                   \\ 
&Layer 6                                                                                                                                                                                                                             & \multicolumn{1}{r|}{11.60\%} & \multicolumn{1}{r|}{7.49\%}    & \multicolumn{1}{r|}{3.49\%}         & \multicolumn{1}{r|}{25.30\%}     & \multicolumn{1}{r|}{8.70\%}     & \multicolumn{1}{r|}{9.99\% }   & \multicolumn{1}{r|}{87.73\%} & \multicolumn{1}{r|}{87.80\%}      & \multicolumn{1}{r|}{87.95\%}         & \multicolumn{1}{r|}{87.62\%}     & \multicolumn{1}{r|}{87.90\%}   & \multicolumn{1}{r|}{86.73\%}                              \\ 
                     \cline{2-14}
& \begin{tabular}[c]{@{}l@{}}\{4,3,2\\1,0,5,6\}\end{tabular}                                                                   & \multicolumn{1}{r|}{\textbf{64.72\% }} & \multicolumn{1}{r|}{\textbf{63.29\% }}    & \multicolumn{1}{r|}{\textbf{26.29\%} }         & \multicolumn{1}{r|}{\textbf{44.18\%}}     & \multicolumn{1}{r|}{\textbf{37.57\%}}  & \multicolumn{1}{r|}{\textbf{17.65\%}}     & \multicolumn{1}{r|}{\textbf{84.46\%} } & \multicolumn{1}{r|}{\textbf{84.67\%} }      & \multicolumn{1}{r|}{\textbf{85.52\% }}         &    \multicolumn{1}{r|}{\textbf{86.04 \%}}    &    \multicolumn{1}{r|}{\textbf{85.21 \%}}         &    \multicolumn{1}{r|}{\textbf{85.91 \%}}                         \\ 

& ARLS \cite{bartoldson2024adversarialrobustnesslimitsscalinglaw}                                                                                                                                                                                                                         & \multicolumn{1}{r|}{34.68\%} & \multicolumn{1}{r|}{35.20\%}    & \multicolumn{1}{r|}{13.21\%}      & \multicolumn{1}{r|}{\cellcolor{lightgray}}     & \multicolumn{1}{r|}{\cellcolor{lightgray}}    & \multicolumn{1}{r|}{\cellcolor{lightgray}}      & \multicolumn{1}{r|}{84.73\%} & \multicolumn{1}{r|}{85.62\%}      & \multicolumn{1}{r|}{84.10\%}         & \multicolumn{1}{r|}{\cellcolor{lightgray}}     & \multicolumn{1}{r|}{\cellcolor{lightgray}}    & \multicolumn{1}{r|}{\cellcolor{lightgray}}                              \\ 
\cline{2-14}
 & \begin{tabular}[c]{@{}l@{}}ARLS \cite{bartoldson2024adversarialrobustnesslimitsscalinglaw}   \\ + \sysname{} \end{tabular}                                                                                                                                                                                                                          & \multicolumn{1}{r|}{55.71 \%} & \multicolumn{1}{r|}{45.20\%}    & \multicolumn{1}{r|}{23.14\%}         & \multicolumn{1}{r|}{\cellcolor{lightgray}}    & \multicolumn{1}{r|}{\cellcolor{lightgray}}    & \multicolumn{1}{r|}{\cellcolor{lightgray}}      & \multicolumn{1}{r|}{85.13\%} & \multicolumn{1}{r|}{84.69\%}      & \multicolumn{1}{r|}{84.71\%}         & \multicolumn{1}{r|}{\cellcolor{lightgray}}     & \multicolumn{1}{r|}{\cellcolor{lightgray}}               & \multicolumn{1}{r|}{\cellcolor{lightgray}}                                \\ 
\hline

  \end{tabular}}

\caption{\small  {\bf Repair actions per layer  for CIFAR10 and MNIST}. For each $layer_i$ and attack ($att_i$) we use \sysname{}($att_i$) to produce actions on $layer_i$ and evaluate their impact on the model's  Benign accuracy and its accuracy against $att_i$.  Results highlighted in \cellcolor{MyColor2}{cyan} represent the best performing layer results in terms of adversarial accuracy. 
} 
\label{tab:summary}
\end{table*}

\begin{table}[!h]
\centering
\scalebox{.47}{
\begin{tabular}{|l|l|l|l|rlll|rlll|}
\hline
\textbf{\LARGE Model} & \textbf{ \LARGE Layers}  & \begin{tabular}[c]{@{}l@{}}\textbf{ \LARGE Layer Type} \\ (\# Neurons) \end{tabular} & \begin{tabular}[c]{@{}l@{}}\textbf{\LARGE Ac Fu}\end{tabular} & \multicolumn{4}{c|}{\begin{tabular}[c]{@{}c@{}}\textbf{\LARGE Accuracy}\\ \textbf{\LARGE under Attack}\end{tabular}}  & \multicolumn{4}{c|}{\begin{tabular}[c]{@{}c@{}}\textbf{\LARGE Accuracy}\\ \textbf{\LARGE on Benign}\end{tabular}}\\ 

\hline

\multirow{6}{*}{ \textbf{ \LARGE Cuckoo}} & \multicolumn{3}{r|}{ \cellcolor{lightgray}  }                                                              & \multicolumn{4}{c|}{\begin{tabular}[c]{@{}c@{}}\textbf{\ \sysname{}}\\ \textbf{(Bit-Flip)}\end{tabular}}                                                        & \multicolumn{4}{c|}{\begin{tabular}[c]{@{}c@{}}\textbf{\ \sysname{}}\\ \textbf{(Bit-Flip)}\end{tabular}}                                                                                                                   \\ \cline{2-12}

 & \multicolumn{3}{c|}{No Actions}             &   \multicolumn{4}{c|}{  52.58\%}          &  \multicolumn{4}{c|}{95.4 \%}               \\ \cline{2-12}
&Layer 0               & \textbf{Dense (32)  }    & \textbf{Relu  }         &  \multicolumn{4}{c|}{\cellcolor{Mycolor2}  \textbf{87.85\% }    }        &   \multicolumn{4}{c|}{ \cellcolor{Mycolor2} \textbf{85.50\% }   }                  \\ 
& Layer 1             & Dense (18)      & Relu              &   \multicolumn{4}{c|}{ 57.61\%  }           &  \multicolumn{4}{c|}{  92.68\%  }                  \\ 
& Layer 2               & Dense  (16)     & Relu                 &  \multicolumn{4}{c|}{54.03 \%}             &   \multicolumn{4}{c|}{93.54\%  }                \\ 
\cline{2-12}
&\begin{tabular}[c]{@{}l@{}}\{0,1,2\}\end{tabular} &  \multicolumn{2}{c|}{All Layers}          &   \multicolumn{4}{c|}{ \textbf{87.85\%  }}          & \multicolumn{4}{c|}{\textbf{94.96\%    }}          
          \\ 
          \cline{2-12}
& ARLS \cite{bartoldson2024adversarialrobustnesslimitsscalinglaw}               & -    & -                 &  \multicolumn{4}{c|}{90.05 \%}             &   \multicolumn{4}{c|}{94.5\%  }                
\\
\cline{2-12}
&\begin{tabular}[c]{@{}l@{}}ARLS + \\ \sysname{} \end{tabular} & \multicolumn{2}{c|}{-}                   & \multicolumn{4}{c|}{\textbf{91.05\%}    }          &   \multicolumn{4}{c|}{ \textbf{94.59\%}  }             \\
          \hline

\multirow{14}{*}{ \textbf{ \LARGE Ember}} & \multicolumn{3}{r|}{ \cellcolor{lightgray}  }                                                                          & \multicolumn{4}{c|}{\begin{tabular}[c]{@{}c@{}}\textbf{\ \sysname{}}\\ \textbf{(Emb-Att)}\end{tabular}}          & \multicolumn{4}{c|}{\begin{tabular}[c]{@{}c@{}}\textbf{\ \sysname{}}\\ \textbf{(Emb-Att)}\end{tabular}}                                                                                                 \\ \cline{2-12}                                                   
&\multicolumn{3}{c|}{No Actions}          & \multicolumn{4}{c|}{ 0.05\%  }            & \multicolumn{4}{c|}{94.88\%  }                    \\ \cline{2-12}
&Layer 0                & \textbf{Dense (2381) }     & \textbf{Relu  }                 & \multicolumn{4}{c|}{\cellcolor{Mycolor2}  \textbf{77.16\%}     }         & \multicolumn{4}{c|}{  \cellcolor{Mycolor2}  \textbf{90.03\%} }               \\ 
&Layer 1  & Dense  (128)    & -               & \multicolumn{4}{c|}{45.40\%   }             & \multicolumn{4}{c|}{93.14\% }                   \\ 
&Layer 2  & BatchNorm1d (64)  & Relu         & \multicolumn{4}{c|}{ 57.21\%   }          & \multicolumn{4}{c|}{ 90.12\% }             \\ 
&Layer 3  & Dropout (64)   & -                 & \multicolumn{4}{c|}{ 57.21\% }           & \multicolumn{4}{c|}{ 90.12\% }                      \\ 
&Layer 4  & Dense  (32)    & \multicolumn{1}{r|}{-}                          & \multicolumn{4}{c|}{ 43.44\%   }         & \multicolumn{4}{c|}{92.54\%  }                 \\ 
&Layer 5  & BatchNorm1d (32) & Relu            & \multicolumn{4}{c|}{37.55\% }             & \multicolumn{4}{c|}{ \textbf{94.37\% } }            \\ 
&Layer 6  & Dropout (32)   & \multicolumn{1}{r|}{-}                   & \multicolumn{4}{c|}{ 44.87\%  }           & \multicolumn{4}{c|}{93.23\% }             \\ 
&Layer 7  & Dense (16)   & \multicolumn{1}{r|}{-}                  & \multicolumn{4}{c|}{ 44.47 \%}              & \multicolumn{4}{c|}{93.21\%  }               \\ 
&Layer 8  & BatchNorm1d (16)  & Relu          &    \multicolumn{4}{c|}{38.27\% }             & \multicolumn{4}{c|}{90.25\%}                  \\ 
&Layer 9  & Dropout (16)   &                  & \multicolumn{4}{c|}{ 45.71\%   }          & \multicolumn{4}{c|}{93.07\%  }            \\ 
\cline{2-12}
&\begin{tabular}[c]{@{}l@{}}\{0,2,3,9,1,\\6,7,4,5,8\} \end{tabular} & \multicolumn{2}{c|}{All Layers}                   & \multicolumn{4}{c|}{\textbf{84.34\%}    }          &   \multicolumn{4}{c|}{ \textbf{94.56\%}  }             \\  \cline{2-12}
& ARLS \cite{bartoldson2024adversarialrobustnesslimitsscalinglaw}               & -    & -                 &  \multicolumn{4}{c|}{86.31 \%}             &   \multicolumn{4}{c|}{94.25\%  }                \\
\cline{2-12}
&\begin{tabular}[c]{@{}l@{}}ARLS + \\ \sysname{} \end{tabular} & \multicolumn{2}{c|}{-}                   & \multicolumn{4}{c|}{\textbf{87.12\%}    }          &   \multicolumn{4}{c|}{ \textbf{94.01\%}  }             \\
\hline

 \end{tabular}}
\caption{\small \bf Repair actions per layer  for Cuckoo and Ember. Ac Fu = Activation Function.}
\label{tab:summary2}
\vspace{-2.3em}
\end{table}

To address \textbf{RQ2} and \textbf{RQ3}, we assess \sysname{}'s effectiveness in identifying repair actions for robustness enhancement. First, we use the actions evaluation approach (\S \ref{subsec:act-eval}) to determine the most effective sequence of cumulative actions generated by Algorithm \ref{alg:Action-gen}. We then investigate the impact of these actions on model performance for both adversarial and benign inputs. Additionally, we examine the transferability of \sysname{}'s actions across different attacks.  Finally, we provide detailed results of actions performed per-layer for each model, as shown in Tables \ref{tab:summary} and \ref{tab:summary2}.

\begin{figure*}[t!]
    \centering
    \captionsetup[subfloat]{labelformat=empty}
    
    \subfloat[]{\includegraphics[width=0.33\textwidth, ]{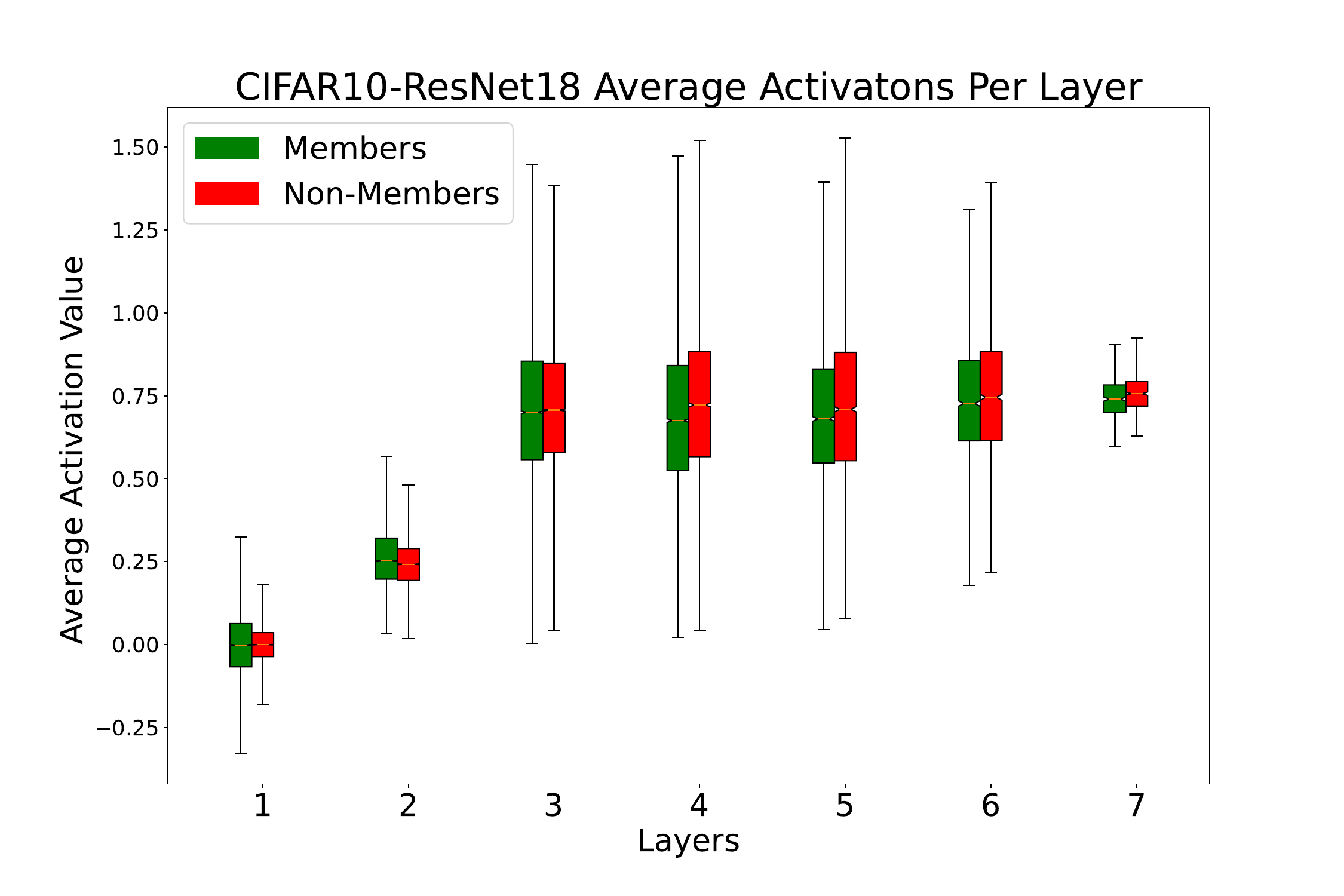}}
     \subfloat[]{\includegraphics[width=0.33\textwidth]{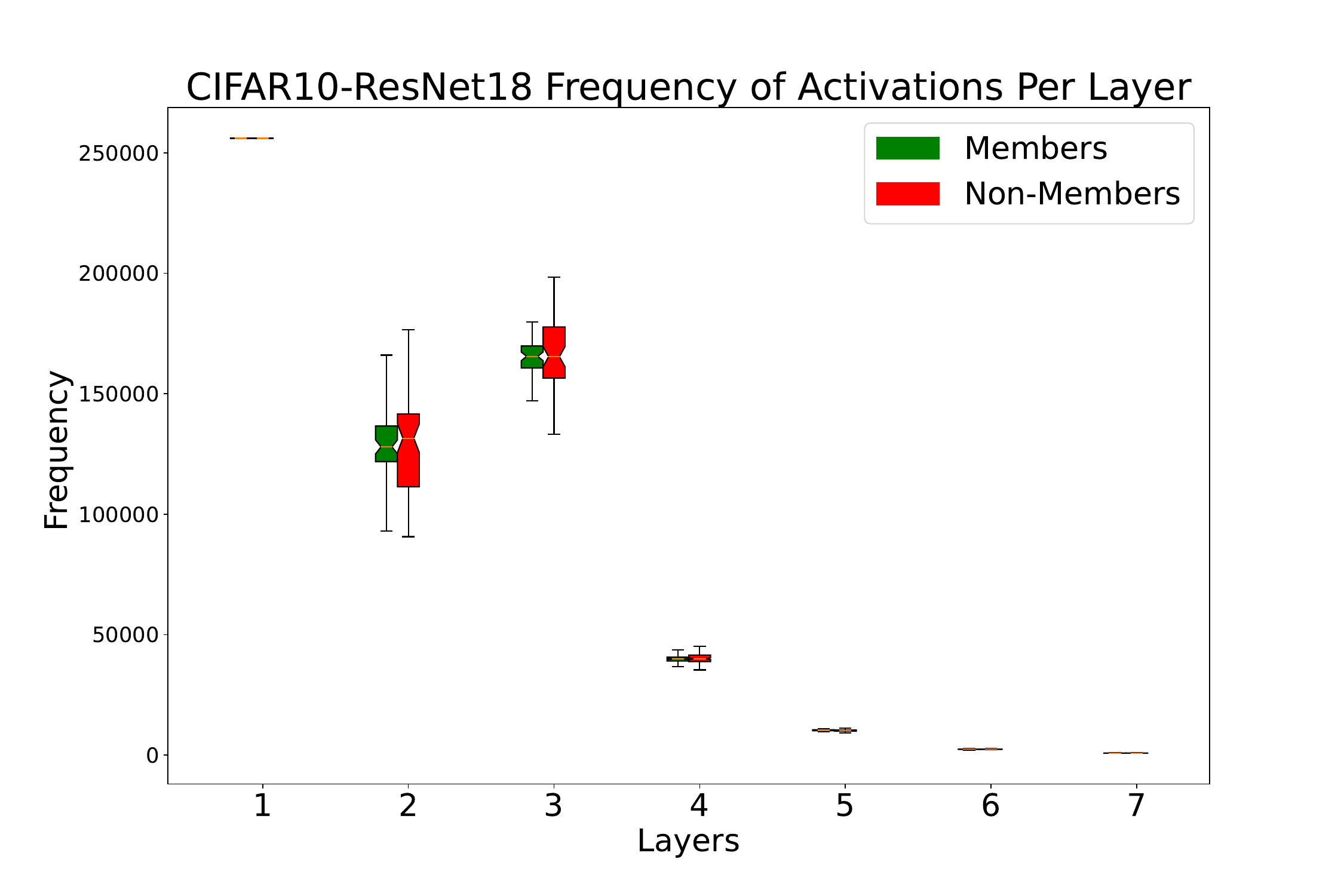}}
     \subfloat[]{\includegraphics[width=0.33\textwidth]{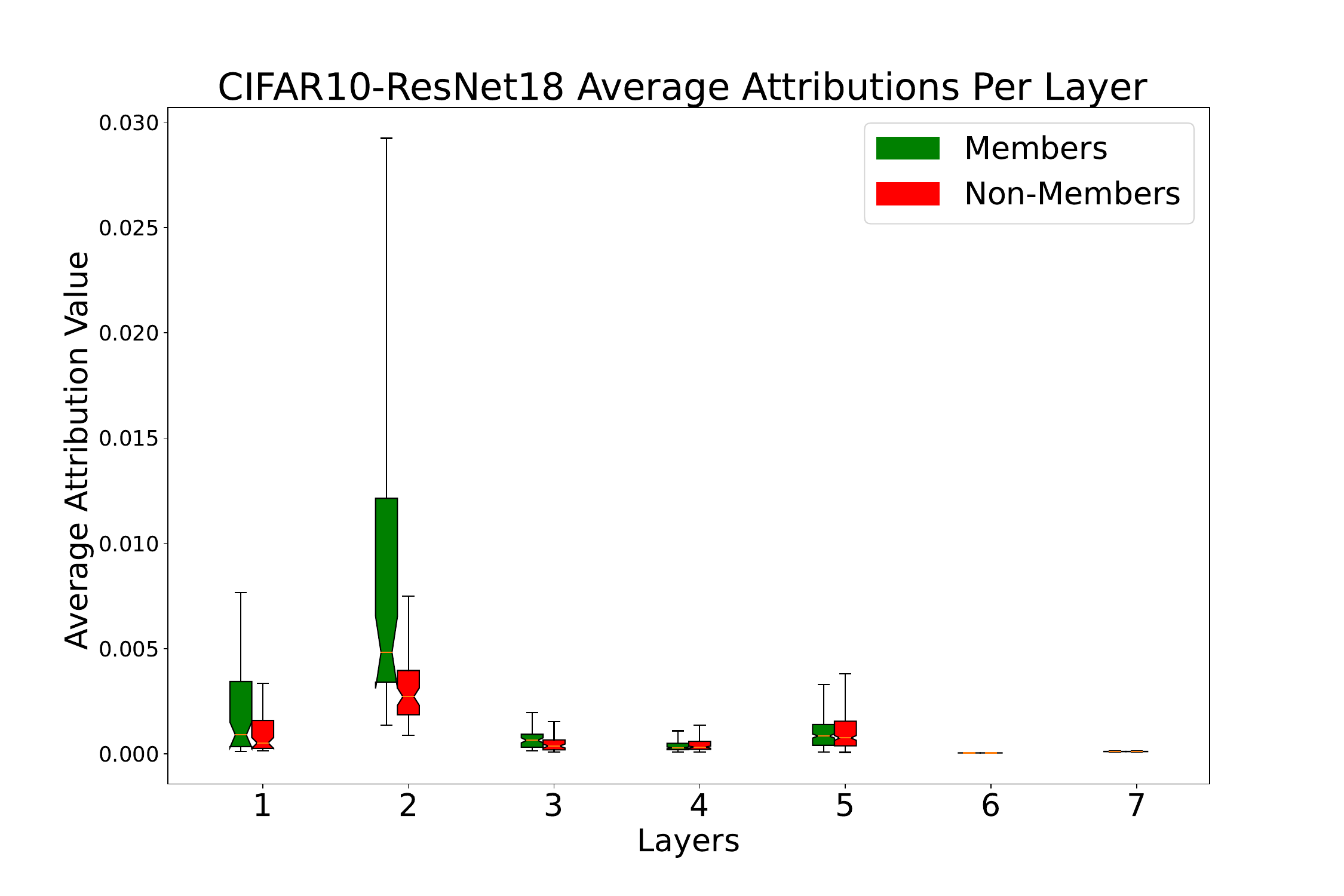}}
    \vspace{-2.5em}
    \subfloat[ ]{\includegraphics[width=0.33\textwidth]{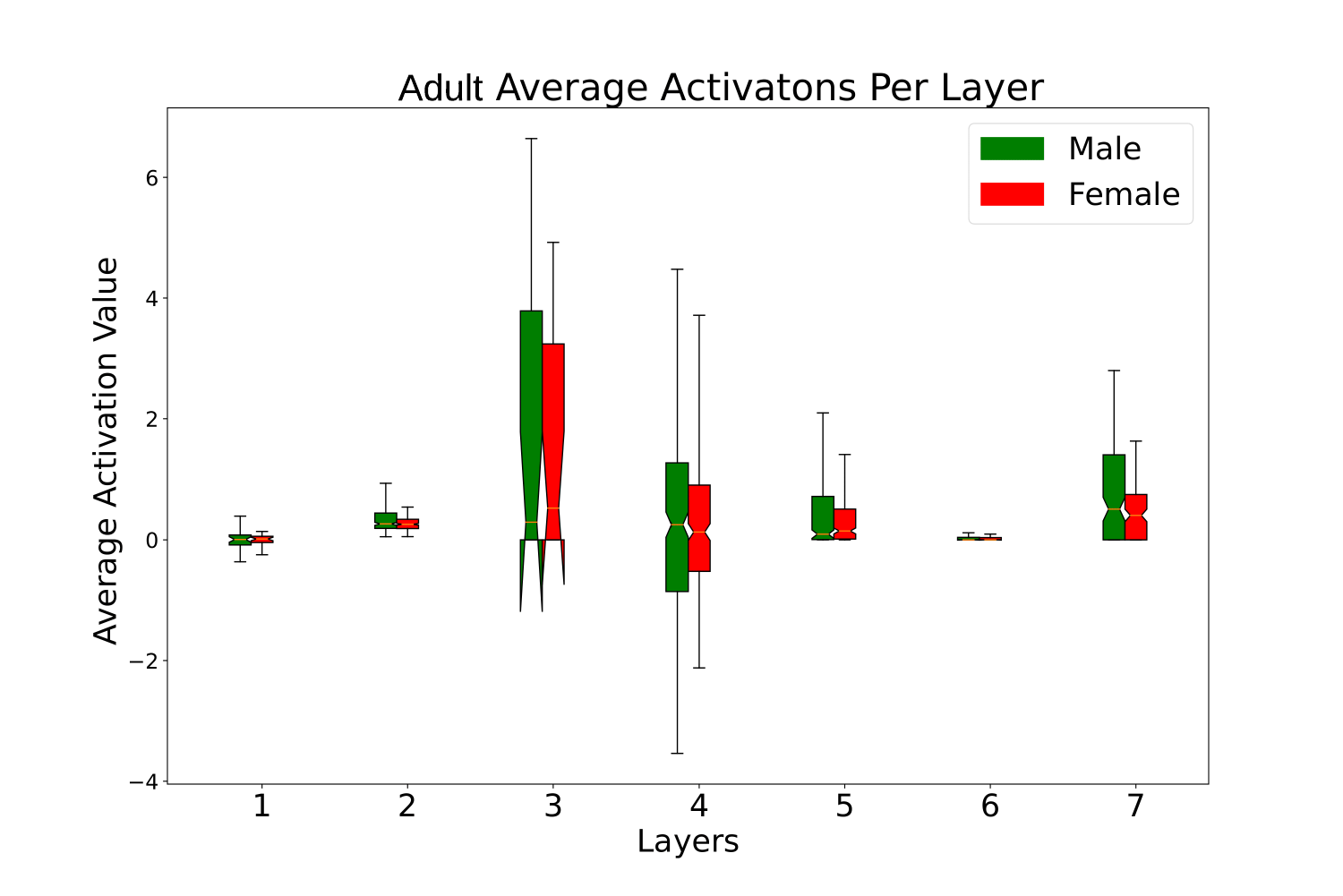}}
     \subfloat[]{\includegraphics[width=0.33\textwidth]{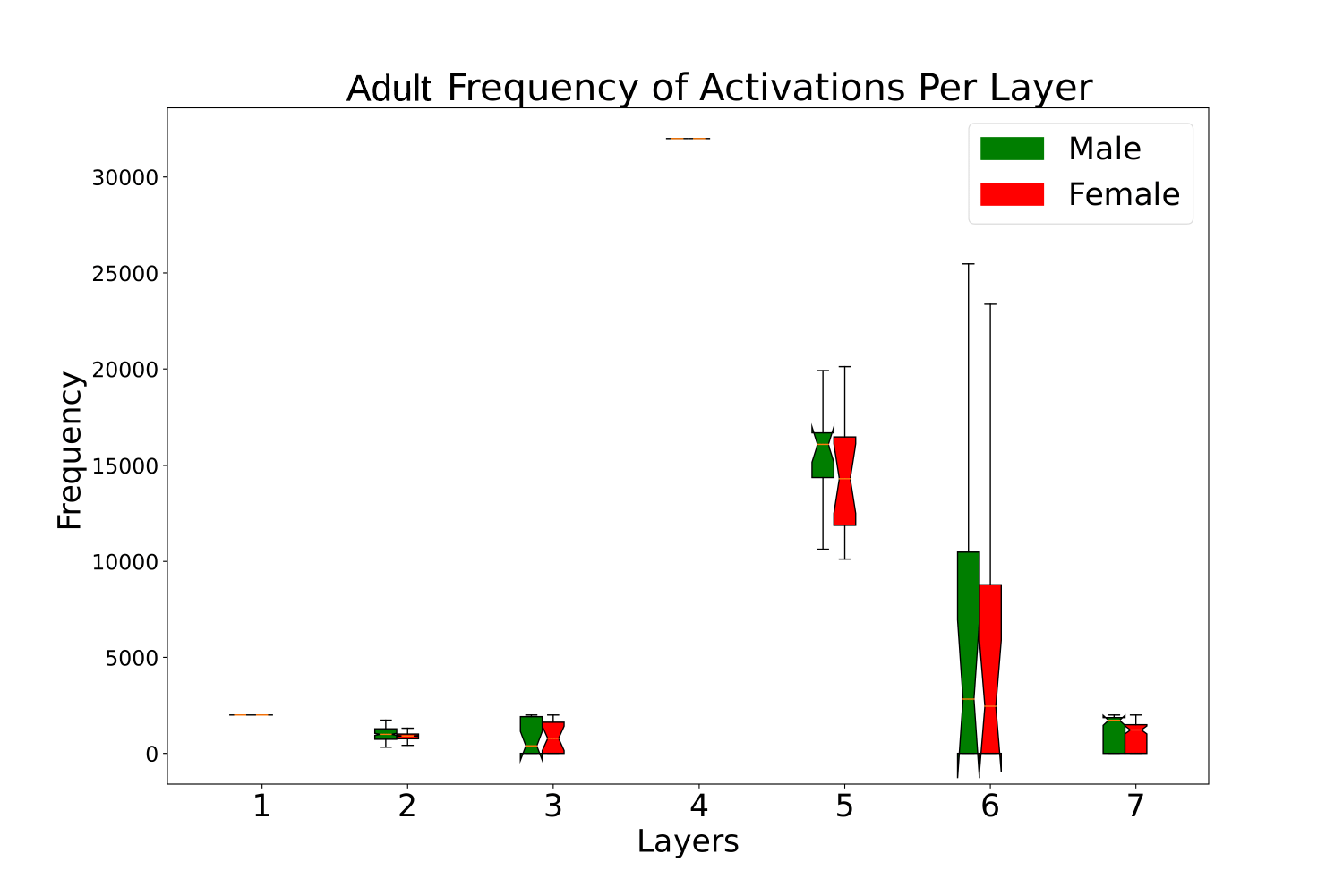}}
     \subfloat[]{\includegraphics[width=0.33\textwidth]{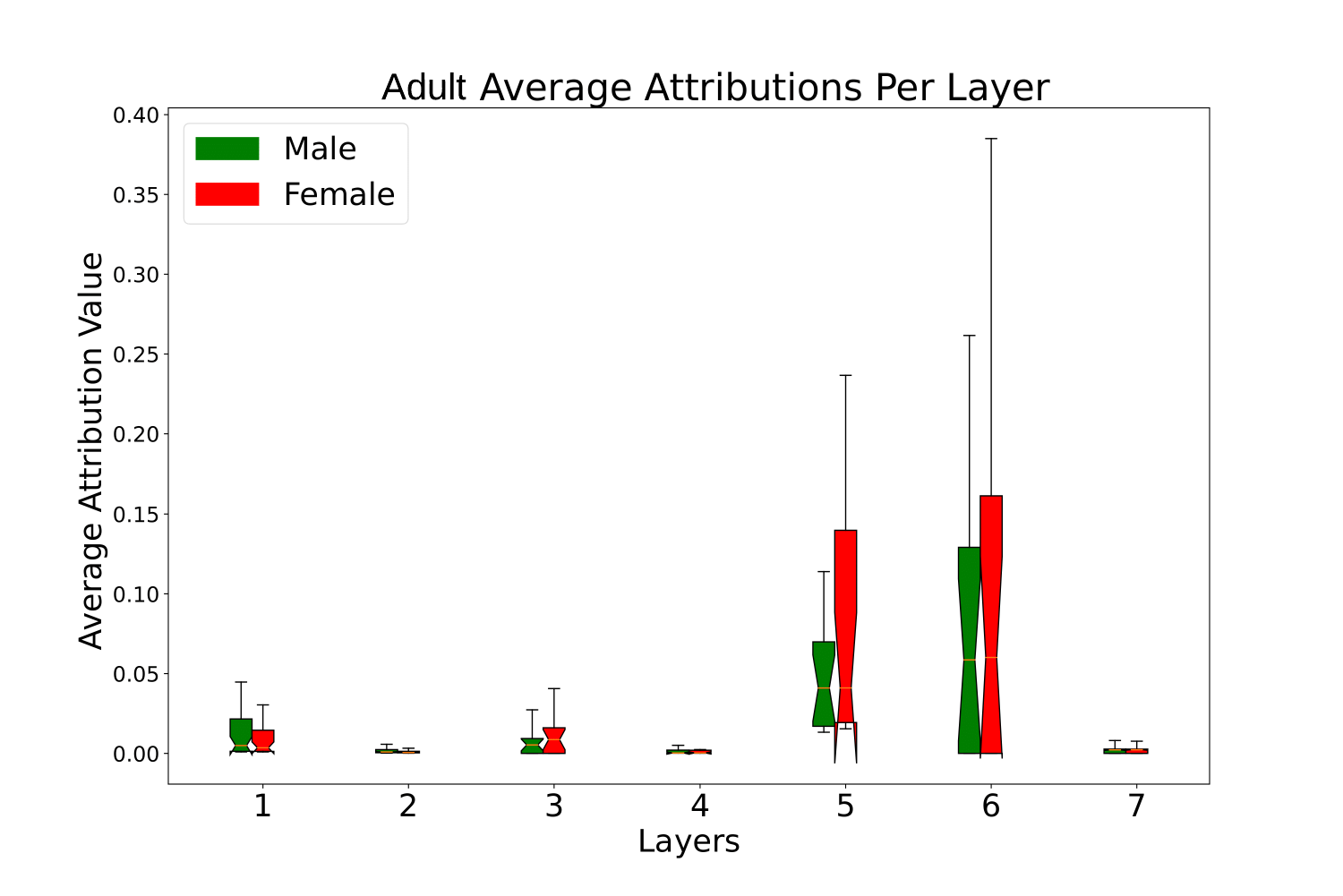}}
         \vspace{-1.4em}
      \caption{\small \bf Preliminary results on \sysname{}'s adaptability to privacy auditing (top row) and fairness analysis (bottom row).}
    \label{fig:charac-priv-fairness}
  
\end{figure*}

\textbf{Model Repair Effectiveness}:
To evaluate \sysname{}'s effectiveness in generating node-level repair actions for robustness, we applied model characterization and action generation on MNIST-DNN and CIFAR10-ResNet18 under various attacks (rows 4–8 in Table \ref{tab:tranf}). Each \sysname{} instance (denoted as "\sysname{} (\texttt{[Attack]})") is evaluated on both benign and all studied attacks (columns 2–18). For Cuckoo-DNN and Ember-DNN, we evaluate against Bit-flip and Emb-Att (row 9). Table \ref{tab:tranf} reports the best cumulative action sequence results across all layers.

We first examine model performance against the attack used to guide \sysname{} (bold values). For MNIST-DNN, we observe notable gains: +67.21\% (FGSM), +71.06\% (PGD), +22.28\% (APGD-DLR), +71.99\% (SPSA), +18.31\% (Square), and +25.35\% (SIT), with minimal benign accuracy loss. Cuckoo-DNN and Ember-DNN also show improvements of +35.27\% and +80.39\%, respectively.

CIFAR10-ResNet results show less impact due to high parameter counts and activation diversity, complicating the selection of reference values via $Reference\_Dist\_Agg$ (Algorithm \ref{alg:Action-gen}, line 24). To address this, we introduce an adversarial input detector to apply \sysname{} actions selectively. On ResNet18, we observe improvements of +55.46\% (FGSM), +60.15\% (PGD),+22.14\% (APGD-DLR), +20.13\% (SPSA), +34.12\% (Square), and +16.95\% (SIT). These gains highlight the value of IPG-driven characterization via \sysname{}. Accuracy loss remains negligible for MNIST-DNN, Cuckoo-DNN, and Ember-DNN, and is limited to ~1–2\% for CIFAR10-ResNet18. We further evaluate this for CIFAR10-CNN (Table \ref{tab:cifar10-results} in the  Appendix).
\textbf{\em \sysname{} repair actions significantly improve robustness: on average, adversarial accuracy increases by about \textbf{55\%}, often with negligible impact on benign performance.}

\textbf{Does \sysname{} Advance State-of-the-Art Defenses?} 
To evaluate \sysname{}'s contribution to advancing defenses, we augment it with ARLS~\cite{bartoldson2024adversarialrobustnesslimitsscalinglaw}, an adversarial training defense tested against white-box attacks (state-of-the-art as of this writing). We also compare \sysname{} across all datasets and three white-box attacks (FGSM, PGD, and A-PGD). On Ember and Cuckoo datasets, \sysname{} performs comparably to ARLS (Table \ref{tab:summary2}). However, on image datasets, \sysname{} outperforms ARLS significantly, averaging an 18\% gain on CIFAR10-CNN and CIFAR10-ResNet18 (Table \ref{tab:summary}). ARLS performs better on MNIST, with a 6\% average advantage.
\textbf{\em Overall, combining ARLS~\cite{bartoldson2024adversarialrobustnesslimitsscalinglaw} with \sysname{} notably improves adversarial robustness, underscoring the value of IPG-guided characterization in enhancing existing defenses}.

\textbf{Inter-setting Transferability of Repair Actions}:
Now, we shift our focus to assessing whether repair actions on one attack (\sysname{}($Att_i$)) transfer to other attacks $Att_j$. While \sysname{}'s repair actions are most effective against the specific attack they target, they also offer partial robustness gains against other attacks, highlighting some degree of inter-attack transferability. This transfer is especially notable for simpler attacks like FGSM and PGD, with instances such as \sysname{}(FGSM) on MNIST-DNN significantly boosting robustness against PGD. \textbf{\em Although transferability decreases with attack complexity, repair actions generated for one attack often generalize to other attacks, 
delivering partial robustness improvements across diverse attacks}.

\textbf{Per-Layer Analysis:} 
Table \ref{tab:tranf} shows improved performance following repair actions applied to selected nodes across all layers. This optimization proceeds in two phases: first, identifying the best action sequence for each layer independently (\S \ref{subsec:act-eval}); second, applying those actions across layers in an order that maximizes performance. To find the most effective ordering, we exhaustively explored all combinations and ranked layers by $Cum\_TS$ (Figure \ref{fig:cum-act}), with optimal configurations summarized in the final row of Tables \ref{tab:summary} and \ref{tab:summary2} for each model.
We also evaluate the effect of actions applied to individual layers by reporting $Cum\_TS$ (filtering for improvements), accuracy on benign inputs, and accuracy under attack (Figure \ref{fig:cum-act}). Results show that even acting on a single layer can yield substantial robustness gains, up to +77\% accuracy on adversarial data and $Cum\_TS = 71.07\%$ (Layer 0 of Ember-DNN), with minimal benign accuracy loss.

Additionally, we report the impact of actions performed on each layer, focusing on the resulting $Cum\_TS$ that filters only the improving actions (Figure \ref{fig:cum-act}), accuracy on benign input, and accuracy on adversarial input. In tables \ref{tab:summary} and \ref{tab:summary2}, we observe, across all models, that acting on just one layer is sufficient to start noticing robustness improvement, which can reach up to $+77\%$ more accuracy on adversarial data and $Cum\_TS = 71.07\%$ (Layer 0 of Ember-DNN in figure \ref{fig:cum-act}), reflecting a substantial increase in robustness, compared to a minimal loss in accuracy on benign data ($-4\%$ without the adversarial detector). Additionally, tables \ref{tab:summary} and \ref{tab:summary2} indicate that \sysname{}'s characterization led to identifying influential repair actions on all layers regardless of their type (e.g., ReLu, Conv2D, etc), dimension and order. Nevertheless, it is noteworthy that actions determined on early layers (e.g., layers 0, 1, etc) can be way more influential than actions implemented on final layers of the model. In particular, on average across all models, actions performed only on one layer induce $\approx 55\%$ more accuracy under attack. It contributes in $\approx 85\%$ of the overall improvement brought about by all layers, on average in all models. This is explained by the immediate impact of adversarial perturbations on the first layers, which propagates to subsequent layers during the inference computation.
 \textbf{\em Repairs applied to early layers yield the largest gains in robustness, 
sometimes up to +77\% accuracy against adversarial examples. The early layers contribute the majority of the overall improvement in robustness.}


\subsection{Adaptability to Other Repair 
Goals}\label{subsec:adaptability}
 To assess \sysname{}'s adaptability beyond robustness-enhancing repair, we characterize IPGs for privacy- and fairness-relevant repair goals. Figure \ref{fig:charac-priv-fairness} shows preliminary results on \sysname{}'s adaptability to privacy auditing (top row) and fairness analysis (bottom row).

\textbf{Adaptability to Privacy as Repair Goal:}  Due to memorization and overfitting, DNNs are vulnerable to privacy threats such as membership inference \cite{MemInfeAttack}. To characterize DNN inference dynamics for privacy enhancing repair, we extract IPGs of a CIFAR10-ResNet18 model on members and non-members of its training data. 

The top row of Figure \ref{fig:charac-priv-fairness} shows box-plots for \sysname{}'s empirical and structural characterization approaches. We observe that, similar to the robustness enhancing case studies, \sysname{} identifies layers of the DNN where IPGs of members (green) and non-members (red) are distinct, suggesting that \textbf{\em privacy-enhancing repair actions can be applied to minimize the distinction between IPGs for members and non-members}. 

\textbf{Adaptability to Fairness Analysis:} Fairness is critical in DNNs used for high-stakes applications for decisions such as loan approval and predictive diagnostics. DNNs may reflect biases from their training data and the data processing pipeline, leading to unfair decisions against protected attributes like gender or race \cite{fairness}. To explore \sysname{}'s potential for fairness analysis, we train a DNN on the UCI Adult~\cite{misc_adult_2} dataset to predict income. The model has accuracy disparity between male (83\%) and female (91\%). Across these two groups, we characterize IPGs of the DNN's inferences empirically and structurally. The bottom row of Figure \ref{fig:charac-priv-fairness} shows box-plots of distinctions between IPG-guided characterization for each layer. Based on these differences, \textbf{\em the DNN can be repaired towards eliminating accuracy disparity or error-rate disparity between demographic groups in a dataset}.

\section{Conclusion}\label{sec: concl}
This paper presented \sysname{}, a task- and model-agnostic system that enables DNN deployers to capture and analyze the inference-time computational flow of a model via Inference Provenance Graphs (IPGs). \sysname{} systematically characterizes IPG across multiple settings, both structurally and empirically, and leverages the insights to guide targeted model repairs aligned with specific repair goals. 
Through extensive case studies focused on adversarial robustness-enhancing repair goal, \sysname{} demonstrates its effectiveness in enhancing model resilience by identifying and modifying key nodes and edges, particularly in early layers, with negligible impact on benign accuracy. 
Beyond robustness, \sysname{} also shows potential in mitigating privacy risks and reducing performance disparities, highlighting the broader applicability of inference provenance-driven characterization and repair of DNNs.

\section*{Acknowledgments}
We thank the anonymous reviewers for their insightful feedback
that improved the paper. This work was supported by the National
Science Foundation (NSF) awards CNS-2238084 and OAC-2103596.

\bibliographystyle{IEEEtran}
\bibliography{main}
\section{Appendix}\label{sec:appendix}
    
    
\subsection{Repair Actions Evaluation}
\begin{figure}
\vspace{0.2cm}
    \centering
     \vspace{-0.25cm}
       \captionsetup[subfloat]{labelformat=empty}
    \captionsetup[sub]{subrefformat=empty}
     \subfloat[]{\includegraphics[width=0.24\textwidth]{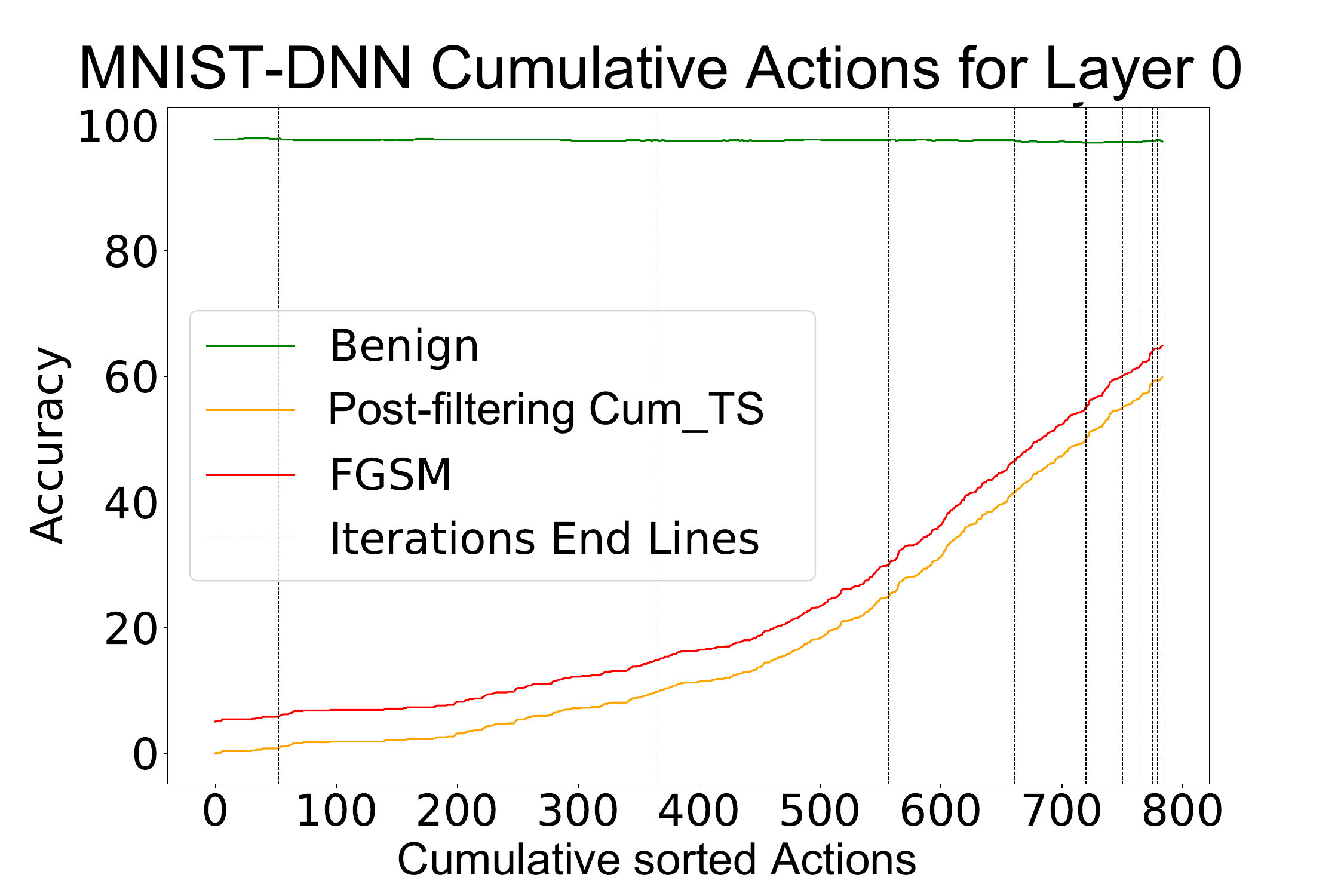}}
       \subfloat[]{\includegraphics[width=0.24\textwidth]{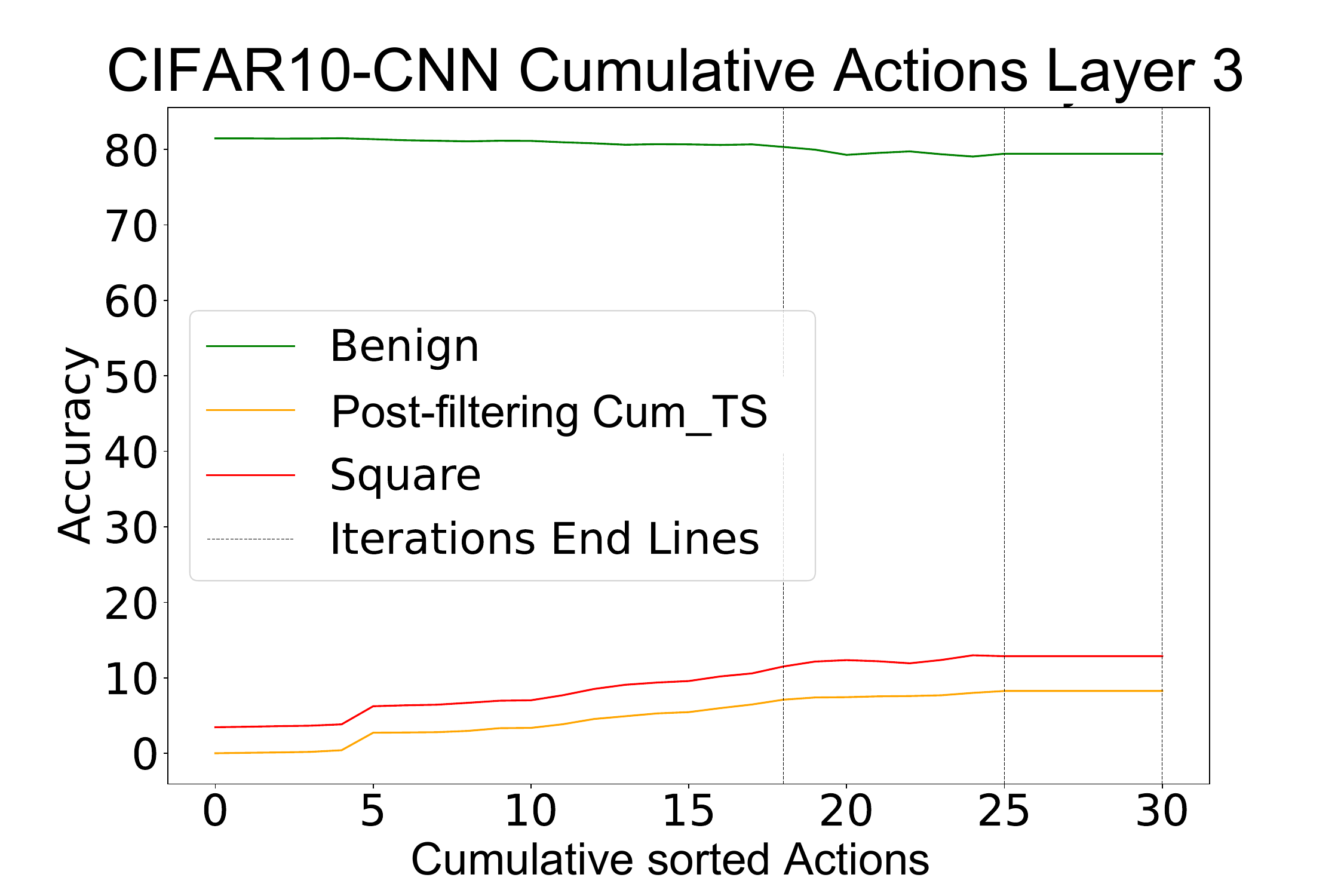}}
     \vspace{-0.5cm}
\\
          \subfloat[]{\includegraphics[width=0.24\textwidth]{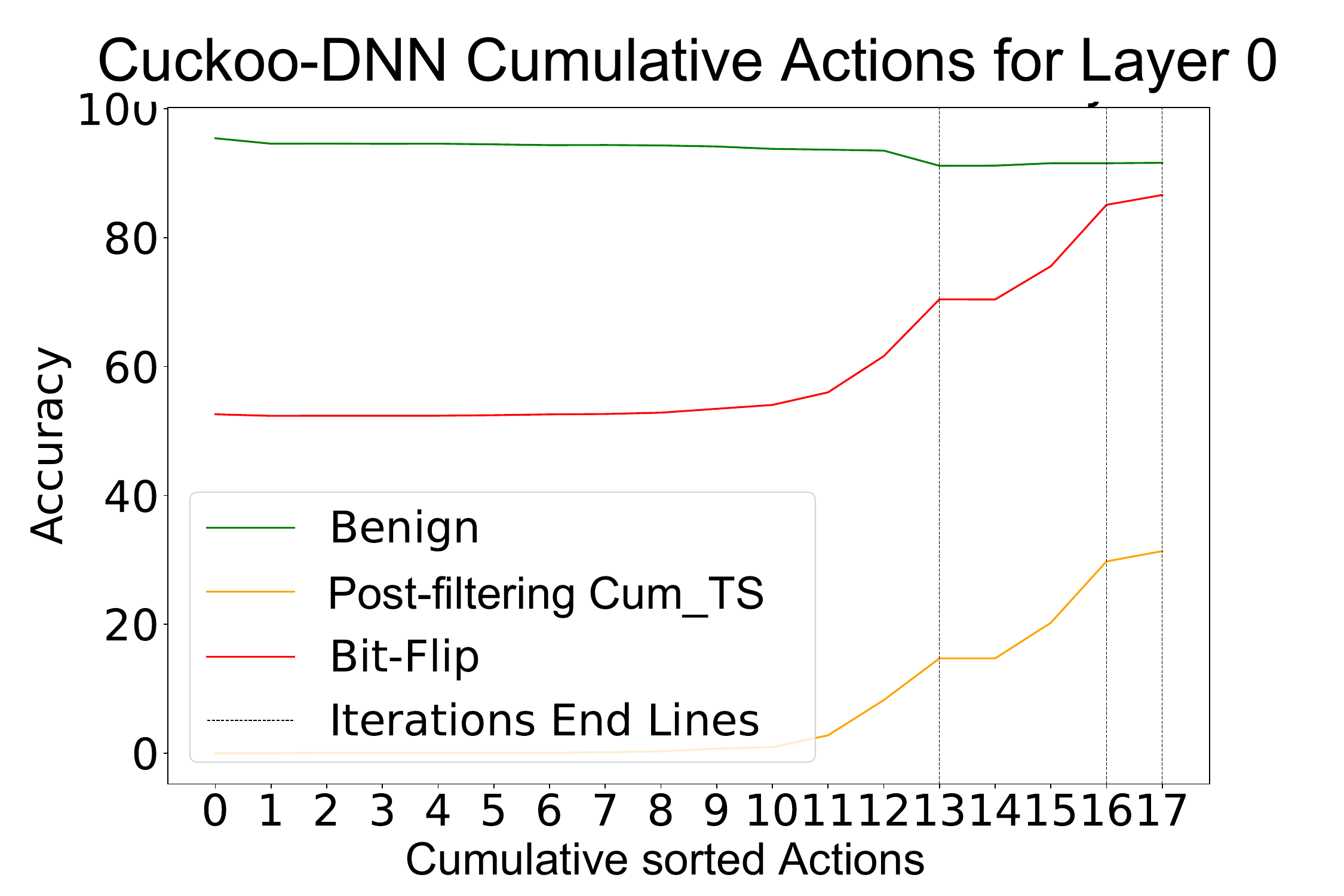}}
     \subfloat[]{\includegraphics[width=0.24\textwidth]{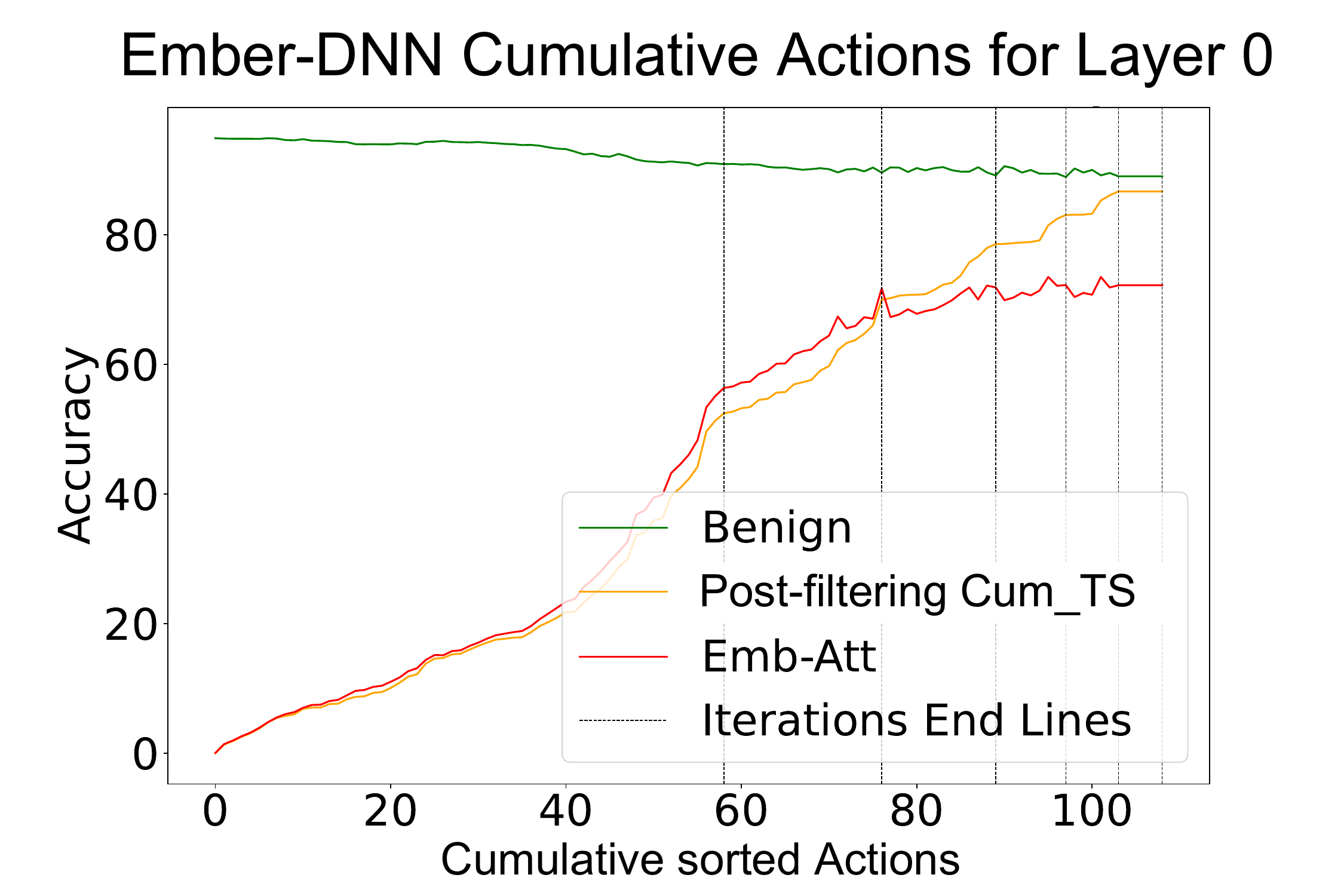}}\\
      \subfloat[]{\includegraphics[width=0.24\textwidth]{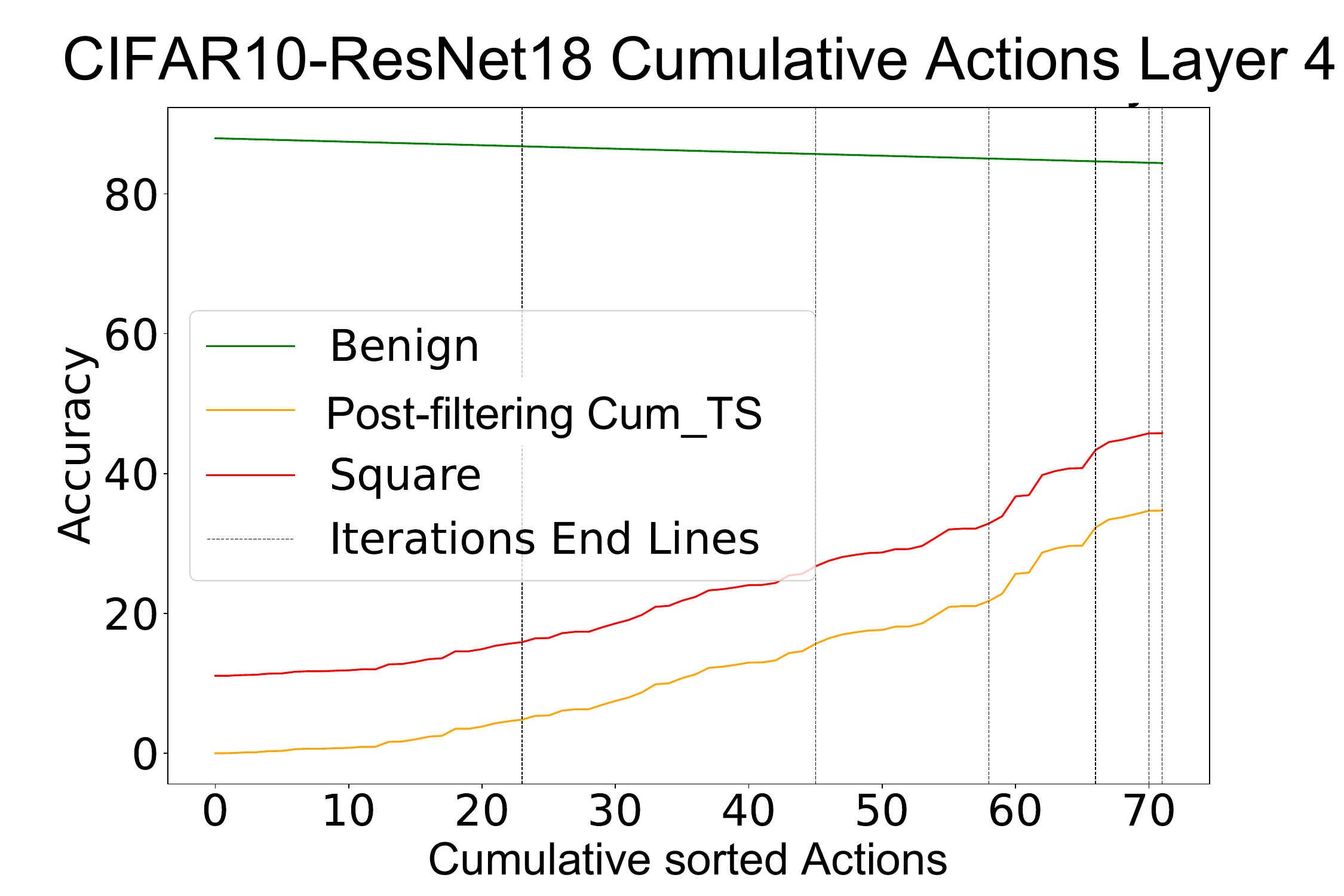}}
     \vspace{-0.75cm}
    
    \caption{Cumulative repair actions per model. Vertical lines denote the end of a step (iteration) in the evaluation methodology (described in \S \ref{subsec:act-eval}). All plots showcase the post-filtering and ordering cumulative actions of each step, for the most performing layer.}
    \label{fig:cum-act}
   
\end{figure}

Figure \ref{fig:cum-act} shows the cumulative actions (x-axis) and their impact on model accuracy (y-axis) after each iteration (vertical lines), post-filtering and ordering. For clarity, we highlight the best-performing layer per model: layer 0 for MNIST-DNN, Cuckoo-DNN, and Ember-DNN; layer 3 for CIFAR10-CNN; and layer 4 (ResNetBlock) for CIFAR10-ResNet18. In all five graphs, accuracy on adversarial data (red lines) steadily improves with cumulative graph-specific actions, while benign accuracy drops slightly ($\le$3\%). Similarly, the trade-off curves (orange lines) increase in tandem with adversarial accuracy, showing that Algorithm \ref{alg:Action-gen} helps \sysname{} identify effective robustness-enhancing actions. A more detailed analysis of other layers appears in Tables \ref{tab:summary} and \ref{tab:summary2}. Next, we evaluate model repair effectiveness across attacks. Additional results on CIFAR10-CNN are shown in Table \ref{tab:cifar10-results}.
\subsection{\textbf{Confidence Parameter $\beta$}}
Node activation distributions in some model layers often have multiple peaks. Selecting the highest peak as $Reference\_Dist\_Agg()$ can alter others, reducing model utility. To address this, $\beta$, computed per node (line 23), is the average absolute difference between activation values across settings (e.g., benign and adversarial). This parameter indicates how far apart the distributions in different settings are and serves as a mask per sample in our actions.$\beta$ indicates distribution separation and masks actions on samples where activation exceeds $\beta$.
\subsection{Sensitivity vs. Commutative Trade-Off}\label{subsec:alpha-vs-tradeoff}
\begin{figure}
    \centering
     \vspace{-0.25cm}
     \subfloat[MNIST]{\includegraphics[width=0.25\textwidth]{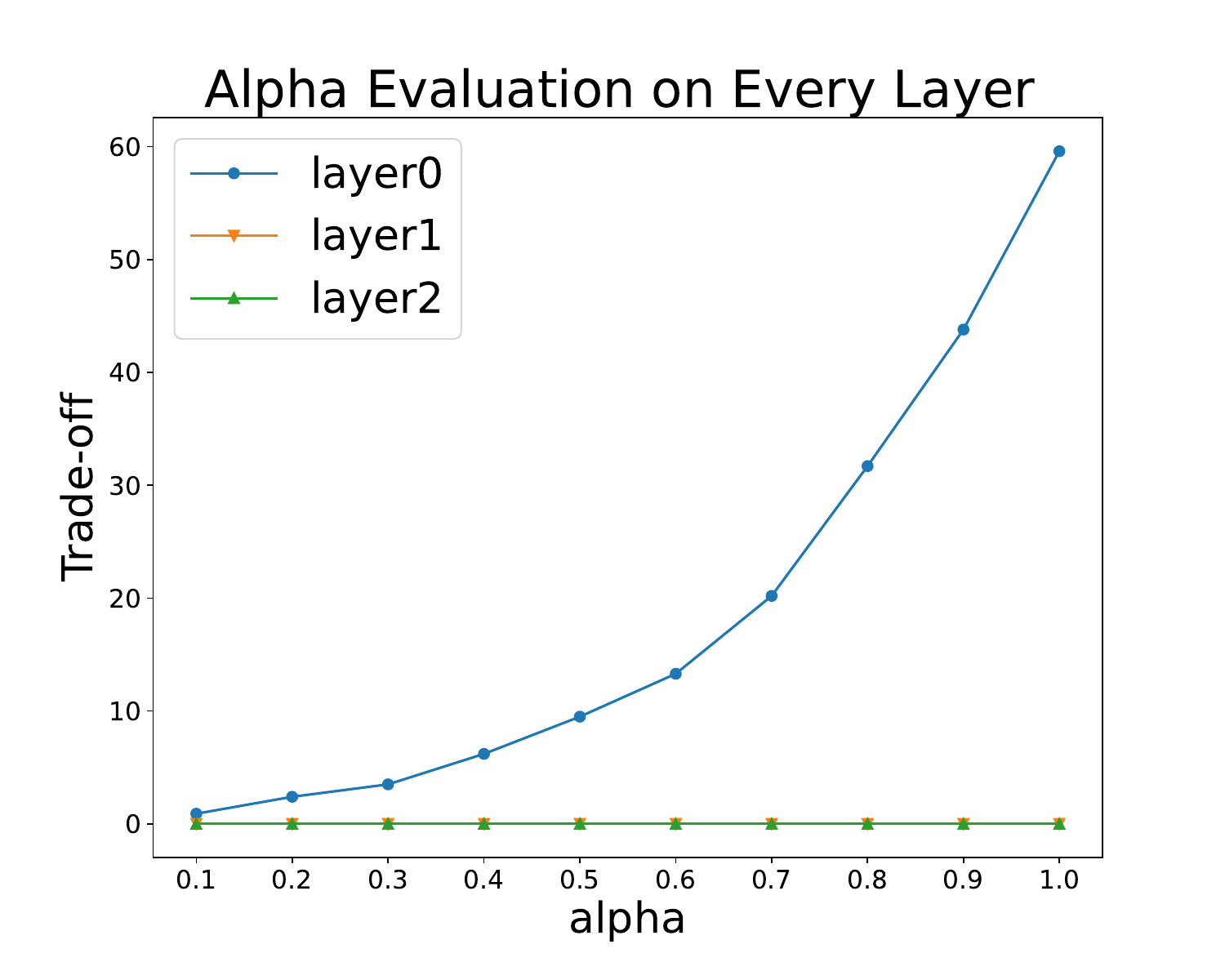}}
     \subfloat[CIFAR-10-CNN]{\includegraphics[width=0.25\textwidth ]{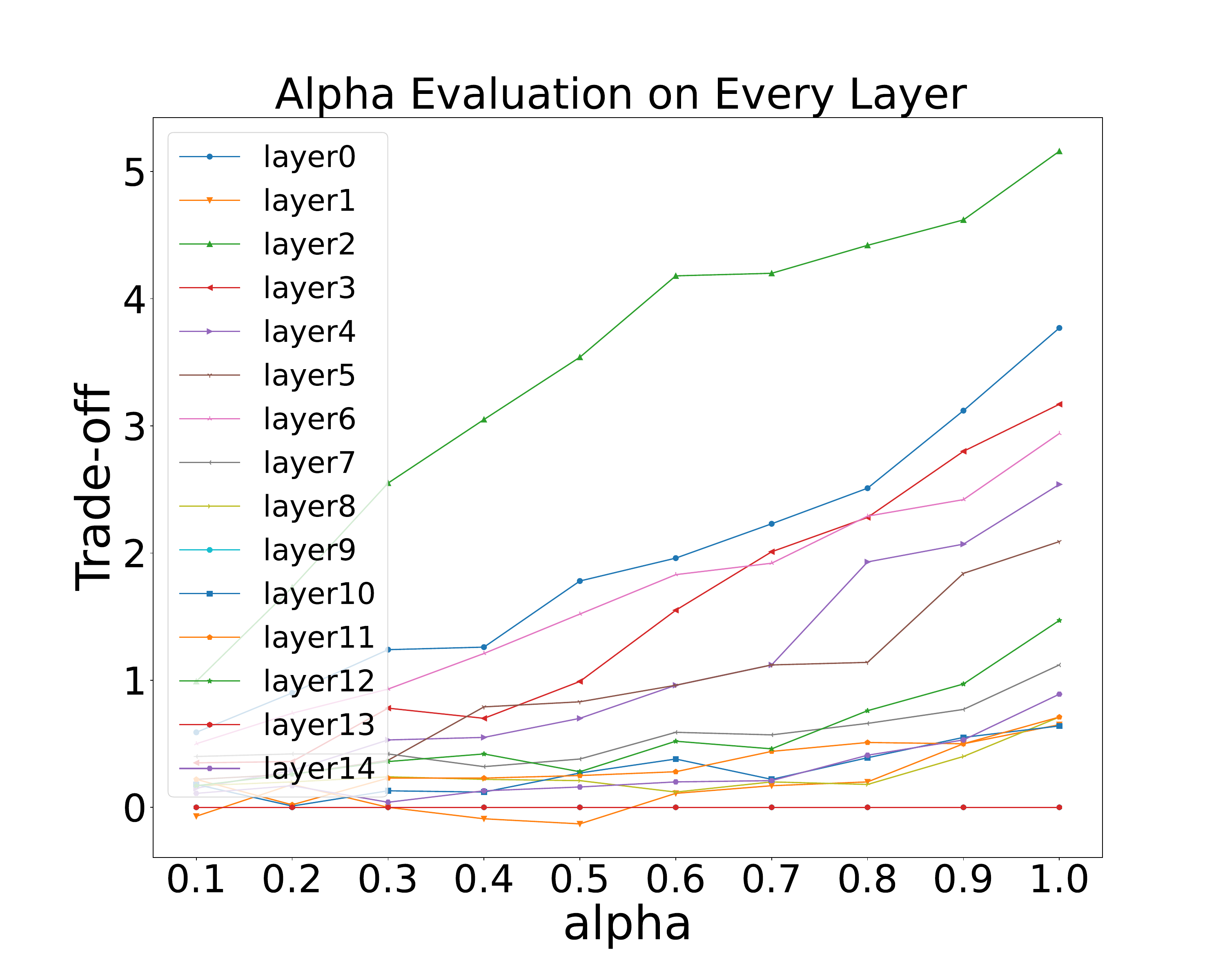}}\\
     \vspace{-0.20cm}
          \subfloat[Cuckoo ]{\includegraphics[width=0.25\textwidth ]{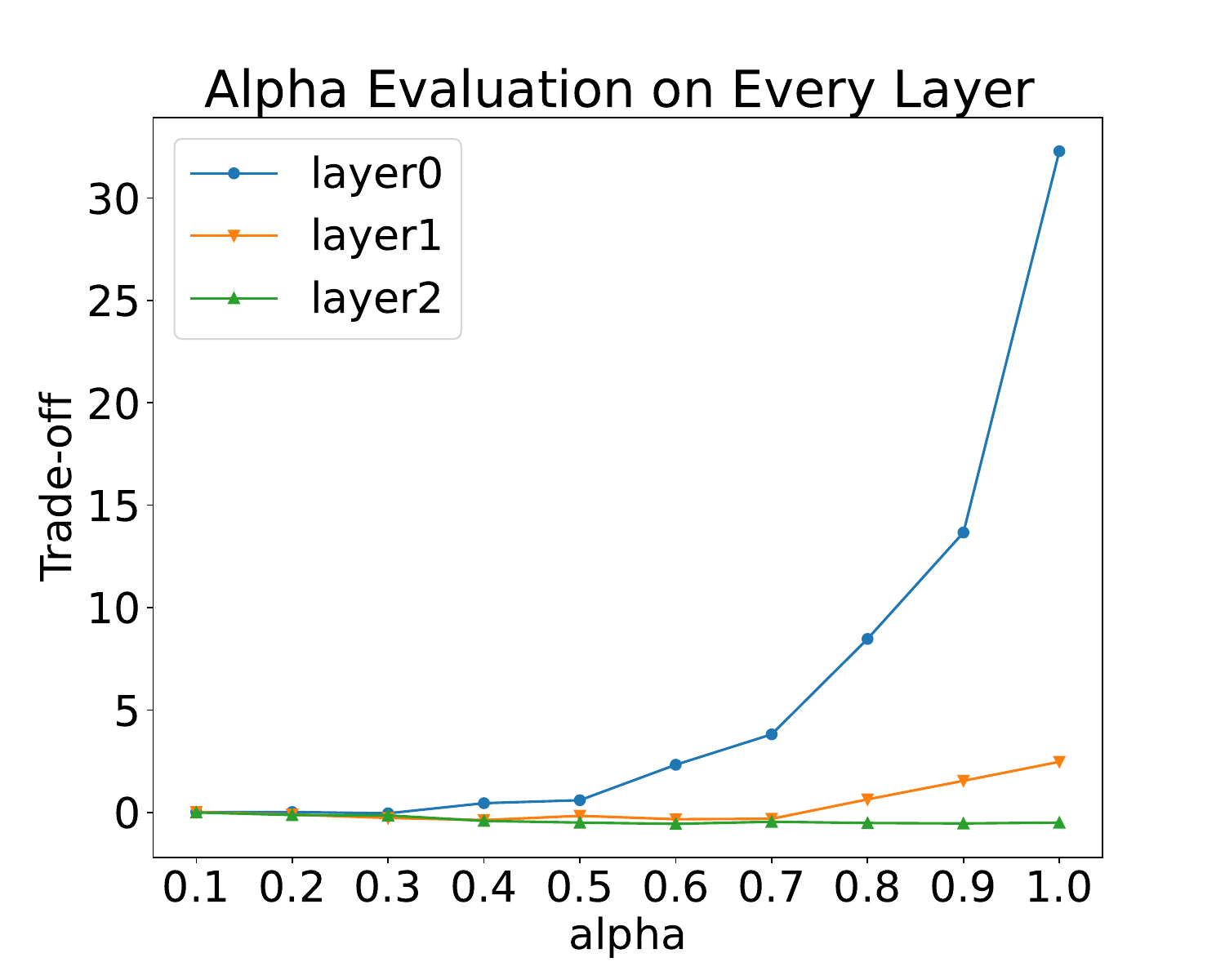}}
     \subfloat[Ember]{\includegraphics[width=0.25\textwidth]{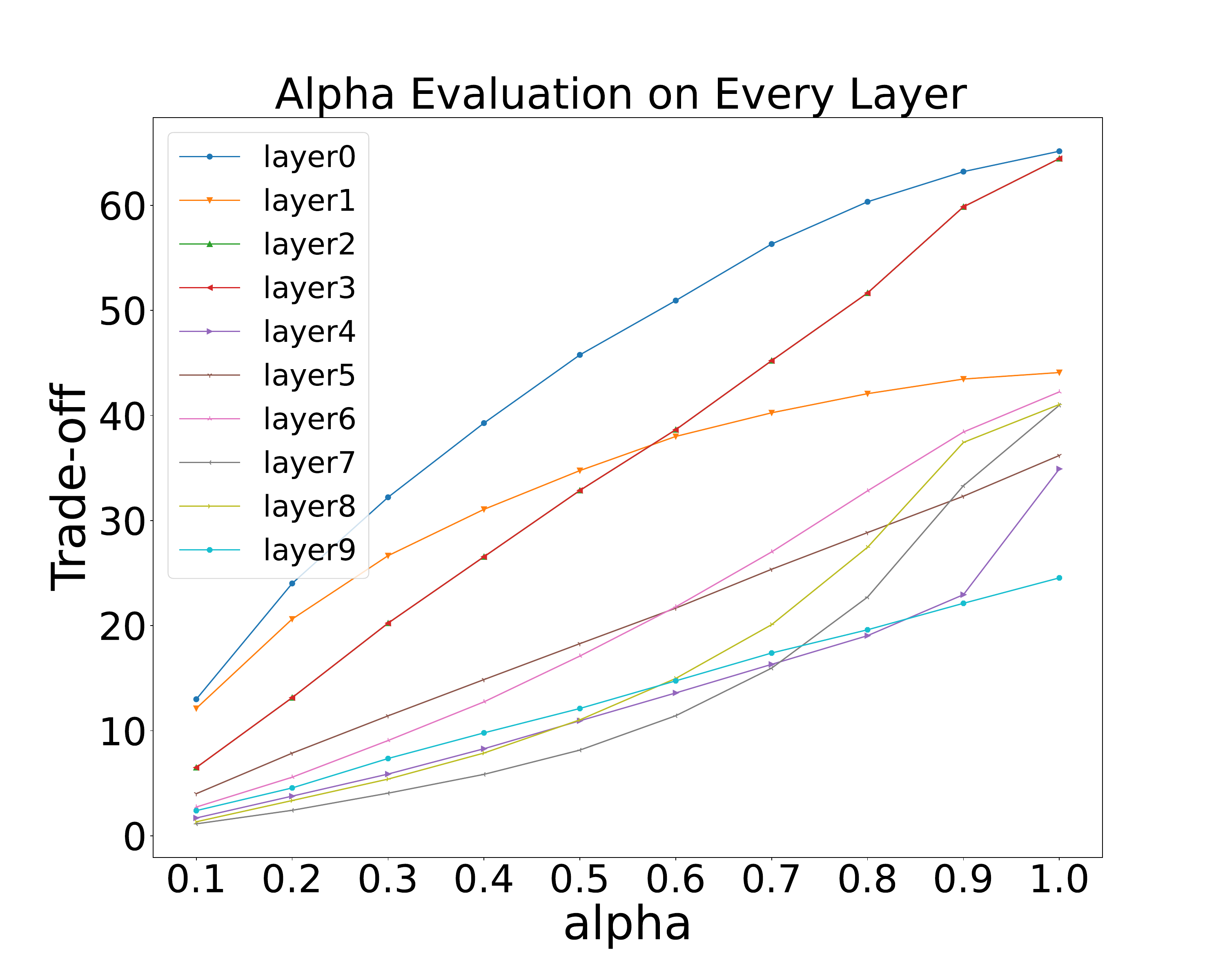}}\\
     \vspace{-0.20cm}
     \subfloat[CIFAR10-ResNet18]{\includegraphics[width=0.25\textwidth]{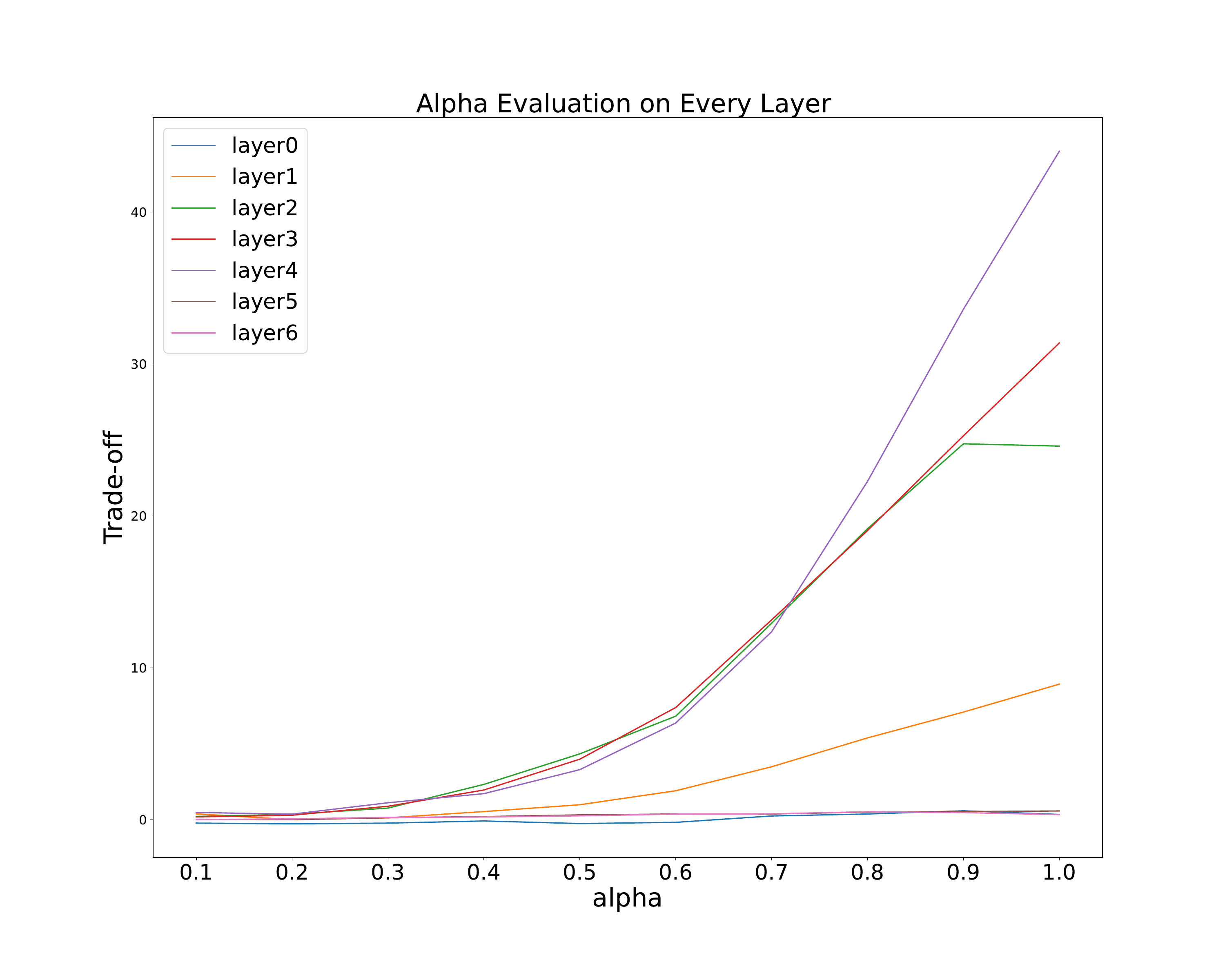}}
         \caption{\small  Action sensitivity ($\alpha$) vs. cumulative trade-off per model.}
    \label{fig:alpha}
    \vspace{-1em}
\end{figure}

Figure \ref{fig:alpha} shows the per-model plots of actions' sensitivity $\alpha$ and cumulative trade-off across repair actions for each layer.

\subsection{Computation Overhead}
We measured the computational overhead incurred by \sysname{} for each dataset and adversarial attack. The computational cost is a one-time expense that remains feasible for model deployers utilizing standard commodity hardware. Our performance overhead analysis revealed that the time incurred by \sysname{} varies depending on the dataset, the complexity of the DNN architecture, and the computational demands of the adversarial attack. Accordingly, the overhead ranged from 2 minutes to 1400 minutes, with an average of 400 minutes.

When keeping the adversarial attack constant while varying the model architecture and dataset, the overhead remained largely consistent. This suggests that the upper bound of 1400 minutes is primarily dictated by the complexity of the adversarial attack rather than the components of \sysname{}. For instance, under identical settings, a less computationally intensive attack required only 30 minutes, highlighting that the majority of the computational burden arises from the attack's complexity rather than our framework.
\begin{table*}[t!] 
\scalebox{0.72}{ 
\begin{tabular}{|l|l|l|l|l|l|l|l|l|l|l|l|l|l|}
\hline
\textbf{\large Model} & \textbf{ \large Layers}  & \multicolumn{6}{c|}{\textbf{\large Accuracy under Attack}}  & \multicolumn{6}{c|}{\begin{tabular}[c]{@{}c@{}}\textbf{\large Accuracy} \textbf{\large on Benign}\end{tabular}} \\ 

\hline

\multirow{7}{*}{ \textbf{ \large MNIST}} &\multicolumn{1}{c|}{ \cellcolor{lightgray} }                                                                                                                                   

& \multicolumn{1}{r|}{\textbf{FGSM}} 
& \multicolumn{1}{r|}{\textbf{PGD}} 
&\multicolumn{1}{r|}{\begin{tabular}[c]{@{}c@{}}\textbf{APGD}\\ \textbf{DLR} \end{tabular}}
& \textbf{SPSA}  
& \textbf{Square}
& \textbf{SIT}

& \multicolumn{1}{r|}{\begin{tabular}[c]{@{}c@{}}\textbf{\sysname{}}\\ \textbf{(FGSM)} \end{tabular}}
&\multicolumn{1}{r|}{\begin{tabular}[c]{@{}c@{}}\textbf{\sysname{}}\\ \textbf{(PGD)} \end{tabular}}
&\multicolumn{1}{r|}{\begin{tabular}[c]{@{}c@{}}\textbf{\sysname{}}\\ \textbf{APGD}\\ \textbf{-DLR} \end{tabular}}
& \multicolumn{1}{r|}{\begin{tabular}[c]{@{}c@{}}\textbf{\sysname{}}\\ \textbf{(SPSA)} \end{tabular}}
& \multicolumn{1}{r|}{\begin{tabular}[c]{@{}c@{}}\textbf{\sysname{}}\\ \textbf{(Square)} \end{tabular}}
& \multicolumn{1}{r|}{\begin{tabular}[c]{@{}c@{}}\textbf{\sysname{}}\\ \textbf{(SIT)} \end{tabular}}
         \\ 

\multirow{10}{*}{ \begin{tabular}[c]{@{}l@{}}\textbf{ \large CIFAR10} \\ \large \textbf{-CNN} \end{tabular}}& \multicolumn{1}{c|}{No Actions}                                                      & \multicolumn{1}{r|}{11.84\%} & \multicolumn{1}{r|}{3.96\%}    & \multicolumn{1}{r|}{2.54\%}         &    \multicolumn{1}{r|}{13.5\% }   &    \multicolumn{1}{r|}{3.45\% }  &    \multicolumn{1}{r|}{0.00\% }  & \multicolumn{1}{r|}{81.47\%} & \multicolumn{1}{r|}{81.47\%} & \multicolumn{1}{r|}{81.47\%} & \multicolumn{1}{r|}{81.47\%}     & \multicolumn{1}{r|}{81.47\%}  & \multicolumn{1}{r|}{81.47\%}  \\

\cline{2-14}

& Layer 0                                                                                                                                                                                                                                    & \multicolumn{1}{r|}{26.52\%} & \multicolumn{1}{r|}{15.36\%}    & \multicolumn{1}{r|}{4.93\%}         & \multicolumn{1}{r|}{20.87\%}     & \multicolumn{1}{r|}{10.88\%}   & \multicolumn{1}{r|}{10.00\%}    & \multicolumn{1}{r|}{74.51\%} & \multicolumn{1}{r|}{78.93\%}      & \multicolumn{1}{r|}{81.36\%}         &   \multicolumn{1}{r|}{80.62 \%}             &   \multicolumn{1}{r|}{76.83 \%}          &   \multicolumn{1}{r|}{80.12 \%}                            \\ 
&Layer 1                                                                                                                                                                                                                                                                                            & \multicolumn{1}{r|}{27.45\%} & \multicolumn{1}{r|}{20.34\%}    & \multicolumn{1}{r|}{12.53\%}         & \multicolumn{1}{r|}{\cellcolor{Mycolor2}\textbf{21.59\%}}    & \multicolumn{1}{r|}{11.66\%}  & \multicolumn{1}{r|}{10.47\%}     & \multicolumn{1}{r|}{81.48\%} & \multicolumn{1}{r|}{79.16\%}      & \multicolumn{1}{r|}{81.14\%}         &       \multicolumn{1}{r|}{\cellcolor{Mycolor2}\textbf{80.75\%} }            &       \multicolumn{1}{r|}{80.31\% }   &       \multicolumn{1}{r|}{80.67\% }                          \\ 


&Layer 2                                                                                                                                                                                                                                                                                   & \multicolumn{1}{r|}{14.29\%} & \multicolumn{1}{r|}{21.36\%}    & \multicolumn{1}{r|}{4.17\%}         &  \multicolumn{1}{r|}{19.49 \% }  &  \multicolumn{1}{r|}{17.56 \% }  &  \multicolumn{1}{r|}{10.23 \% }   & \multicolumn{1}{r|}{78.99\%} & \multicolumn{1}{r|}{79.09\%}      & \multicolumn{1}{r|}{81.49\%}         &      \multicolumn{1}{r|}{77.86\%}       &      \multicolumn{1}{r|}{80.85\%}        &      \multicolumn{1}{r|}{79.23\%}                                   \\ 
&Layer 3                                                                                                 & \multicolumn{1}{r|}{13.84\%} & \multicolumn{1}{r|}{18.68\%}    & \multicolumn{1}{r|}{3.34\%}         & \multicolumn{1}{r|}{19.28\% }
& \multicolumn{1}{r|}{17.47\% } & \multicolumn{1}{r|}{9.49\% } & \multicolumn{1}{r|}{78.79\%} & \multicolumn{1}{r|}{79.55\%}      & \multicolumn{1}{r|}{81.18\%}         &   \multicolumn{1}{r|}{  79.49\% }         &   \multicolumn{1}{r|}{  77.66\% }              &   \multicolumn{1}{r|}{  78.42\% }                      \\ 
&Layer 4                                                                                                                                                                                                                          & \multicolumn{1}{r|}{16.91\%} & \multicolumn{1}{r|}{17.16\%}    & \multicolumn{1}{r|}{3.23\%}         & \multicolumn{1}{r|}{17.94\%}      & \multicolumn{1}{r|}{16.74\%}   & \multicolumn{1}{r|}{10.01\%}    & \multicolumn{1}{r|}{81.02\%} & \multicolumn{1}{r|}{79.15\%}      & \multicolumn{1}{r|}{81.36\%}         &      \multicolumn{1}{r|}{79.7\% }      &      \multicolumn{1}{r|}{80.74\% }    &      \multicolumn{1}{r|}{79.75\% }                                    \\ 
&Layer 5                                                                                                                                                                                                                                                                 & \multicolumn{1}{r|}{\cellcolor{Mycolor2}\textbf {30.08\%}} & \multicolumn{1}{r|}{\cellcolor{Mycolor2}29.26\%}    & \multicolumn{1}{r|}{\cellcolor{Mycolor2}\textbf{14.64}\%}         &   \multicolumn{1}{r|}{ 18.51\%}  &   \multicolumn{1}{r|}{20.16\%}  &   \multicolumn{1}{r|}{10.00\%}    & \multicolumn{1}{r|}{\cellcolor{Mycolor2}\textbf{80.82}\%} & \multicolumn{1}{r|}{\cellcolor{Mycolor2}\textbf{79.75}\%}      & \multicolumn{1}{r|}{\cellcolor{Mycolor2}\textbf{81.31\%}}         &   \multicolumn{1}{r|}{79.19\%}   &   \multicolumn{1}{r|}{80.45\%}       &   \multicolumn{1}{r|}{79.18\%}                              \\ 
&Layer 6                                                                                                                                                                                                                                                                    & \multicolumn{1}{r|}{13.92\%} & \multicolumn{1}{r|}{4.39\%}    & \multicolumn{1}{r|}{3.49\%}         & \multicolumn{1}{r|}{15.2\%}  & \multicolumn{1}{r|}{12.52\%}  & \multicolumn{1}{r|}{11.37\%}      & \multicolumn{1}{r|}{77.75\%} & \multicolumn{1}{r|}{79.94\%}      & \multicolumn{1}{r|}{81.24\%}         & \multicolumn{1}{r|}{79.29\%}  & \multicolumn{1}{r|}{78.96\%}   & \multicolumn{1}{r|}{79.25\%}                                 \\ 
&Layer 7                                                                                                                                                                                                                                                                                        & \multicolumn{1}{r|}{14.78\%} & \multicolumn{1}{r|}{4.61\%}    & \multicolumn{1}{r|}{3.66\%}      & \multicolumn{1}{r|}{  {17.87}\%}   & \multicolumn{1}{r|}{ \cellcolor{Mycolor2}\textbf {25.25}\%} & \multicolumn{1}{r|}{\cellcolor{Mycolor2}\textbf{12.07}\%}       & \multicolumn{1}{r|}{81.27\%} & \multicolumn{1}{r|}{79.83\%}      & \multicolumn{1}{r|}{81.33\%}         &     \multicolumn{1}{r|}{79.79 \%}   &     \multicolumn{1}{r|}{\cellcolor{Mycolor2} \textbf{81.48 \%}}   &     \multicolumn{1}{r|}{\cellcolor{Mycolor2}\textbf {80.71} \%}                                 \\ 
&Layer 8                                                                                                                                                                                                                                                                      & \multicolumn{1}{r|}{12.86\%} & \multicolumn{1}{r|}{4.05\%}    & \multicolumn{1}{r|}{3.66\%}         & \multicolumn{1}{r|}{14.62 \% }     & \multicolumn{1}{r|}{13.44 \% } & \multicolumn{1}{r|}{9.98 \% }     & \multicolumn{1}{r|}{81.23\%} & \multicolumn{1}{r|}{79.21\%}      & \multicolumn{1}{r|}{81.53\%}         &   \multicolumn{1}{r|}{79.9\%}        &   \multicolumn{1}{r|}{80.37\%}             &   \multicolumn{1}{r|}{78.25\%}                           \\
&Layer 9                                                                                                                                                                                                                                   & \multicolumn{1}{r|}{11.84\%} & \multicolumn{1}{r|}{4.12\%}    & \multicolumn{1}{r|}{2.63\%}         & \multicolumn{1}{r|}{3.45\% }        & \multicolumn{1}{r|}{13.5\% }   & \multicolumn{1}{r|}{0.00\% }     & \multicolumn{1}{r|}{ {81.47\%}} & \multicolumn{1}{r|}{ {81.47\%}}      & \multicolumn{1}{r|}{ {81.47\%}}         &    \multicolumn{1}{r|}{  {81.47\%}}    &    \multicolumn{1}{r|}{  {81.47\%}}                   &    \multicolumn{1}{r|}{  {81.47\%}}                            \\ 
&Layer 10                                                                                    & \multicolumn{1}{r|}{13.55\%} & \multicolumn{1}{r|}{6.98\%}    & \multicolumn{1}{r|}{3.20\%}         &  \multicolumn{1}{r|}{16.48 \% }   &  \multicolumn{1}{r|}{17.76\% }   &  \multicolumn{1}{r|}{10.12\% }    & \multicolumn{1}{r|}{81.37\%} & \multicolumn{1}{r|}{81.23\% }      & \multicolumn{1}{r|}{81.41\% }         &   \multicolumn{1}{r|}{77.72\%  }     &   \multicolumn{1}{r|}{81.11\%  }     &   \multicolumn{1}{r|}{80.60\%  }                               \\ 
&Layer 11                                                                                                                                                                                                                       & \multicolumn{1}{r|}{12.98\%} & \multicolumn{1}{r|}{6.64\% }    & \multicolumn{1}{r|}{3.2\% }         &  \multicolumn{1}{r|}{17.65\%}     &  \multicolumn{1}{r|}{13.21\%} &  \multicolumn{1}{r|}{10.00\%}    & \multicolumn{1}{r|}{81.28\% } & \multicolumn{1}{r|}{80.13\% }      & \multicolumn{1}{r|}{81.49\% }         &  \multicolumn{1}{r|}{76.24\% }      &  \multicolumn{1}{r|}{80.78\% }   &  \multicolumn{1}{r|}{79.20\% }                                        \\ 
&Layer 12                                                                                 & \multicolumn{1}{r|}{12.78\%}  & \multicolumn{1}{r|}{5.21\% }    & \multicolumn{1}{r|}{5.61\% }         & \multicolumn{1}{r|}{14.69\%}      & \multicolumn{1}{r|}{13.09\%}   & \multicolumn{1}{r|}{10.56\%}       & \multicolumn{1}{r|}{81.4\% }  & \multicolumn{1}{r|}{78.41\% }      & \multicolumn{1}{r|}{81.53\% }         & \multicolumn{1}{r|}{81.18\%}     & \multicolumn{1}{r|}{81.23\%}     & \multicolumn{1}{r|}{80.26\%}                                             \\ 
&Layer 13                                                                         & \multicolumn{1}{r|}{11.84\%} & \multicolumn{1}{r|}{3.96\% }    & \multicolumn{1}{r|}{2.63\% }         & \multicolumn{1}{r|}{13.5\% }  & \multicolumn{1}{r|}{3.45\% }  & \multicolumn{1}{r|}{3.45\% }       & \multicolumn{1}{r|}{81.47\% } & \multicolumn{1}{r|}{81.47\% }      & \multicolumn{1}{r|}{81.47\% }         &     \multicolumn{1}{r|}{81.47\%}   &     \multicolumn{1}{r|}{81.47\%}   &     \multicolumn{1}{r|}{81.47\%}                        \\ 
&Layer 14                                                                                                                                                                                                                                  & \multicolumn{1}{r|}{12.65\% } & \multicolumn{1}{r|}{4.7\% }    & \multicolumn{1}{r|}{6.12\% }         & \multicolumn{1}{r|}{13.88\%}  & \multicolumn{1}{r|}{9.94\%}  & \multicolumn{1}{r|}{11.20\%}       & \multicolumn{1}{r|}{81.29\% } & \multicolumn{1}{r|}{81.05\% }      & \multicolumn{1}{r|}{{81.48\% }}         &     \multicolumn{1}{r|}{81.61\%}         &     \multicolumn{1}{r|}{80.90\%}   &     \multicolumn{1}{r|}{79.23\%}                                 \\ 
\hline
\cline{2-14}
& \begin{tabular}[c]{@{}l@{}}\{2,4,5,7,1,10,\\11,3,8,9,6,0\}\end{tabular}                                                                      & \multicolumn{1}{r|}{\textbf{40.25\% }} & \multicolumn{1}{r|}{\textbf{36.33\% }}    & \multicolumn{1}{r|}{\textbf{20.45\%} }         & \multicolumn{1}{r|}{\textbf{27.18\%}}  & \multicolumn{1}{r|}{\textbf{26.22\%}}   & \multicolumn{1}{r|}{\textbf{17.94\%}}      & \multicolumn{1}{r|}{\textbf{80.84\%} } & \multicolumn{1}{r|}{\textbf{80.23\%} }      & \multicolumn{1}{r|}{\textbf{80.12\% }}         &    \multicolumn{1}{r|}{\textbf{80.21\%}}   &    \multicolumn{1}{r|}{\textbf{81.03 \%}}  &    \multicolumn{1}{r|}{\textbf{80.54 \%}}                                      \\ 
& ARLS \cite{bartoldson2024adversarialrobustnesslimitsscalinglaw}                                                                                                                                                                                                                               & \multicolumn{1}{r|}{29.01\%} & \multicolumn{1}{r|}{11.1\%}    & \multicolumn{1}{r|}{10.1\%}         & \multicolumn{1}{r|}{\cellcolor{lightgray}}    & \multicolumn{1}{r|}{\cellcolor{lightgray}}   & \multicolumn{1}{r|}{\cellcolor{lightgray}}       & \multicolumn{1}{r|}{79.02\%} & \multicolumn{1}{r|}{80.64\%}      & \multicolumn{1}{r|}{80.41\%}         & \multicolumn{1}{r|}{\cellcolor{lightgray}}     & \multicolumn{1}{r|}{\cellcolor{lightgray}}             & \multicolumn{1}{r|}{\cellcolor{lightgray}}                \\ 
\cline{2-14}
 & \begin{tabular}[c]{@{}l@{}}ARLS \cite{bartoldson2024adversarialrobustnesslimitsscalinglaw}   \\ + \sysname{} \end{tabular}                                                                                                                                                                                                                          & \multicolumn{1}{r|}{40.19\%} & \multicolumn{1}{r|}{30.41\%}    & \multicolumn{1}{r|}{19.63\%}         & \multicolumn{1}{r|}{\cellcolor{lightgray}}    & \multicolumn{1}{r|}{\cellcolor{lightgray}}    & \multicolumn{1}{r|}{\cellcolor{lightgray}}      & \multicolumn{1}{r|}{80.13\%} & \multicolumn{1}{r|}{80.32\%}      & \multicolumn{1}{r|}{80.71\%}         & \multicolumn{1}{r|}{\cellcolor{lightgray}}     & \multicolumn{1}{r|}{\cellcolor{lightgray}}               & \multicolumn{1}{r|}{\cellcolor{lightgray}}       
       \\
\hline

 \end{tabular}}

\caption{CIFAR10-CNN results}
\label{tab:cifar10-results}

\end{table*}

\subsection{Models Architecture}\label{subsec:modelarchi}
Table \ref{tab:model-arch} shows the details of model architecture for MNIST, CIFAR10-CNN, and CIFAR10-ResNet18 models used in our case studies.

\begin{table}[htbp!]
\scalebox{0.75}{
    \centering
    \begin{tabular}{|c|c|c|c|}
        \hline
        \textbf{Model} & \textbf{Layers} & \textbf{Layer Type} (\# Neurons) & \textbf{Activation Function} \\
        \hline
        \multirow{4}{*}{MNIST} 
        & Layer 0 & flatten (784) & - \\
        & Layer 1 & Dense (350) & Relu \\
        & Layer 2 & Dense (50) & Relu \\
   
        \hline
        \multirow{16}{*}{CIFAR10-CNN} 
        & Layer 0 & Conv2d (64) & - \\
        & Layer 1 & BatchNorm2d (64) & Relu \\
        & Layer 2 & Conv2d (64) & - \\
        & Layer 3 & BatchNorm2d (64) & Relu \\
        & Layer 4 & MaxPool2d (64) & - \\
        & Layer 5 & Dropout (64) & - \\
        & Layer 6 & Conv2d (128) & - \\
        & Layer 7 & BatchNorm2d (128) & Relu \\
        & Layer 8 & Conv2d (128) & - \\
        & Layer 9 & BatchNorm2d (128) & Relu \\
        & Layer 10 & MaxPool2d (128) & - \\
        & Layer 11 & Dropout (128) & - \\
        & Layer 12 & Flatten (1152) & - \\
        & Layer 13 & Linear (1024) & Relu \\
        & Layer 14 & Linear (1024) & - \\
 
        \hline
        \multirow{8}{*}{CIFAR10-ResNet18} 
        & Layer 0 & Conv2d (64) & - \\
        & Layer 1 & BatchNorm2d (64) & Relu \\
        & Layer 2 & ResnetBlock (64) & - \\
        & Layer 3 & ResnetBlock (128) & - \\
        & Layer 4 & ResnetBlock (256) & - \\
        & Layer 5 & ResnetBlock (512) & - \\
        & Layer 6 & AvgPool (512) & - \\
   
        \hline
    \end{tabular}}
    \caption{Models architectures used.}
    \label{tab:model-arch}
\end{table}

    \subsection{Datasets}\label{subsec: datasets}
    We use four datasets from two domains. From the malware classification domain, we use two complementary datasets, one based on static analysis, and the other on execution behavioral analysis. From image classification, we use the handwritten digits recognition dataset MNIST for a proof of concept and CIFAR10 for a fairly complicated model architecture. We selected these two as representative domains because (a) malware detection is a naturally adversarial domain where adversarial robustness to adversarial examples is expected and (b) image classification has been heavily used in recent adversarial ML literature \cite{Carlini-list}. We describe these datasets next.
    
    \textbf{CuckooTraces.} Cuckoo-Traces  have been introduced in a previous work for malware detection \cite{cuckoo-data}. It includes 40K Windows PE files with 50\% malware (collected from VirusShare~\cite{virusshare}) and the other 50\% benign PEs (collected from a public goodware site~\cite{cnet}). We used 70\% of the dataset as a training set for the target black-box model, and the remaining 30\% as evasion test. Each sample is represented as a binary feature vector. Each feature indicates the presence/absence of behavioral features captured up on execution of each PE in the Cuckoo Sandbox \cite{cuckoo}. Behavioral analysis of 40K PEs resulted in 1549 features, of which 80 are API calls, 559 are I/O system files, and 910 are loaded DLLs.
    
    \textbf{EMBER.} To assess \sysname{} on complementary (static analysis-based) malware dataset, we use EMBER~\cite{EMBER2018}, a benchmark dataset of malware and benign PEs released with a trained LightGBM with 97.3\% test accuracy. EMBER consists of 2351 features extracted from 1M PEs using a static binary analysis tool LIEF~\cite{lief}. The training set contains 800K samples composed of 600K labeled samples with 50\% split between benign and malicious PEs and 200K unlabeled samples, while the test set consists of 200K samples, again with the same ratio of label split. VirusTotal~\cite{virustotal} was used to label all the samples. The feature groups include: PE metadata, header information, byte histogram, byte-entropy histogram, string information, section information, and imported/exported functions.
    
    \textbf{MNIST~\cite{MNIST}}. This dataset comprises of 60K training and 10K test images of handwritten digits. The classification task is to identify the digit corresponding to each image. Each 28x28 gray-scale sample is encoded as a vector of pixel intensities in the interval [0, 1].
    
    \textbf{CIFAR10~\cite{CIFAR10}.} It consists of 60K 32x32 color images divided into 10 classes, with 6,000 images per class. The classes represent airplanes, automobiles, birds, cats, deer, dogs, frogs, horses, ships, and trucks. The dataset is split into 50K training images and 10K test images.
    
    
     \subsection{Attacks Used in Case Studies}
     \label{subsec:attacks}
     \textbf{Fast Gradient Sign Method \cite{FGSM}}. This attack creates an adversarial example quickly and in one step. The activation grows by $\boldsymbol{\theta}^{\top}\boldsymbol{x'} = \boldsymbol{\theta}^{\top}\boldsymbol{x} + \boldsymbol{\theta}^{\top}\boldsymbol{\delta}$ due to the adversarial perturbation, when taking into account the dot product of the weight vector $\boldsymbol{\theta}$ and an adversarial example (i.e., $\boldsymbol{x'} = \boldsymbol{x} + \boldsymbol{\delta}$). According to Goodfellow et al.~[20], $\boldsymbol{\delta} = \text{sign}(\boldsymbol{\nabla_{\theta}})$ should be assigned in order to maximise this increase, subject to the maximum perturbation constraint $\|\boldsymbol{\delta}\|_{\infty} < \epsilon$. The ideal perturbation, given a sample $\boldsymbol{x}$, is as follows:
     The formula for $\boldsymbol{\delta}^* = \epsilon \,\operatorname{sign}\!\big(\nabla_{\boldsymbol{x}} J(\boldsymbol{\theta}, \boldsymbol{x}, y_{\mathrm{target}})\big)$.\\
    
     \textbf{Projected Gradient Descent (PGD) \cite{PGD}}. The PGD attack is an iterative algorithm that starts with an input $x$ and an initial perturbation $\delta$, usually zero or a random point within the allowed perturbation range. The steps are as follows:
    
     \begin{enumerate}
         \item Compute the gradient of the loss with respect to the input image: $\nabla_x J(\theta, x + \delta, y_{\text{target}})$.
         \item Update the perturbation: $\delta \leftarrow \delta + \alpha \cdot \text{sign}(\nabla_x J(\theta, x + \delta, y_{\text{target}}))$.
         \item Project the perturbation back onto the $\epsilon$-ball around the original image $x$ to ensure it remains within the specified bound: $\delta \leftarrow \text{Proj}_{\epsilon}(\delta)$.
         \item Repeat the above steps for a specified number of iterations or until convergence.
     \end{enumerate}
    
     The projection step ensures that the perturbation does not exceed the specified $\epsilon$-limit. The update at each iteration $t$ can be formally written as:
    
     \[
     \delta_{t+1} = \text{Proj}_{\epsilon} \left( \delta_t + \alpha \cdot \text{sign}(\nabla_x J(\theta, x + \delta_t, y_{\text{target}})) \right)
     \]\\
    
     \textbf{Auto Projected Gradient Descent (Auto PGD) with DLR Loss Function \cite{APGD-DLR}}. 
     AutoPGD is an adversarial attack algorithm that iteratively adjusts perturbations using a DLR loss function and adaptive step sizes. The DLR loss is defined for targeted attacks as:
     \[
     L_{\text{DLR}}(z, y) = - \frac{z_y - \max\{z_i : i \neq y\}}{z_{\text{max2}} - z_{\text{min2}}}
     \]
     Adaptive perturbation update with step size \( \eta \) is applied as follows:
    
     \[
     \delta_{\text{new}} = \text{Proj}_{\epsilon}(\delta_{\text{old}} + \eta \cdot \text{sign}(\nabla_x L_{\text{DLR}}))
     \]
    
     where \( \text{Proj}_{\epsilon} \) projects the perturbation \( \delta \) onto the \( \epsilon \)-ball around the original input, ensuring it does not exceed the maximum allowed perturbation \( \epsilon \). This approach enhances the search for effective adversarial examples compared to conventional PGD.\\ 
    
     \textbf{SPSA \cite{uesato2018adversarial}}. 
     Simultaneous Perturbation Stochastic Approximation (SPSA) is an attack that adapts gradient-free optimization methods for generating adversarial examples. This approach examines adversarial risk as an indicator of a model's performance against the most challenging inputs. However, calculating the exact adversarial risk is computationally infeasible, so SPSA instead utilizes commonly employed attacks (like those previously discussed) and adversarial evaluation metrics to establish a practical surrogate objective that approximates the true adversarial risk. \\
    
     \textbf{Square \cite{andriushchenko2020square}}. 
     The Square Attack is a black-box adversarial technique that introduces structured noise in the form of squares to generate adversarial examples. It initializes by adding vertical stripes and then proceeds by sampling square-shaped perturbations. The perturbation update is sampled uniformly within the allowed \( \ell_\infty \) norm bound \( \epsilon \), and applied to the image as follows:
    
     \[
     \delta_{\text{new}} = \text{Proj}_{\ell_\infty}(\delta_{\text{old}} + \text{Uniform}(-\epsilon, \epsilon) \cdot h)
     \]
    
     where \( h \) is the side length of the square, and \( \text{Proj}_{\ell_\infty} \) ensures that the perturbation remains within the \( \ell_\infty \) norm constraints. This attack exploits the sensitivity of convolutional networks to high-frequency perturbations.

 \textbf{SIT \cite{wang2023structure}}.
 The Structure Invariant Transformation (SIT) attack improves black-box transferability by applying \emph{local}, block-wise input transformations that preserve the global object structure while greatly increasing view diversity. At each iteration, the image is randomly partitioned into \(s \times s\) blocks; each block independently receives one of several transformations (vertical/horizontal shift or flip, \(180^\circ\) rotation, pixel-value scaling, additive noise, resize+resample, DCT high-frequency masking, or pixel dropout). Let \(X=\{x_i\}_{i=1}^{N}\) be the set of \(N\) SIT-transformed views used for gradient estimation. The update proceeds by averaging gradients over these views and using momentum (e.g., MI-FGSM) within an \(\ell_\infty\) budget:
 \[
 \bar{g}_{t+1} \;=\; \frac{1}{N}\sum_{i=1}^{N} \nabla_x J(x_i, y), 
 \qquad
 g_{t+1} \;=\; \mu\, g_t + \frac{\bar{g}_{t+1}}{\lVert \bar{g}_{t+1}\rVert_1},
 \]
 \[
 x_{t+1} \;=\; \operatorname{Proj}_{\mathcal{B}_\infty(x_0,\epsilon)\cap[0,1]}\!\Bigl(x_{t} + \alpha \cdot \operatorname{sign}(g_{t+1})\Bigr),
 \]
 where \(\mu\) is the momentum decay, \(\alpha=\epsilon/T\) is the step size, and \(\operatorname{Proj}\) enforces both the \(\ell_\infty\) constraint and valid pixel range. By maximizing transformation diversity at the block level without breaking semantics, SIT attains substantially higher black-box success rates on both CNNs and Vision Transformers.

    \subsection{Attacks Parameters}
Table \ref{tab:attack_hyperparameters} shows the details of attack hyperparmaters used for our case studies.

\begin{table}[H]
\centering
\begin{tabular}{ll}
\hline
Attack & Hyper-Parameters \\ \hline
FGSM\cite{FGSM}   
       & $\epsilon = 0.3, \text{norm}= \text{Linf}$ . \\ \hline
PGD  \cite{PGD} & rho =0.01: parameter for decreasing \\
       & the step size $\epsilon = 0.3, \text{norm}= \text{inf},$  \\
       & $\text{maxIter} = 40$ . \\ \hline
APGD-DLR \cite{APGD-DLR}   &  rho =0.01: parameter for decreasing  \\
       & the step size $\epsilon = 0.3, \text{norm}= \text{inf},$ \\
       & $\text{maxIter} = 40$ . loss="dlr",\\ \hline
SPSA \cite{uesato2018adversarial}  & $eps=0.3,nb\_iter=100,$  \\
       & $norm = \text{Linf}$ \\ \hline
Square \cite{andriushchenko2020square}  & $norm=\text{Linf}, eps=0.3,$ \\
       & $n\_queries=100$ . \\ \hline
SIT \cite{wang2023structure}  & $norm=\text{Linf}, eps=0.3,$ \\
       & $n\_copies=100, alpha=0.03, epoch=10$ \\ \hline
\end{tabular}
\caption{Attack Hyper-Parameters.}
\label{tab:attack_hyperparameters}
\end{table}

\end{document}